\documentclass[12pt]{article}
\usepackage{amsmath}
\usepackage{amsfonts}
\usepackage{latexsym}
\usepackage{verbatim}
\usepackage{epsfig}
\usepackage{graphicx}
\usepackage{xcolor}
\usepackage{url}
\usepackage{verbatim}
\usepackage{bbm}
\usepackage{bbold}
\definecolor{darkgreen}{rgb}{0.,0.6,0.2}
\definecolor{dark_purple}{rgb}{102, 0, 102}
\definecolor{aqua}{rgb}{0.0, 1.0, 1.0}

\begin{document}

\newcounter{INDEX}
\setcounter{section}{0}
\setcounter{subsection}{0}
\setcounter{equation}{0}
%

\newcommand{\C}{{\mathbb C}}
\newcommand{\R}{{\mathbb R}}
\newcommand{\Z}{{\mathbb Z}}
\newcommand\be{\begin{equation}}                                                                                         

\newcommand{\klaus}{\textcolor{magenta}}
\newcommand{\nonklaus}{\textcolor{blue}}

\title{Elaboration on the kinetic approach of Derbenev and Kondratenko to spin-polarized beams in electron storage rings}

\author{
  K. Heinemann$^a$, D.T. Abell$^b$, J. Agudelo$^a$, D.P. Barber$^a$, O. Beznosov$^d$,\\
  J.P. Devlin$^e$, J.A. Ellison$^a$,
  E. Hamwi$^e$, B. Nash$^f$, M. Vogt$^c$\\
$^a$ \small{University of New Mexico, New Mexico, USA}\\
$^b$ \small{University of Maryland College Park, Maryland, USA}\\
$^c$ \small{Deutsches~Elektronen--Synchrotron (DESY), Germany}\\
$^d$ \small{Los Alamos National Laboratory (LANL), New Mexico, USA}\\
$^e$ \small{Cornell University, New York, USA}\\ 
$^f$  \small{Emily Griffith Technical College, Colorado, USA}}


\date{\today}

\maketitle

\tableofcontents

\begin{abstract}

We present a detailed account of the kinetic approach for describing the 
effect of synchrotron radiation on electron and positron spin polarization in storage rings. 
This approach was introduced in 1974 by Derbenev and Kondratenko and was 
extended by us since 2001. The kinetic approach is much less frequently utilized but it is more
general than the original non-kinetic approach of Derbenev and Kondratenko from 1972 
since the kinetic approach is not centered on the invariant spin field. As with the non-kinetic approach
the kinetic approach covers the radiative depolarization effect, the Sokolov-Ternov effect 
and its Baier-Katkov correction as well as the kinetic polarization effect but it enables the 
calculation of corrections to the original Derbenev-Kondratenko formulas and thereby 
provides estimates of the reliability of the latter. 
The kinetic approach is applicable to storage rings with energies 
from a few GeV up to the energies of the FCC-ee and CEPC and beyond. The kinetic approach 
is based on the spin-orbit Wigner functions which lead to the so-called Bloch equation
for the polarization density which is a generalization of Fokker-Planck equations to spin motion.
In turn, as discovered in 2019, the Bloch equation is based on stochastic ordinary differential 
equations which can be used to develop Monte-Carlo spin tracking codes covering the key
effects beyond the radiative depolarization effect. These 
stochastic ordinary differential equations lead to a new viewpoint of the physical effects, in particular
the kinetic polarization effect.

\end{abstract}

\section{Introduction}

\subsection{Generalities}

We first make some general remarks on the dynamics of spin-polarized beams.
Spin dynamics for electrons and positrons in the electric and magnetic fields in particle accelerators and storage rings is influenced
by two main effects, namely spin precession following the
Thomas-Bargmann-Michel-Telegdi (T-BMT)
equation (see the second equation in (\ref{eq:new2.44vbnewnewred}))
and polarization build-up during spin flipping by the emission of synchrotron radiation 
(the Sokolov-Ternov effect) \cite{BR23}.
Moreover,  the stochastic emission of the photons comprising the 
synchrotron radiation causes particle recoil and thereby injects noise
into the otherwise deterministic motion.  This injects noise via the particle momentum
and the electric and magnetic fields in the T-BMT precession vector field (see (\ref{eq:new2.99anew}))
and thereby noise into the spin motion. 
One calls this spin diffusion or radiative depolarization
which, if left unchecked, causes the beam to be depolarized.
However, the Sokolov-Ternov effect counteracts the radiative depolarization effect and may even lead, 
after a sufficiently large but practically feasible time, to
a polarization close to the
equilibrium polarization. 

Among the accelerators we have in mind in this work 
are the electron resp. positron
storage rings in the future machines, the FCC-ee at CERN, the EIC at Brookhaven National Lab and the
CEPC.

Analytical calculations of the equilibrium polarization 
usually rely on the formalism of Derbenev and Kondratenko exhibited
in \cite{DK72,DK73,Man87-1,Man87-2}.
That approach exploits use of parameters applicable to electron and positron storage 
rings that have been in use so far and it provides a framework 
for understanding how to maximize the attainable polarization \cite{BR23}. 
However with the pioneering, \cite{DK72}, it was already clear that
this approach has the potential
drawback of relying on the assumption that on average the spin vectors of the particles in a bunch
are aligned along the so-called invariant spin field 
(see related remarks at the end of Section 4 where also the concept of radiative invariant
spin field is outlined).

\subsection{Outline of the kinetic approach}

We now turn to the kinetic approach.
This approach does not rely on the use of the invariant spin field which opens the possibility of exploring new parameter regimes.
The kinetic approach has been suggested by Derbenev and Kondratenko in 1974, in Russian
(the English translation is \cite{DK75}),
where they consider the orbital density (= density in orbital phase space), $\rho[W]$, and the polarization density, $\vec{\cal P}[W]$, with their time evolution in the presence
of synchrotron radiation.
The present work, which is an outgrowth of discussions between the authors, is an exposition 
of the kinetic approach with the aim of making this much less practiced
but more general approach 
better known and more transparent.

The dynamics of the kinetic approach is described in \cite{DK75} by a
linear PDE system consisting of the orbital Fokker-Planck equation for the $\R$-valued
orbital density $\rho[W]$ (see (\ref{eq:new02.010new0})) 
and an equation of the Fokker-Planck
type for the $\R^3$-valued polarization density $\vec{\cal P}[W]$, namely
the so-called full Bloch equation (see (\ref{eq:new02.010new20})).
Thus Sections 3 and 4 contain all the dynamical information given to us from \cite{DK75}.
The above $[W]$-notation is justified by the fact that
$\rho[W],\vec{\cal P}[W]$ are the building blocks of the so-called spin-$1/2$ Wigner function
(sometimes called the spin-orbit Wigner function),
$W$, see (\ref{eq:new02.10}), whereby the description of a bunch 
in terms of $\rho[W],\vec{\cal P}[W]$ is equivalent to a
description in terms of $W$.

As is shown in Section 5.1, the PDE system for $\rho[W],\vec{\cal P}[W]$ is equivalent to
a linear PDE system (see (\ref{eq:new002.100})) for  the $\C^{2\times 2}$-valued function
$W$ reflecting the fact that the description in terms of
$\rho[W],\vec{\cal P}[W]$ is equivalent to the perhaps more elegant (as evidenced by
Section 5)
description in terms of $W$ (both descriptions are defined in Sections 3-5).
Note that, (\ref{eq:new02.010new0}), (\ref{eq:new02.010new20}) 
(but not (\ref{eq:new002.100})) were presented
in \cite{DK75} and that 
neither  (\ref{eq:new02.010new0}), (\ref{eq:new02.010new20}) nor (\ref{eq:new002.100})
were derived in \cite{DK75} (and they are not derived here as well).
However (\ref{eq:new02.010new0}), (\ref{eq:new02.010new20}), (\ref{eq:new002.100})
were derived, from QED, in Kondratenko's thesis \cite{Kon82}
which underlies \cite{DK75} (QED =Quantum Electrodynamics).
For more remarks on \cite{Kon82}, see Section 5.5.
Also in Section 5.2 we will characterize the physically meaningful
$\rho[W],\vec{\cal P}[W]$ and $W$ by imposing, in addition to the evolution equations, 
so-called statistical conditions. These statistical conditions
are convenient mathematical abstractions (used in Sections 5,6,12-16)
of what a realistic bunch is.

Following \cite{DK75} 
we use in this work cartesian coordinates, i.e., the so-called laboratory frame.
Moreover we use the same units as in \cite{DK75} 
which one may interprete as Gaussian cgs 
units with $c=1$ (for remarks on Gaussian cgs units without
$c=1$, see the discussions around 
(\ref{eq:new2.44vcdmnewcgs}), (\ref{eq:new2.44gcgsOL})).
By dealing with a kinetic approach we have
$W=W(t,q)$ and $\rho[W]=\rho[W](t,q),\vec{\cal P}[W]=\vec{\cal P}[W](t,q)$ where $t\in\R$ is time and 
$q= \left( \begin{array}{c}  
\vec{r} \\  
\vec{\pi}  \end{array}\right)\in\R^6$ with $\vec{r}\in\R^3$ being the position vector and 
$\vec{\pi}\in\R^3$ being the momentum vector of the particle.
In the present work, $\vec{r},\vec{\pi}$ and $q$, as all vectors, are column vectors (thus
$\vec{r}^T,\vec{\pi}^T,q^T$ are row vectors, where $T$ denotes the operation of transposition).
The polarization vector of the whole bunch (at time, $t$) is given by
\begin{eqnarray}                                                                
&& \hspace{-10mm}\vec{P}[W](t)=\int_{\R^6} \; \vec{\cal P}[W](t,q)  d^6q  \; ,
\nonumber 
\end{eqnarray}
and the polarization by $|\vec{P}[W](t)|$ where $|\cdot|$ is the Euclidean norm.
See (\ref{eq:new02.14}) too. 
Note that in contrast to the kinetic approach, in a fluid approach the focus would 
be on the position space, $\R^3$, not on the orbital 
phase space, $\R^6$.

For numerical computations one generally does not work
in the cartesian coordinates but in machine coordinates., i.e., in a so-called beam frame. 
Thus the equations of the present work are not directly used when it comes to numerical work.
However there is a straightforward and time-tested method
to transform every equation from cartesian coordinates to machine coordinates. We will come back to this
point in Section 10 (see the discussion after (\ref{eq:new2.44vcdm})).

The kinetic approach encapsulates, as the aforementioned non-kinetic approach, 
the radiative depolarization effect due to spin diffusion,
the Sokolov-Ternov effect (plus the Baier-Katkov correction) as well as the so-called kinetic polarization effect. By design, also a wide range of other effects can be included, e.g.,
resonant depolarization via oscillating magnetic fields or
the weak-strong beam-beam effect.
Note that the terminology `kinetic' has different meanings in the phrases 
`kinetic polarization effect' and `kinetic approach': The phrase
`kinetic approach' does not allude to the kinetic polarization effect but to the fact 
that $q$ is a phase space vector (recall that the non-kinetic and kinetic approaches cover
the kinetic polarization effect).

Sometimes it is convenient to write the polarization density as $\rho[W]\vec {\cal P}_{loc}[W]$
where $\vec{\cal P}_{loc}[W]$ is the so-called local polarization vector field.
From the definition of  $\vec{\cal P}_{loc}[W]$, it is 
straightforward to show that the polarization density is proportional to the 
density of spin angular momentum in phase space. This, in turn, 
is what permits the polarization density to obey an equation of the Fokker-Planck type, namely the 
aforementioned full Bloch equation.
For more details on the local polarization vector field, see Section 6.

Whenever it is illuminating, e.g., in Sections 11 and 14,
we will comment on the so-called reduced setup
(see (\ref{eq:new02.010new30a}) resp. (\ref{eq:new02.010new30b}), (\ref{eq:new02.010new30c})). The reduced setup is important since it is the special case       that suffices for studying the radiative depolarization.
For the evolution equation of $W$ in the
reduced setup, see (\ref{eq:new2.100b}). 
As an aside we note that the description in terms 
of (\ref{eq:new2.100b}) 
is equivalent to the description in terms of the orbital Fokker-Planck equation, (\ref{eq:new02.010new0}),
and the so-called reduced Bloch equation (see (\ref{eq:new2.100c})).
In fact (\ref{eq:new02.010new0}), (\ref{eq:new2.100c}) and (\ref{eq:new2.100b})
are built around
just the orbital Fokker-Planck operator, $l_{orb}$, 
(see (\ref{eq:new02.010a}))      
and the T-BMT precession vector field, 
$\vec{\Omega}_{TBMT}$, reflecting the aforementioned
fact that the radiative depolarization effect results from the interplay between the orbital noise and the T-BMT precession.

\subsection{Outline of the extension of the kinetic approach}

We now outline an extension which was developed since 2001.
While in \cite{DK75} the bunch is described in terms of 
$\rho[W]$ and $\vec{\cal P}[W]$ (and thus implicitly in terms of $W$)
we have developed, starting with \cite{BH01},
an extension of the kinetic approach which reached a first milestone
with \cite{HABBE19}. 
In this extension the bunch is described in terms of the so-called
spin-orbit density, $f=f(t,q,\vec{s})$, whose time evolution is determined by a single Fokker-Planck
equation on the spin-orbit phase space, $\R^9$, called the
full spin-orbit Fokker-Planck equation, see (\ref{eq:new2.44vanewnew}),
which was introduced in \cite{HABBE19}
(for the precise definition of spin-orbit densities, see Section 12).
The description 
in terms of $f$, which is the subject matter of Sections 7-16,
is just an extension, not a modification, of the kinetic approach 
since the dynamical content of (\ref{eq:new2.44vanewnew})
is the same as the dynamical content of the time evolution,  
(\ref{eq:new02.010new0}), (\ref{eq:new02.010new20}), of
$\rho[W],\vec{\cal P}[W]$ resp. $W$.
In fact, using that $f$ uniquely determines $W$ (via (\ref{eq:new2.44r}))
we present (\ref{eq:new2.44vanewnew})
(and derive it from (\ref{eq:new02.010new0}), (\ref{eq:new02.010new20})) in Sections 7-10.
To our knowledge 
no derivation of (\ref{eq:new2.44vanewnew}) was
previously published (not even in \cite{HABBE19})
so that the importance of Sections 7-10 is underscored.
Note that our search of a Fokker-Planck equation for $f$, which started in 2001,
was encouraged by $\rho[W]$ satisfying a Fokker-Planck equation (see Section 3)
and by 
the full Bloch equation for $\vec{\cal P}[W]$ being of Fokker-Planck type (see Section 4).
While these features of  $\rho[W],\vec{\cal P}[W]$ were an encouragement for 
extending the kinetic approach, they were not a guarantee (see the discussions after
(\ref{eq:new2.44vcdc0new}), (\ref{eq:new2.44vanewnew})).

The extension of the kinetic approach involves 
not only the full spin-orbit Fokker-Planck equation.
In fact since (\ref{eq:new2.44vanewnew})
is a Fokker-Planck equation it has an associated Ito
SDE system (see (\ref{eq:new2.44vbnewnew}))
which has the same dynamical content as
(\ref{eq:new2.44vanewnew})
and which we call the full spin-orbit SDE system (SDE=stochastic ODE).
The interesting aspect of (\ref{eq:new2.44vbnewnew}) is 
being neither a PDE nor a system of PDEs and
so is potentially of great practical interest since (\ref{eq:new2.44vbnewnew}) may be 
numerically solved, e.g., by a standard SDE solver or
by the time-tested technique of particle tracking.
In fact this aspect was our key reason to become interested in the extension of the kinetic
approach via the description in terms of $f$. In other words it was the search for
the full spin-orbit SDE system, and not so much the search for the
full spin-orbit Fokker-Planck equation,
which captured our imagination since 2001.

One possible approach of numerically solving
(\ref{eq:new2.44vbnewnew}) is to
extend an existing Monte-Carlo 
spin tracking code (developed for computing the so-called radiative
depolarization time, e.g., a code from Bmad) to (\ref{eq:new2.44vbnewnew})
(Bmad is a software library at Cornell University).
The existing Monte-Carlo 
spin tracking codes we are aware of are not 
modeled after (\ref{eq:new2.44vbnewnew}) but
after the, simpler, reduced spin-orbit SDE system (see (\ref{eq:new2.44vbnewnewred})).
Note that (\ref{eq:new2.44vbnewnewred}) is obtained by applying the reduced setup to
(\ref{eq:new2.44vbnewnew}) and that (\ref{eq:new2.44vbnewnewred}) is
an Ito SDE system associated with the reduced spin-orbit Fokker-Planck equation
(see (\ref{eq:new2.44vanewnewredfinal})) the latter being obtained  
by applying the reduced setup to (\ref{eq:new2.44vanewnew}).
Note also that (\ref{eq:new2.44vbnewnewred}) (as well as (\ref{eq:new2.44vanewnewredfinal})) 
contains all the information of the
radiative depolarization effect (see Section 11).
For more comments on numerically solving (\ref{eq:new2.44vbnewnew})
see the discussion after (\ref{eq:new2.44vcdm}). 

Although the extension of the kinetic approach does not modify the dynamical
content of the kinetic approach it allows us, in Sections, 13-16,
to share insights into the statistics and dynamics 
of spin-polarized beams in electron and positron storage rings which go
beyond \cite{DK75} (and even beyond \cite{BH01,HABBE19}).
For example, the extension of the kinetic approach leads to a derivation of the 
Baier-Katkov-Strakhovenko equation in Section 15.1
which is very different from the original
derivation in \cite{Bai69} and in \cite{BKS70} 
and thereby leads to new insights, e.g., the so-called
generalized Baier-Katkov-Strakhovenko equation 
(see Section 15.2).
Moreover the extension of the kinetic approach allows us to locate each of 
the aforementioned main physical effects, which are covered
by the kinetic and non-kinetic approaches, to certain terms of the full spin-orbit SDE system
and of the full spin-orbit Fokker-Planck equation (see Section 16).
In particular, and perhaps surprisingly, the kinetic polarization effect resides in the two white-noise
terms of the full spin-orbit SDE system. 
These two terms are the orbital white-noise term,
${\cal B}_{orb}(t,Q)\nu(t)$, and the spin white-noise term, $\vec{\cal B}_{spin}(t,Q)  \nu(t)$ 
(both displayed in the expression, (\ref{eq:new2.44vbnewnew}),
of the full spin-orbit SDE system). Note that the spin white-noise term
is one of the features of the kinetic approach which comes to light only thanks to the
extension of the kinetic approach (in contrast,  the orbital white-noise term
is a feature of the kinetic approach which is evident by
\cite{DK75} since it is the white-noise term in the
Ito SDE system, see (\ref{eq:new2.44uanew}), associated with the orbital Fokker-Planck equation).
Note also, by Section 16, that, among the main physical effects, the kinetic
polarization effect is the only one which depends on the spin white-noise term.

As an aside we note that the emphasis in the present work is on associated Ito SDE systems
(not on associated Stratonovich SDE systems) as indicated by
the full spin-orbit SDE system being an Ito SDE system associated with 
(\ref{eq:new2.44vanewnew}).  For completeness we thus define and discuss around
(\ref{eq:new2.44vbnewnewOL}) a Stratonovich SDE system 
associated with (\ref{eq:new2.44vanewnew}).
A major lesson from (\ref{eq:new2.44vbnewnewOL}) 
is that if one neglects (which is very common in applications)
the quantum radiation force field, $\vec{\cal Q}$, then Ito
and Stratonovich are the same (for the definition of $\vec{\cal Q}$, see Section 3).
Thus in these applications the terminologies `Ito' and `Stratonovich' are superfluous.

This completes the outline of the kinetic approach to the dynamics of spin-polarized
beams in electron and positron storage rings.
We hope that
the present work will lead to progress in (and better
understanding of) this field of Accelerator Physics.

\subsection{The motivation of the kinetic approach}

Let us finally mention the motivation behind the kinetic approach.
One obvious motivation underlying \cite{DK75} was to shed light on the accuracy of
the Derbenev-Kondratenko formulas, e.g., a formula for the equilbrium polarization
(these formulas are the central result of
\cite{DK72,DK73,Man87-1,Man87-2}).
For example, using the evolution equations of $\rho[W]$ and $\vec{\cal P}[W]$,
terms are introduced in
\cite{DK75} which are somehow to be merged
with the Derbenev-Kondratenko formulas, 
aimed at accounting for so-called
uncorrelated resonance crossings,
see, e.g., \cite{Kon74,DKS79}.
Such  terms
are of special interest for circular colliders with very high electron or positron energies 
like the future machines, the FCC-ee and the CEPC (or even the EIC) but their validity and 
use in the
form presented requires clarification.
For a recent critical assessment of 
these  terms in the context of the CEPC,
see \cite{XDBWWG23}.
While the Derbenev-Kondratenko formulas and any
correction terms are not our main concern here, we hope that
the present work will strengthen understanding of these two notions.

Finally, the extension of the kinetic approach has its own motivation namely 
to obtain an SDE system which carries the dynamical content of the kinetic approach (as mentioned
above, this system turned out to be
the full spin-orbit SDE system).
In other words, the motivation behind the extension is drawn from `the world of
Monte-Carlo spin tracking' not from `the world of Derbenev-Kondratenko formulas'.
Historically speaking, the emancipation from the Derbenev-Kondratenko formulas
began around 1984 with J. Kewisch's PhD thesis whose outcome was the first 
Monte-Carlo spin tracking code, SITROS \cite{Kew85}.
In fact with SITROS it became possible to compute the radiative depolarization time
without using the Derbenev-Kondratenko formulas.
We hope that with the  full spin-orbit SDE system this emancipation process has reached a next stage
since it even promises to compute the equilibrium polarization 
without using the Derbenev-Kondratenko formulas (and without using the BKS equation).

\section{The spin-$1/2$ Wigner function: Basic facts}

In this section we give a basic outline of $W$ 
needed for Sections 3 and 4 (in Section 5 we take a closer look at $W$).

A bunch can be described by a so-called spin-$1/2$
Wigner function, $W$, which is a $\C^{2\times 2}$-valued function
that we write, for cartesian coordinates, in the form 
\begin{eqnarray}                                                            
&&  \hspace{-10mm} W(t,q) = \frac{1}{2}\biggl(  I_{2\times 2}\rho[W](t,q)
+ \sigma_i {\cal P}_i[W](t,q)\biggr) \; ,
\label{eq:new02.10}                                                           
\end{eqnarray}
where $I_{2\times 2}$ is the unit $2\times 2$-matrix and 
\begin{eqnarray}                                                            
&&  \vec{\cal P}[W] \equiv  \left( \begin{array}{c}  
{\cal P}_1[W] \\
{\cal P}_2[W]  \\
{\cal P}_3[W] 
\end{array}\right) \; , \quad 
\vec{\sigma} \equiv  \left( \begin{array}{c}  
\sigma_1 \\
\sigma_2 \\
\sigma_3
\end{array}\right) \; ,
\label{eq:new02.10000b}                                                           
\end{eqnarray}
with the so-called Pauli matrices,
$\sigma_1,\sigma_2,\sigma_3$ defined by
\begin{eqnarray}                                                            
&& \hspace{-20mm} \sigma_1:= \left( \begin{array}{cc} 0 & 1 \\
1 & 0  \end{array}\right) \; , \quad \sigma_2:= \left( \begin{array}{cc} 0 & -i \\
i & 0  \end{array}\right) \; , \quad \sigma_3:= \left( \begin{array}{cc} 1 & 0 \\
0 & -1  \end{array}\right) \; .
\label{eq:new02.10P}                                                           
\end{eqnarray}
We use in the present work the summation convention for repeated lower indices, e.g.,
$\sigma_i {\cal P}_i[W]\equiv \sigma_1{\cal P}_1[W]+\sigma_2{\cal P}_2[W]+\sigma_3{\cal P}_3[W]$
and $q_i q_i \equiv q_1 q_1+\cdots+q_6q_6$ (note that (\ref{eq:new02.10}) is our first,
but not last, use of this convention).
Note also that (\ref{eq:new02.10}) is eq. 1 in \cite{HABBE19}, albeit in
different notation.

For later reference we note, by (\ref{eq:new02.10P}) and for $i,j=1,2,3$, that
\begin{eqnarray}                                                            
&&  Tr_{2\times 2}[\sigma_i]=0 \; , \quad
Tr_{2\times 2}[\sigma_i\sigma_j]
=2\delta_{i,j} \; ,
\label{eq:new02.10000a}                                                           
\end{eqnarray}
where $\delta_{i,j}$ is the Kronecker symbol and $Tr_{2\times 2}$ symbolizes
the trace operation over $2\times 2$-matrices.
Note, by (\ref{eq:new02.10}), (\ref{eq:new02.10000a})     
and for $t\in\R,q\in\R^6$, 
\begin{eqnarray}                                                            
&&   \hspace{-20mm} \rho[W](t,q)
 = Tr_{2\times 2}[W(t,q)] \; , \quad 
\vec{\cal P}[W](t,q)=\left( \begin{array}{c} 
Tr_{2\times 2}[\sigma_1 W(t,q)]\\
Tr_{2\times 2}[\sigma_2 W(t,q)]\\
Tr_{2\times 2}[\sigma_3 W(t,q)] 
\end{array}\right) \; .
\label{eq:new02.10a}                                                           
\end{eqnarray}
We will use (\ref{eq:new02.10}), (\ref{eq:new02.10a}) in Section 5 to show that
the description of the bunch in terms of $W$ is equivalent to a description 
in terms of $\rho[W]$ and $\vec{\cal P}[W]$ (the latter description is utilized in \cite{DK75}).

The description of the bunch in terms of $W$, which is the subject matter of
Section 5 below, involves an evolution equation for $W$ (see (\ref{eq:new002.100})) 
and the so-called statistical conditions on $W$ (see
(\ref{eq:new02.10c}), (\ref{eq:new02.10d}), (\ref{eq:new02.10k})) leading to the notion of the
spin-$1/2$ Wigner function of a bunch (or shortly, the notion of physically meaningful $W$).
Roughly speaking, the notion of physically meaningful is a convenient abstraction of the
notion of a realistic bunch without the need
to pinpoint what a realistic bunch exactly is.
In fact in Sections 5,6,12-16 we will see the usefulness of this notion.

To obtain the evolution equation for
$W$ in Section 5.1 we first have to provide, in Sections 3 and 4,
the evolution equations for
$\rho[W]$ and $\vec{\cal P}[W]$ as they are given to us from \cite{DK75}.

\section{The orbital Fokker-Planck equation}

In this section we outline the dynamics of $\rho[W]$ (recall from Section 1.2 that
$\rho[W]$ is $\R$-valued).

The evolution equation for $\rho[W]$ is given to us from \cite{DK75} as the following linear PDE:
\begin{eqnarray}                                                            
&&     \hspace{-10mm}                                                                  
\frac{\partial \rho[W]}{\partial t} 
+ \frac{\partial}{\partial r_j} (v_j \rho[W])
+ \frac{\partial}{\partial \pi_j} ( {\cal F}_j \rho[W])  = St \;\rho[W] \; ,
\label{eq:new02.010new0}                                                           
\end{eqnarray}
where the velocity vector field, $\vec{v}$, the Lorentz force field, $\vec{\cal F}$,  
and the linear operator $St$ will be defined below in this section. 
%

Before we define $\vec{v},\vec{\cal F}$ and $St$ let us make some remarks on 
(\ref{eq:new02.010new0}). First of all, the rhs of (\ref{eq:new02.010new0}) is its radiative part,
see the discussion after (\ref{eq:new02.010new8}).
Secondly, for the theoretical origins of (\ref{eq:new02.010new0}), see the
discussion after (\ref{eq:new02.010new8}).
Thirdly, we call (\ref{eq:new02.010new0}) (and every equation which only differs 
from (\ref{eq:new02.010new0}) in terms of notation)
the orbital Fokker-Planck equation (see also the
discussions after (\ref{eq:new2.44e}) and 
(\ref{eq:new02.010a})).
Fourthly, the orbital Fokker-Planck equation was presented, but 
not derived, in \cite{DK75} (except for some remarks).
Instead it was derived, from QED, in \cite{Kon82} which is a work which underlies \cite{DK75}
(for more remarks on \cite{Kon82} see Section 5.3).
Fifthly, in Section 5.2 we will determine which solutions 
$\rho[W]$ of (\ref{eq:new02.010new0}) we will call the orbital density of a bunch.

We now define $\vec{v},\vec{\cal F}$ and $St$.
The $\R^3$-valued function, $\vec{v}$, is the velocity vector field, i.e.,
\begin{eqnarray}                                                            
&& \hspace{-10mm}
\left( \begin{array}{c} 
v_1(q)\\
v_2(q)\\
v_3(q) 
\end{array}\right) 
\equiv \vec{v}(q):=\frac{\vec{\pi}}{m\gamma(q)} \; , \quad   \gamma(q):=
\sqrt{ \frac{1}{1-|\vec{v}(q)|^2}} \; ,
\label{eq:new2.44b} 
\end{eqnarray}
reflecting the fact that $\vec{\pi}$ is the particle's kinetic momentum vector.
Note that $m$ is the mass of the particle and $|\cdot|$ is the Euclidean norm. 
Note also that $\vec{v}$ determines the time derivative of the position vector, see (\ref{eq:new2.44ubnew})
(which is the same in the radiationless case, namely the first equality of
(\ref{eq:new2.44uaanewnew})).
The $\R^3$-valued function, $\vec{\cal F}$, is the Lorentz force field for 
the particle in the external electromagnetic field, i.e.,
\begin{eqnarray}     
&& \hspace{-10mm}  \left( \begin{array}{c} 
{\cal F}_1(t,q)\\
{\cal F}_2(t,q)\\
{\cal F}_3(t,q) 
\end{array}\right) 
\equiv 
\vec{{\cal F}}(t,q):= e
\biggl(  \vec{E}(t,\vec{r}) +
\vec{v}(q)\times\vec{B}(t,\vec{r})\biggr) \; ,
\label{eq:new2.44c} 
\end{eqnarray}
where $e$ is the charge of the particle and where $\vec{E}$ and $\vec{B}$ are 
the external electric and magnetic fields.
The linear operator $St$ is defined by  
\begin{eqnarray}
&& St:= -\frac{\partial}{\partial \pi_j} {\cal C}_j 
+\frac{1}{2}\frac{\partial}{\partial \pi_i} {\cal E}_{i,j}\frac{\partial}{\partial \pi_j} \; ,
\label{eq:new02.010new1}                                          
\end{eqnarray}
where the $\R^3$-valued function, $\vec{\cal C}$, and
the $\R^{3\times 3}$-valued function
${\cal E}$ are defined by
\begin{eqnarray}
&& \hspace{-15mm} \left( \begin{array}{c} 
{\cal C}_1(t,q)\\
{\cal C}_2(t,q)\\
{\cal C}_3(t,q) 
\end{array}\right) 
\equiv 
 \vec{{\cal C}}(t,q):=-\frac{2}{3}\frac{e^2}{m}
\gamma^3(q)
|\vec{a}_{\cal F}(t,q)|^2\vec{\pi} \; , 
\label{eq:new2.44d}\\
&& \hspace{-15mm} \left( \begin{array}{ccc} 
{\cal E}_{1,1}(t,q) &  {\cal E}_{1,2}(t,q) & {\cal E}_{1,3}(t,q) \\
{\cal E}_{2,1}(t,q) &  {\cal E}_{2,2}(t,q) & {\cal E}_{2,3}(t,q) \\
{\cal E}_{3,1}(t,q) &  {\cal E}_{3,2}(t,q) & {\cal E}_{3,3}(t,q) 
\end{array}\right) 
\equiv {\cal E}(t,q)
\nonumber\\
&& \hspace{-15mm} 
:=\frac{55}{24\sqrt{3}}\lambda(t,q)\vec{\pi}\vec{\pi}^T \; , \quad
\lambda(t,q):=\hbar\frac{e^2}{m^2}\gamma^5(q)
|\vec{a}_{\cal F}(t,q)|^3 \; .
\label{eq:new2.44g} 
\end{eqnarray}
Here $\vec{a}_{\cal F}$ is the acceleration field of 
the nonradiative particle, i.e.,
\begin{eqnarray}
&& \hspace{-15mm} \vec{a}_{\cal F}(t,q)= \frac{e}{m\gamma(q)}
\biggl( \vec{v}(q)\times\vec{B}(t,\vec{r})
+\vec{E}(t,\vec{r}) - \vec{v}(q)v_i(q)E_i(t,\vec{r})\biggr) \; .
\label{eq:new02.010new3}
\end{eqnarray}
Note that $\vec{a}_{\cal F}$ is denoted in \cite{DK75} by $\dot{\bf v}$.
Note also that $\vec{\cal C}$ is the classical radiation-force field, see the discussion after
(\ref{eq:new02.010new11}).
As is common and also (tacitly) practiced in \cite{DK75} we neglect 
electric bremsstrahlung effects, i.e., we only account for magnetic bremsstrahlung, that is, 
synchrotron radiation
and so (\ref{eq:new02.010new3}) simplifies to
\begin{eqnarray}   
&&\vec{a}_{\cal F}(t,q):=\frac{e}{m\gamma(q)}
\biggl(\vec{v}(q)\times\vec{B}(t,\vec{r})\biggr) \; ,
\label{eq:new2.44e} 
\end{eqnarray}
which, in the present work, is the definition of  $\vec{a}_{\cal F}$.
This completes our definition of $\vec{v},\vec{\cal F}$ and $St$.

The expression, (\ref{eq:new02.010new0}), of the orbital Fokker-Planck equation
is not
always the most convenient one and so we will present in this section three more expressions, namely
(\ref{eq:new02.010new8}), (\ref{eq:new02.010new0final})     
and (\ref{eq:new02.010}).
To derive (\ref{eq:new02.010new8}) from (\ref{eq:new02.010new0}) we 
rewrite the rhs of (\ref{eq:new02.010new0}). We thus
compute, for $i=1,2,3$,
\begin{eqnarray}
&&\hspace{-10mm} 
{\cal E}_{i,j}\frac{\partial}{\partial \pi_j} = \frac{\partial}{\partial \pi_j} 
{\cal E}_{i,j} - 2{\cal Q}_i \; ,
\label{eq:new02.010new4}
\end{eqnarray}
where the $\R^3$-valued function $\vec{\cal Q}$ is the quantum radiation-force field 
defined, for $i=1,2,3$, by
\begin{eqnarray}
&& {\cal Q}_i:=
\frac{1}{2}\frac{\partial {\cal E}_{i,j}}{\partial \pi_j} \; .
\label{eq:new02.010new5}
\end{eqnarray}
For later reference we note that one can show,
by using (\ref{eq:new2.44b}), (\ref{eq:new2.44g}), 
(\ref{eq:new2.44e}),
(\ref{eq:new02.010new5}), that
\begin{eqnarray}
&& \vec{{\cal Q}}(t,q)=  \frac{55}{48\sqrt{3}}
\biggl( 6+ \frac{1}{\gamma^2(q)}\biggr)\lambda(t,q)\vec{\pi} \; .
\label{eq:new2.44f} 
\end{eqnarray}
For the terminology used for $\vec{\cal Q}$, see the discussion after
(\ref{eq:new02.010new11}).
From (\ref{eq:new02.010new4}) we get
\begin{eqnarray}
&& \hspace{-15mm} 
\frac{\partial}{\partial \pi_i} {\cal E}_{i,j} \frac{\partial}{\partial \pi_j} 
= \frac{\partial}{\partial \pi_i} \frac{\partial}{\partial \pi_j} 
{\cal E}_{i,j} - 2 \frac{\partial}{\partial \pi_i} {\cal Q}_i\; ,
\nonumber
\end{eqnarray}
resulting, by (\ref{eq:new02.010new1}), in:                                          
\begin{eqnarray}
&& \hspace{-30mm}
St = -\frac{\partial}{\partial \pi_j} \biggl( {\cal C}_j + {\cal Q}_j\biggr)
+\frac{1}{2}\frac{\partial}{\partial \pi_i} \frac{\partial}{\partial \pi_j} 
{\cal E}_{i,j}\; .
\label{eq:new02.010new7}                         
\end{eqnarray}
With (\ref{eq:new02.010new7}) one can write (\ref{eq:new02.010new0}) as
\begin{eqnarray}                                                            
&&     \hspace{-10mm}                                                                  
\frac{\partial \rho[W]}{\partial t} 
+ \frac{\partial}{\partial r_j} (v_j \rho[W])
+ \frac{\partial}{\partial \pi_j} ( {\cal F}_j \rho[W])  
\nonumber\\
&& \hspace{-10mm} = -\frac{\partial}{\partial \pi_j} \biggl( ({\cal C}_j + {\cal Q}_j)\rho[W]\biggr)
+\frac{1}{2}\frac{\partial}{\partial \pi_i} \frac{\partial}{\partial \pi_j} 
({\cal E}_{i,j}\rho[W])\; .
\label{eq:new02.010new8}                                                           
\end{eqnarray}
It is clear by its derivation that (\ref{eq:new02.010new8}) is the same as (\ref{eq:new02.010new0}).

Before we will write the orbital Fokker-Planck equation in the forms of
(\ref{eq:new02.010new0final}), (\ref{eq:new02.010}) we make some comments on 
(\ref{eq:new02.010new8}). 
First of all we
comment on the theoretical origins of (\ref{eq:new02.010new8}) which are rooted in QED.
According to the discussion after eq. 1 in \cite{DK75}, $-\frac{\partial}{\partial \pi_j} {\cal C}_j$, 
accounts for the radiation friction whereas
$(1/2)\frac{\partial}{\partial \pi_i} {\cal E}_{i,j} \frac{\partial}{\partial \pi_j}$ and thus:
$(1/2)\frac{\partial}{\partial \pi_i} \frac{\partial}{\partial \pi_j} 
{\cal E}_{i,j} - \frac{\partial}{\partial \pi_i} {\cal Q}_i$
accounts for the quantum fluctuations of the radiation.
Thus, by (\ref{eq:new02.010new7}),  
$St$ is responsible for all radiative terms of the orbital Fokker-Planck equation, 
(\ref{eq:new02.010new8}), i.e.,
the rhs is the radiative part of (\ref{eq:new02.010new8}) (and thus the rhs of (\ref{eq:new02.010new0}) is 
the radiative part of (\ref{eq:new02.010new0})).
We now take a look at the lhs 
of (\ref{eq:new02.010new8}). Neglecting the radiative effects, i.e., neglecting the
rhs of (\ref{eq:new02.010new8}) we get
\begin{eqnarray}                                                            
&&     \hspace{-10mm}                                                                  
\frac{\partial \rho[W]}{\partial t} 
+ \frac{\partial}{\partial r_j} (v_j \rho[W])
+ \frac{\partial}{\partial \pi_j} ( {\cal F}_j \rho[W])  = 0 \; ,
\label{eq:new02.010new13}                                                           
\end{eqnarray}
which, by the above, is  the Liouville equation \cite{wikiLT}
describing the orbital motion of a nonradiating
particle in the external electromagnetic field.
Thus the orbital Fokker-Planck equation is conditioned by the 
combination of Hamiltonian and radiative dynamics.
Secondly, we remark that
the orbital Fokker-Planck equation has a prehistory dating back at least as far as 1954.
In fact it is possible to rederive it by using
\cite{Dir38,Sch54} and some simple theoretical assumptions.

After these comments on (\ref{eq:new02.010new8})
we will next write the orbital Fokker-Planck equation in the forms,
(\ref{eq:new02.010new0final}), (\ref{eq:new02.010}).
We thus define the $\R^6$-valued functions,
${\cal D}_{orb,nrad},{\cal D}_{orb,rad},{\cal D}_{orb},{\cal B}_{orb}$, by
\begin{eqnarray}                                                            
&& {\cal D}_{orb,nrad}(t,q):=\left( \begin{array}{c} \vec{v}(q)\\
\vec{\cal F}(t,q) \end{array}\right) \; , 
\label{eq:new2.44anew} \\
&& {\cal D}_{orb,rad}(t,q):=\left( \begin{array}{c}  \vec{0} \\
\vec{\cal C}(t,q)+\vec{\cal Q}(t,q)\end{array}\right) \; ,
\label{eq:new2.44anewnew} \\
&&  {\cal D}_{orb}:={\cal D}_{orb,nrad}+{\cal D}_{orb,rad} \; ,
\label{eq:new2.44a} \\
&& {\cal B}_{orb}(t,q):=
\sqrt{ \frac{55}{24\sqrt{3}}}\sqrt{\lambda(t,q)}
\left( \begin{array}{c} \vec{0} \\ \vec{\pi}
\end{array}\right) \; .
\label{eq:new2.44h} 
\end{eqnarray}
It follows from the above that ${\cal D}_{orb,rad}$ is the radiative and
${\cal D}_{orb,nrad}$ is the nonradiative part of ${\cal D}_{orb}$.
Because of (\ref{eq:new02.010new8}) and by the discussion after (\ref{eq:new02.010new8}) 
we call $\vec{\cal C} +\vec{\cal Q}$ the radiation-force field.
It follows from (\ref{eq:new2.44g}),
(\ref{eq:new2.44anew}), (\ref{eq:new2.44anewnew}), 
(\ref{eq:new2.44a}), (\ref{eq:new2.44h}) that
\begin{eqnarray}                                                            
&& \hspace{-15mm} \frac{\partial}{\partial q_k} {\cal D}_{orb,k}
= 
\frac{\partial}{\partial r_j} v_j 
+ \frac{\partial}{\partial \pi_j} \biggl( {\cal F}_j  + {\cal C}_j + {\cal Q}_j\biggr)  \; , \quad
\frac{\partial}{\partial q_k} \frac{\partial}{\partial q_l} 
{\cal B}_{orb,k}  {\cal B}_{orb,l} 
=\frac{\partial}{\partial \pi_j} \frac{\partial}{\partial \pi_i} 
{\cal E}_{i,j}  \; ,
\nonumber
\end{eqnarray}
and so we can write (\ref{eq:new02.010new8}) as
\begin{eqnarray}                                                            
&&     \hspace{-20mm}                                                                  
\frac{\partial \rho[W]}{\partial t} 
= - \frac{\partial}{\partial q_k}({\cal D}_{orb,k}\rho[W])
+\frac{1}{2}\frac{\partial}{\partial q_k} \frac{\partial}{\partial q_l} 
({\cal B}_{orb,k}  {\cal B}_{orb,l}\rho[W])  \; .
\label{eq:new02.010new0final}                                                           
\end{eqnarray}
We also write (\ref{eq:new02.010new0final}) more compactly as  
\begin{eqnarray}                                                            
&&  \frac{\partial \rho[W]}{\partial t}  = l_{orb}\rho[W]\; ,
\label{eq:new02.010}                                                           
\end{eqnarray}
where the linear operator $l_{orb}$ is defined by
\begin{eqnarray}                                                            
&&           \hspace{-15mm}                                                               
l_{orb}:=-  \frac{\partial}{\partial r_j} v_j 
-\frac{\partial}{\partial \pi_j} \biggl( {\cal F}_j  + {\cal C}_j + {\cal Q}_j\biggr)  
+\frac{1}{2}\frac{\partial}{\partial \pi_i} \frac{\partial}{\partial \pi_j} {\cal E}_{i,j}
\nonumber\\
&&
= - \frac{\partial}{\partial q_k} {\cal D}_{orb,k}
+\frac{1}{2}\frac{\partial}{\partial q_k} \frac{\partial}{\partial q_l} 
{\cal B}_{orb,k}  {\cal B}_{orb,l} 
\nonumber\\
&&
= - \frac{\partial}{\partial q_k} {\cal D}_{orb,k}
+\frac{1}{2}\frac{\partial}{\partial q_k} \frac{\partial}{\partial q_l} 
({\cal B}_{orb} {\cal B}_{orb}^T)_{k,l} 
= -  \frac{\partial}{\partial r_j} v_j 
- \frac{\partial}{\partial \pi_j} {\cal F}_j  + St  \; ,
\label{eq:new02.010a}                                                           
\end{eqnarray}
and where in the fourth equation of (\ref{eq:new02.010a}) we used 
(\ref{eq:new02.010new7}). It is clear by their derivations that (\ref{eq:new02.010new0final}), (\ref{eq:new02.010}) 
are the same as (\ref{eq:new02.010new8}) 
(and thus are the same as (\ref{eq:new02.010new0})).
Note that, (\ref{eq:new02.010}), is eq. 2 
in \cite{HABBE19}.

The key point about 
(\ref{eq:new02.010new0final}) is being an expression in terms of the so-called drift vector field and of the
so-called diffusion matrix field. Thus (\ref{eq:new02.010new0final}) is that 
expression of the orbital Fokker-Planck equation by which it is immediately 
clear that it is a Fokker-Planck equation.
In fact ${\cal D}_{orb}$ is the so-called drift vector field and 
${\cal B}_{orb}{\cal B}_{orb}^T$ is the so-called diffusion matrix field of
(\ref{eq:new02.010new0final}). In particular
${\cal D}_{orb}$ encapsulates the spatial drift via the velocity field, $\vec{v}$, and the momentum drift
via the Lorentz force field, $\vec{\cal F}$,
and via the radiation-force field, $\vec{\cal C}+\vec{\cal Q}$.
Also, ${\cal B}_{orb}{\cal B}_{orb}^T$ encapsulates the diffusion effect due to the
synchrotron radiation.
Since ${\cal B}_{orb}$ is a vector field, every value of 
${\cal B}_{orb}{\cal B}_{orb}^T$, is an outer product matrix.
The so-called noise vector field, ${\cal B}_{orb}$, of (\ref{eq:new02.010new0final})
is not uniquely determined by (\ref{eq:new02.010new0final}) since each of the values of
${\cal B}_{orb}$ can be multiplied by either $1$ or $-1$ without
affecting ${\cal B}_{orb}{\cal B}_{orb}^T$
(our choice of ${\cal B}_{orb}$ is given by
(\ref{eq:new2.44h})). For example, $-{\cal B}_{orb}$ is a noise vector field
of (\ref{eq:new02.010new0final}), too.
For the notions of Fokker-Planck equation, drift vector field and diffusion matrix field 
see, e.g., \cite{Arn74,Gar04}.

In the remaining parts of this section we make further
comments on the orbital Fokker-Planck equation.
First of all, one can show that $1/\lambda$ is of the order of the Sokolov-Ternov polarization time
and thus much larger than the orbital damping time 
(the latter being associated with $\vec{\cal C}$).
Therefore, and because of (\ref{eq:new2.44f}) 
it is very 
common in applications to neglect $\vec{\cal Q}$, i.e.,
to approximate $\vec{\cal C}+\vec{\cal Q}$ by $\vec{\cal C}$.
Nevertheless, in the present work, we do not neglect $\vec{\cal Q}$ except when explicitly
mentioned. In particular
this allows us to display the close relation between $\vec{\cal Q}$ and the 
orbital noise vector field. In fact, by (\ref{eq:new2.44f}), (\ref{eq:new2.44h}), ${\cal B}_{orb}$
is related to $\vec{\cal Q}$ via
\begin{eqnarray}
&& \hspace{-22mm} 
\sqrt{|\vec{\cal Q}(t,q)|} {\cal B}_{orb}(t,q) =
\gamma(q)\sqrt{\frac{2m\sqrt{\gamma^2(q)-1}}{6\gamma^2(q)+1}}
\left( \begin{array}{c} \vec{0} \\  \vec{\cal Q}(t,q) \end{array}\right) \; .
\label{eq:new02.010new11}
\end{eqnarray}
Secondly, $l_{orb}$ is the sum of a zeroth order part $\hbar$ and a first order part in 
$\hbar$, i.e., the sum of
a classical and a quantum term.
In fact, by (\ref{eq:new2.44g}), (\ref{eq:new2.44h}), ${\cal B}_{orb}{\cal B}_{orb}^T$, is of first order $\hbar$. 
Moreover, by (\ref{eq:new2.44b}), (\ref{eq:new2.44c}), (\ref{eq:new2.44d}), (\ref{eq:new2.44f}),
$\vec{v},\vec{\cal F},\vec{\cal C}$ are 
of zeroth order $\hbar$ and $\vec{\cal Q}$ is of first order $\hbar$. 
Thus, by (\ref{eq:new2.44anew}), (\ref{eq:new2.44anewnew}), (\ref{eq:new2.44a}), 
${\cal D}_{orb}$, has a part of
zeroth order $\hbar$, namely $\left( \begin{array}{c} \vec{v}\\
\vec{\cal F}+ \vec{\cal C}\end{array}\right)$ 
and a part of  first order $\hbar$, namely
$\left( \begin{array}{c}  \vec{0} \\  \vec{\cal Q}\end{array}\right)$.
This justifies the terminologies classical resp. quantum radiation-force field
(recall that $\vec{\cal C}+\vec{\cal Q}$ is the radiation-force field).
We conclude, by (\ref{eq:new02.010a}), that $l_{orb}$
has a part of zeroth order $\hbar$, namely
$-  \frac{\partial}{\partial r_j} v_j 
- \frac{\partial}{\partial \pi_j} ({\cal F}_j  +  {\cal C}_j)$ and a
part of first order $\hbar$, namely
$- \frac{\partial}{\partial \pi_j} {\cal Q}_j  
+\frac{1}{2}\frac{\partial}{\partial q_k} \frac{\partial}{\partial q_l} 
{\cal B}_{orb,k}  {\cal B}_{orb,l}$.
Thirdly, the quantum part of $l_{orb}$ is of a
radiative nature (the converse does not hold because of the term,
$-\frac{\partial}{\partial \pi_j}{\cal C}_j$).
Fourthly, recalling that ${\cal B}_{orb}$ is a noise vector field of 
the orbital Fokker-Planck equation, 
the following is an Ito SDE system associated with the orbital Fokker-Planck equation:
\begin{eqnarray}                                                            
&&    Q' = {\cal D}_{orb}(t,Q) +{\cal B}_{orb}(t,Q)\nu(t) \; ,
\label{eq:new2.44uanew}
\end{eqnarray}
where $\nu$ is the one-dimensional (=scalar) white-noise process (recall that 
SDE=stochastic ODE) and where we used (\ref{eq:new02.010new0final}).
Note that we denote dependent variables (like $Q$ in (\ref{eq:new2.44uanew}))
by roman capital letters. For the notion of associated Ito SDE system see, e.g., \cite{Arn74,Gar04}.
Recalling that $-{\cal B}_{orb}$ is a noise vector field 
of (\ref{eq:new02.010new0final}) we note that:
\begin{eqnarray}                                                            
&&    Q' = {\cal D}_{orb}(t,Q) -{\cal B}_{orb}(t,Q)\nu(t) \; ,
\nonumber
\end{eqnarray}
is another Ito SDE system associated with the orbital Fokker-Planck equation.
By (\ref{eq:new2.44anew}), (\ref{eq:new2.44anewnew}),
(\ref{eq:new2.44a}), (\ref{eq:new2.44h}) the SDE system,
(\ref{eq:new2.44uanew}), can be written more explicitly as
\begin{eqnarray}                                                            
&&    \vec{R}' =\vec{v}(Q) \; , 
\label{eq:new2.44ubnew}\\
&& \vec{\Pi}' = \vec{\cal F}(t,Q)+\vec{\cal C}(t,Q)+\vec{\cal Q}(t,Q)
+\sqrt{\frac{55}{24\sqrt{3}}}\sqrt{\lambda(t,Q)}\vec{\Pi}\nu(t) \; .
\label{eq:new2.44uaanew}
\end{eqnarray}
Fifthly,
neglecting all radiative effects, the SDE system,
(\ref{eq:new2.44ubnew}), (\ref{eq:new2.44uaanew}), simplifies to
\begin{eqnarray}                                                            
&&    \vec{R}' =\vec{v}(Q) \; , \quad  \vec{\Pi}' = \vec{\cal F}(t,Q)  \; .
\label{eq:new2.44uaanewnew}
\end{eqnarray}
Note that the solutions $\left( \begin{array}{c} \vec{R}\\\vec{\Pi}\end{array}\right)$ of (\ref{eq:new2.44uaanewnew}) 
are characteristics of the Liouville equation,
(\ref{eq:new02.010new13}), in the sense that:
$\frac{d}{dt} \Biggl( \rho[W]\biggl(t,\vec{R}(t),\vec{\Pi}(t)\biggr)\Biggr) = 0$
%
%
where $\rho[W]$ is any solution of (\ref{eq:new02.010new13}).

\section{The full Bloch equation for the polarization density}

In this section we outline the dynamics of $\vec{\cal P}[W]$ (recall from
Section 1.2 that $\vec{\cal P}[W]$ is $\R^3$-valued).

The evolution equation for $\vec{\cal P}[W]$ is given to us from \cite{DK75} 
as the following linear inhomogeneous PDE system:
\begin{eqnarray}                                                            
&&            \hspace{-10mm}                                                                  
\frac{\partial \vec{\cal P}[W]}{\partial t} 
+ \frac{\partial}{\partial r_j} (v_j \vec{\cal P}[W])
+ \frac{\partial}{\partial \pi_j} ( {\cal F}_j \vec{\cal P}[W])                                                                
- \vec{\Omega}_{TBMT}\times\vec{\cal P}[W]
\nonumber\\
&& = St\; \vec{\cal P}[W]+ \Delta l_{hom}\vec{\cal P}[W]
+ l_{inhom}\rho[W]\; ,
\label{eq:new02.010new20}                                                           
\end{eqnarray}
where $\rho[W]$ is a solution of the orbital Fokker-Planck equation.
The function $\vec{\Omega}_{TBMT}$ and the linear operators $\Delta l_{hom}$ and
$l_{inhom}$ will be defined below in this section (recall that
$\vec{v},\vec{\cal F}$ and $St$ are defined in Section 3 above).

Before we define $\vec{\Omega}_{TBMT},\Delta l_{hom}$ and
$l_{inhom}$ let us make some remarks on 
(\ref{eq:new02.010new20}). 
First of all, the rhs of (\ref{eq:new02.010new20}) is its radiative part
(see the
discussion after (\ref{eq:new02.010new22})).
Secondly, for the theoretical origins of (\ref{eq:new02.010new20}), see the
discussion after (\ref{eq:new02.010new22}).
Thirdly we call (\ref{eq:new02.010new20}) 
(and every equation which only differs 
from (\ref{eq:new02.010new20}) in terms of notation)
the full Bloch equation (see also the discussions after 
(\ref{eq:new02.010new24}) and (\ref{eq:new02.010d})).
Fourthly, the full Bloch equation was presented, but 
not derived, in \cite{DK75} (except for some remarks).
Instead it was derived, from QED, in \cite{Kon82} which is a work which underlies \cite{DK75}
(for more remarks on \cite{Kon82} see Section 5.5).
Fifthly, in Section 5.2 we will determine which solutions $\vec{\cal P}[W]$
of (\ref{eq:new02.010new20}) we will call the polarization density of a bunch.

We now define $\vec{\Omega}_{TBMT},\Delta l_{hom}$ and
$l_{inhom}$. The $\R^3$-valued function,
$\vec{\Omega}_{TBMT}$, is the 
T-BMT precession vector field, i.e.,
\begin{eqnarray}
&&  \hspace{-10mm} 
\vec{\Omega}_{TBMT}(t,q):=
-\frac{e}{m\gamma(q)}\biggl( 1+
\frac{g-2}{2}\gamma(q)\biggr)\vec{B}(t,\vec{r})
+ \frac{e}{m} \frac{g-2}{2}
\frac{\gamma(q)}{1+\gamma(q)}
\vec{v}(q)v_i(q) B_i(t,\vec{r})
\nonumber\\
&& +\frac{e}{m}
\biggl( \frac{g-2}{2} +\frac{1}{1+\gamma(q)}\biggr)
\biggl(\vec{v}(q)\times\vec{E}(t,\vec{r})\biggr) \; ,
\label{eq:new2.99anew}
\end{eqnarray}
where $g$ is the particle's $g$-factor.
The linear operator $\Delta l_{hom}$ is a multiplication operator defined by  
\begin{eqnarray}                                                            
&& \hspace{-10mm} \biggl( \Delta l_{hom}\vec{j}\biggr)(t,q):=
-\frac{5\sqrt{3}}{8}\lambda(t,q)\biggl(  I_{3\times 3}
-\frac{2}{9m^2\gamma^2(q)}\vec{\pi}\vec{\pi}^T\biggr)\vec{j}(t,q) \; ,
\label{eq:new2.99b}                                                           
\end{eqnarray}
where $\vec{j}$ is any function: $\R^7\rightarrow\R^3$.
The linear operator $l_{inhom}$ is defined by  
\begin{eqnarray}                                                            
&& \biggl( l_{inhom}k\biggr)(t,q):=
-\biggl( 1 + \frac{\partial}{\partial \pi_j}\pi_j\biggr)
\lambda(t,q)\frac{1}{m\gamma(\vec{\pi})}
\frac{\vec{\pi}\times\vec{a}_{\cal F}(t,q)}{|\vec{a}_{\cal F}(t,q)|}k(t,q) \; ,
\label{eq:new02.010new21}     
\end{eqnarray}
where $k$ is any differentiable function: $\R^7\rightarrow\R$. 
To rewrite (\ref{eq:new02.010new21}) into a more convenient form we define the $\R^3$-valued function
\begin{eqnarray}                                                            
&&  \hspace{-12mm} 
\vec{\cal B}_{spin}(t,q):=   \sqrt{ \frac{24\sqrt{3}}{55}}\sqrt{\lambda(t,q)}
\frac{1}{m\gamma(q)}
\frac{\vec{\pi}\times\vec{a}_{\cal F}(t,q)}{|\vec{a}_{\cal F}(t,q)|} \; .
\label{eq:new2.99d}                                                          
\end{eqnarray}
Thus we obtain from (\ref{eq:new2.44h}):
\begin{eqnarray}                                                            
&& \hspace{-20mm} \lambda(t,q)\frac{1}{m\gamma(\vec{\pi})}
\frac{\vec{\pi}\times\vec{a}_{\cal F}(t,q)}{|\vec{a}_{\cal F}(t,q)|}
= \sqrt{ \frac{55}{24\sqrt{3}}}\sqrt{\lambda(t,q)}\vec{\cal B}_{spin}(t,q)\; ,
\nonumber\\
&&  \hspace{-20mm} \frac{\partial}{\partial \pi_j} \pi_j
\lambda(t,q)\frac{1}{m\gamma(\vec{\pi})}
\frac{\vec{\pi}\times\vec{a}_{\cal F}(t,q)}{|\vec{a}_{\cal F}(t,q)|}
=\frac{\partial}{\partial \pi_j} \pi_j
\sqrt{ \frac{55}{24\sqrt{3}}}
\sqrt{\lambda(t,q)} \vec{\cal B}_{spin}(t,q)
\nonumber\\
&&  \hspace{-20mm}= 
\frac{\partial}{\partial \pi_j} {\cal B}_{orb,j+3}(t,q)
\vec{\cal B}_{spin}(t,q)
= \frac{\partial}{\partial q_i} {\cal B}_{orb,i}(t,q)
\vec{\cal B}_{spin}(t,q) \; ,
\nonumber                                                        
\end{eqnarray}
which implies, by (\ref{eq:new02.010new21}),  
\begin{eqnarray}                                              && \hspace{-15mm} l_{inhom}k =-\sqrt{ \frac{55}{24\sqrt{3}}}\sqrt{\lambda}\vec{\cal B}_{spin}k
-  \frac{\partial}{\partial q_i} ({\cal B}_{orb,i}
\vec{\cal B}_{spin} k)  \; .
\label{eq:new2.99c}                                           
\end{eqnarray}
It follows from (\ref{eq:new2.99c}) that 
\begin{eqnarray}                                                            
&&   l_{inhom} = l_{inhom,1}+ l_{inhom,2}\; ,
\label{eq:new02.010new22a}                                                           
\end{eqnarray}
where the linear operators $l_{inhom,1},l_{inhom,2}$ are defined by
\begin{eqnarray}                                                            
&&  \hspace{-15mm}    \biggl( l_{inhom,1}k\biggr)(t,q):=- \sqrt{ \frac{55}{24\sqrt{3}}}\sqrt{\lambda(t,q)}
\vec{\cal B}_{spin}(t,q) k(t,q) \; ,
\label{eq:new02.010new23} \\
&& \hspace{-15mm} l_{inhom,2}k:=
- \frac{\partial}{\partial q_j} ({\cal B}_{orb,j}
\vec{\cal B}_{spin}k) \; ,
\label{eq:new02.010new24}                                                           
\end{eqnarray}
where $k$ is any differentiable function: $\R^7\rightarrow\R$.
Note that $l_{inhom,1}$ is a multiplication operator.
This completes the definition of  $\vec{\Omega}_{TBMT},\Delta l_{hom}$ and
$l_{inhom}$.

The expression, (\ref{eq:new02.010new20}), of the full Bloch equation is not
always the most convenient one and so we will present in this section three more expressions, namely
(\ref{eq:new02.010new22}), (\ref{eq:new02.010d}), (\ref{eq:new2.44vcdc0new}).     
It follows from (\ref{eq:new2.99c}) that 
(\ref{eq:new02.010new20}) can be written as
\begin{eqnarray}                                                            
&&    \hspace{-10mm}                                                                  
\frac{\partial \vec{\cal P}[W]}{\partial t} 
+ \frac{\partial}{\partial r_j} (v_j \vec{\cal P}[W])
+ \frac{\partial}{\partial \pi_j} ( {\cal F}_j \vec{\cal P}[W])                                                                
- \vec{\Omega}_{TBMT}\times\vec{\cal P}[W]
\nonumber\\
&& \hspace{-10mm} = St\vec{\cal P}[W]
+ \Delta l_{hom}\vec{\cal P}[W]
+ l_{inhom,1}\rho[W]
+ l_{inhom,2}\rho[W] \; .
\label{eq:new02.010new22}                                                           
\end{eqnarray}
It is clear by its derivation that (\ref{eq:new02.010new22}) 
is the same as (\ref{eq:new02.010new20}). 

Before we write the full Bloch equation in the forms of
(\ref{eq:new02.010d}), (\ref{eq:new2.44vcdc0new})
we make some comments on (\ref{eq:new02.010new22}).
We first comment on the theoretical origins of (\ref{eq:new02.010new22}) which are rooted in QED.
According to the discussion after (\ref{eq:new02.010new8}), $St$ 
accounts for the radiation friction and 
the quantum fluctuations of the radiation.
Moreover, by (\ref{eq:new2.99b}), (\ref{eq:new02.010new23}), (\ref{eq:new02.010new24}) and                                                          
according to the discussion after eq. 2 in \cite{DK75}, 
$\Delta l_{hom}\vec{\cal P}[W]
+ l_{inhom,1}\rho[W]$ comprises the so-called direct action of the
radiation on the polarization whereas
$l_{inhom,2}\rho[W]$ comprises the spin-orbit interaction with
the classical radiation (the latter effect is not to be confused with the radiation friction effect!).
We conclude that the rhs is the radiative part of (\ref{eq:new02.010new22})       
(and thus the rhs of (\ref{eq:new02.010new20})).
We now take a look at the lhs of (\ref{eq:new02.010new22}).
Neglecting the rhs of (\ref{eq:new02.010new22}) 
we get
\begin{eqnarray}                                                            
&&            \hspace{-10mm}                                                                  
\frac{\partial \vec{\cal P}[W]}{\partial t} 
+ \frac{\partial}{\partial r_j} (v_j \vec{\cal P}[W])
+ \frac{\partial}{\partial \pi_j} ( {\cal F}_j \vec{\cal P}[W])                                                                
- \vec{\Omega}_{TBMT}\times\vec{\cal P}[W] = \vec{0}  \; ,
\label{eq:new02.010new20a}                                                           
\end{eqnarray}
which describes, by the above, 
a nonradiative spinning particle moving in the external electromagnetic field.
We finally remark that the full Bloch equation was presented but 
not derived in \cite{DK75}. For the derivation of the full Bloch equation, see the
corresponding remarks in Section 1.2.

After these comments on (\ref{eq:new02.010new22})
we now rewrite (\ref{eq:new02.010new22}) into
the form, (\ref{eq:new02.010d}).
We thus define the linear operators $l_{TBMT},l_{hom}$ by
\begin{eqnarray}                                                            
&& \hspace{-10mm} \biggl( l_{TBMT}\vec{j}\biggr)(t,q):=\vec{\Omega}_{TBMT}(t,q)\times \vec{j}(t,q) \; ,
\label{eq:new2.99a}\\
&&  \hspace{-10mm} l_{hom}:= l_{TBMT}+ \Delta l_{hom} \; ,
\label{eq:new2.100d}                                                           
\end{eqnarray}
where $\vec{j}$ is any function: $\R^7\rightarrow\R^3$.
Note that $l_{TBMT}$ and $l_{hom}$
are multiplication operators.
From (\ref{eq:new02.010a}) we get: 
$-\frac{\partial}{\partial r_j} v_j   - \frac{\partial}{\partial \pi_j} {\cal F}_j
+ St  + \Delta l_{hom}= l_{orb}  + \Delta l_{hom}$ 
so that, by (\ref{eq:new2.99a}), (\ref{eq:new2.100d}), we can write
(\ref{eq:new02.010new22}) as
\begin{eqnarray}                                                            
&&          \hspace{-30mm}                                                                
\frac{\partial \vec{\cal P}[W]}{\partial t} 
= ( l_{orb}+l_{hom})\vec{\cal P}[W]
+ l_{inhom}\rho[W]\; ,
\label{eq:new02.010d}                                                           
\end{eqnarray}
which will be needed in Section 5.1. 
It is clear by its derivation that (\ref{eq:new02.010d}) 
is the same as (\ref{eq:new02.010new22}) 
(and thus is the same as (\ref{eq:new02.010new20})).

Before we write the full Bloch equation in the form of
(\ref{eq:new2.44vcdc0new})
we make some comments on (\ref{eq:new02.010d}).
First of all, (\ref{eq:new02.010d}) 
is eq. 8 in \cite{HABBE19} and so, following \cite{HABBE19}, we
call (\ref{eq:new02.010d}) the full Bloch equation 
(and thus, (\ref{eq:new02.010new20}), (\ref{eq:new02.010new22}), are the full Bloch equation).
Secondly, we now comment on how $\hbar$ appears in the full Bloch equation, (\ref{eq:new02.010d}).
We identified after (\ref{eq:new02.010new11}) 
the parts of $l_{orb}$
which are of zeroth order resp. first order $\hbar$.
Moreover, by (\ref{eq:new2.99anew}), (\ref{eq:new2.99a}),
$l_{TBMT}$ is of zeroth order $\hbar$ whereas, by (\ref{eq:new2.44g}), (\ref{eq:new2.99b}),
$\Delta l_{hom}$ is of first order $\hbar$. 
We thus conclude, by (\ref{eq:new2.100d}), that
$l_{TBMT}$ is the part of $l_{hom}$ which is of zeroth order $\hbar$ 
and that $\Delta l_{hom}$ 
is the part of $l_{hom}$ which is of first order $\hbar$.
Thus, by the discussion after (\ref{eq:new02.010new22}), 
the quantum part of $l_{hom}$ is of a radiative nature and, conversely, the radiative part of $l_{hom}$ 
is of quantum nature.
We conclude, by the discussion after (\ref{eq:new02.010new11}),
that the quantum part of 
$l_{orb}+l_{hom}$ is of a radiative nature
(the converse does not hold because of the term, $- \frac{\partial}{\partial \pi_j}{\cal C}_j$,
in $l_{orb}$). By (\ref{eq:new2.44g}), (\ref{eq:new2.44h}), (\ref{eq:new2.99d}),
$l_{inhom}$ is of first order $\hbar$ and is of a radiative nature. In particular $l_{inhom}$ is of first order $\hbar$ (although it is related via its term, $l_{inhom,2}$, 
to the classical radiation!).

After these comments on (\ref{eq:new02.010d})
we now rewrite the full Bloch equation into
the form, (\ref{eq:new2.44vcdc0new}).
We first define the $\R^3$-valued function
$\vec{\cal D}_{spin,0,ST}$ and the $\R^{3\times 3}$-valued functions\\
${\cal D}_{spin,+,TBMT},{\cal D}_{spin,+,ST},{\cal D}_{spin,+,BK}$:
\begin{eqnarray}                                                            
&&  \hspace{-10mm} \vec{\cal D}_{spin,0,ST}(t,q):=
-\sqrt{ \frac{55}{24\sqrt{3}}}\sqrt{\lambda(t,q)}
\vec{\cal B}_{spin}(t,q) 
=-\frac{1}{m\gamma(q)}\lambda(t,q)
\frac{\vec{\pi}\times\vec{a}_{\cal F}(t,q)}
{|\vec{a}_{\cal F}(t,q)|} \; ,
\label{eq:new02.010new26}\\
&&  \hspace{-10mm} {\cal D}_{spin,+,TBMT}(t,q)\vec{s}:=
\vec{\Omega}_{TBMT}(t,q)\times \vec{s} \; ,
\label{eq:new02.010new26TBMT}\\
&&  \hspace{-10mm} {\cal D}_{spin,+,ST}(t,q)\vec{s}:=
-\frac{5\sqrt{3}}{8}\lambda(t,q) \vec{s} \; ,
\label{eq:new02.010new26ST}\\
&&  \hspace{-10mm} {\cal D}_{spin,+,BK}(t,q):=
\frac{5\sqrt{3}}{36m^2\gamma^2(q)}\lambda(t,q) \vec{\pi}\vec{\pi}^T \; ,
\label{eq:new02.010new26BK}
\end{eqnarray}
where $t\in\R,q\in\R^6,\vec{s}\in\R^3$.
We also define the $\R^{3\times 3}$-valued function
${\cal D}_{spin,+}$ and the $\R^3$-valued function $\vec{\cal D}_{spin}$ by
\begin{eqnarray}                                                            
&&  \hspace{-10mm} {\cal D}_{spin,+}:=
{\cal D}_{spin,+,TBMT}+{\cal D}_{spin,+,ST}+{\cal D}_{spin,+,BK} \; ,
\label{eq:new02.010new27}\\  
&& \hspace{-10mm} 
\vec{\cal D}_{spin}(t,q,\vec{s}):=\vec{\cal D}_{spin,0,ST}(t,q)+{\cal D}_{spin,+}(t,q)\vec{s} \; ,
\label{eq:new2.44vanewnewnew}  
\end{eqnarray}
where $t\in\R,q\in\R^6,\vec{s}\in\R^3$. In Section 15
the notations, $\vec{\cal D}_{spin,0,ST},{\cal D}_{spin,+,ST}$, will be justified in terms of the Sokolov-Ternov effect and the notation, ${\cal D}_{spin,+,BK}$, in terms of the Baier-Katkov correction
while the notation, ${\cal D}_{spin,+,TBMT}$, will be justified in terms of the T-BMT precession
effect.
It follows from (\ref{eq:new02.010new22a}), 
(\ref{eq:new02.010new23}), (\ref{eq:new02.010new24}),
(\ref{eq:new2.99a}), (\ref{eq:new2.100d}), (\ref{eq:new02.010new26}), (\ref{eq:new02.010new27}), 
that for any function $\vec{j}:\R^7\rightarrow\R^3$ and any 
differentiable function $k:\R^7\rightarrow\R$, 
\begin{eqnarray}                                                            
&& \hspace{-20mm} \biggl( l_{hom}\vec{j}\biggr)(t,q)={\cal D}_{spin,+}(t,q)\vec{j}(t,q)\; ,
\label{eq:new2.99anewnew}\\        
&& \hspace{-20mm} 
\biggl( l_{inhom,1}k\biggr)(t,q)=\vec{\cal D}_{spin,0,ST}(t,q)k(t,q) \; ,
\label{eq:new02.010new28}\\   
&& \hspace{-20mm} 
\biggl( l_{inhom}k\biggr)(t,q)=\vec{\cal D}_{spin,0,ST}(t,q)k(t,q)
- \frac{\partial}{\partial q_k} \biggl({\cal B}_{orb,k}(t,q)
\vec{\cal B}_{spin}(t,q)k(t,q)\biggr)\; .
\label{eq:new2.99bnew}                                                           
\end{eqnarray}
With (\ref{eq:new2.99anewnew}), (\ref{eq:new2.99bnew})
at hand we can write the full Bloch equation, (\ref{eq:new02.010d}), as
\begin{eqnarray}                                                            
&&        \hspace{-20mm}                                                                
\frac{\partial \vec{\cal P}[W]}{\partial t} 
= l_{orb}\vec{\cal P}[W]
+{\cal D}_{spin,+}\vec{\cal P}[W]
+\vec{\cal D}_{spin,0,ST}\rho[W]
- \frac{\partial}{\partial q_k} ({\cal B}_{orb,k}
\vec{\cal B}_{spin}\rho[W]) \; .
\label{eq:new2.44vcdc0new}                                                           
\end{eqnarray}
It is clear by the derivation that (\ref{eq:new2.44vcdc0new}) 
is the same as (\ref{eq:new02.010d}) (and thus is the same as
(\ref{eq:new02.010new20}),
(\ref{eq:new02.010new22})). Note that the expression, (\ref{eq:new2.44vcdc0new}), 
of the full Bloch equation will be used in Section 9 below.

The fact that ${\cal B}_{orb}$ occurs twice on the rhs of 
(\ref{eq:new2.44vcdc0new}) (namely in the first and fourth terms)
may get unnoticed upon first reading but it is 
crucial for obtaining the extension of the kinetic approach (recall that this extension
is the topic of Sections 7-16).
A reader who is interested in this
fine point should carefully read the discussion after (\ref{eq:new2.44vanewnew}). 

We now make some more 
remarks on the non-kinetic approach mentioned in Section 1.1. 
The non-kinetic approach is the subject matter of
\cite{DK72,DK73,Man87-1,Man87-2} and it relies on
the assumption that on average the spin vectors in a bunch
are aligned along the invariant spin field. This assumption
corresponds in the kinetic approach to the condition that the polarization density
points along the invariant spin field. However, as already hinted at in \cite{DK75}, 
this condition is inconsistent with the
full Bloch equation (and even inconsistent with the reduced Bloch equation, see
(\ref{eq:new2.100c})).
This inconsistency led to the concept of the radiative invariant spin field,
starting with \cite{Hei97} (see also \cite{BH15}) and still under
investigation by us (the radiative invariant spin field is denoted in \cite{BH15} by $\hat{p}$).
The invariant spin field is an approximation of the radiative invariant spin field
which we believe diminishes the usefulness of the Derbenev-Kondratenko formulas
with growing beam energies like in FCC-ee and CEPC. 
In contrast it is believed that the Derbenev-Kondratenko formulas are
useful (even at those high energies) when the invariant spin field is replaced 
by the radiative invariant spin field.
The aforementioned inconsistency is also studied, for the reduced setup, in Beznosov's PhD thesis
\cite{Bez20}.

We finally introduce the reduced Bloch equation. For that purpose we consider
the so-called reduced setup, defined by
\begin{eqnarray}                                                            
&&   l_{hom} = l_{TBMT} \; ,  \quad  l_{inhom}= 0 \; .
\label{eq:new02.010new30a}                                                           
\end{eqnarray}
In the reduced setup the full Bloch equation simplifies to
\begin{eqnarray}                                                            
&&                         
\hspace{-20mm}                                                 
\frac{\partial \vec{\cal P}[W]}{\partial t} = l_{orb}\vec{\cal P}[W]
+\vec{\Omega}_{TBMT}\times\vec{\cal P}[W]\; .
\label{eq:new2.100c}                                                           
\end{eqnarray}
Note that (\ref{eq:new2.100c}) is obtained from (\ref{eq:new2.99a}), (\ref{eq:new02.010d}), (\ref{eq:new02.010new30a}).
Following \cite{HABBE19} we call (\ref{eq:new2.100c})
the reduced Bloch equation. This justifies the terminology of reduced setup.

We recall from Section 1 that the orbital Fokker-Planck equation, in combination
with (\ref{eq:new2.100c}), contains all the information needed to study the
radiative depolarization effect (by the discussion before, (\ref{eq:new02.010new11}),
this is even true if one neglects $\vec{\cal Q}$).
Note also, by
(\ref{eq:new2.99a}), (\ref{eq:new2.44vanewnewnew}), (\ref{eq:new2.99anewnew}), 
(\ref{eq:new2.99bnew}), that (\ref{eq:new02.010new30a}) can be written as
\begin{eqnarray}                                                            
&& \vec{\cal D}_{spin}(t,q,\vec{s}) = \vec{\Omega}_{TBMT}\times\vec{s} \; ,
\label{eq:new02.010new30b}\\        
&& \vec{\cal B}_{spin}(t,q) = \vec{0} \; ,
\label{eq:new02.010new30c}
\end{eqnarray}
where $t\in\R,q\in\R^6,\vec{s}\in\R^3$.

For later reference we call the general situation, where the full Bloch equation holds, i.e.,
where we ignore (\ref{eq:new02.010new30a}) resp. 
(\ref{eq:new02.010new30b}), (\ref{eq:new02.010new30c}),
the full setup.

\section{The spin-$1/2$ Wigner function: The statistical conditions and the evolution equation}

We here proceed as follows.
In Section 5.1  we first derive, from Sections 3 and 4,
the evolution equation, (\ref{eq:new002.100}), for $W$ and then
we show equivalence, i.e., that (\ref{eq:new002.100}) is equivalent to the PDE system,
(\ref{eq:new02.010}), (\ref{eq:new02.010d}), for
$\rho[W],\vec{\cal P}[W]$. 
In Section 5.2 we define the 
so-called statistical conditions on $W$ leading us to the 
notion  of the spin-$1/2$ Wigner function of a bunch (or shortly, the notion of physically
meaningful $W$) which in turn is equivalent to the notions of
the orbital density and polarization density of a bunch.
In Section 5.3 we define observables and in Section 5.4 we study
how the statistical conditions on $W$ fit to the evolution equation, (\ref{eq:new002.100}).
Section 5.5 is devoted to making comments on $W$, e.g., 
on the theoretical origins of $W$ and on the reduced setup.

\subsection{The evolution equation}

To derive an evolution equation for $W$ we first note, by (\ref{eq:new02.10}), that 
\begin{eqnarray}                                                            
&&  \hspace{-20mm} 
\frac{\partial W}{\partial t} 
= \frac{1}{2}\biggl( I_{2\times 2}\; \frac{\partial \rho[W]}{\partial t} 
+ \sigma_i  \frac{\partial {\cal P}_i[W]}{\partial t} \biggr) \; .
\label{eq:new02.010new31}   
\end{eqnarray}
Using (\ref{eq:new02.10}), (\ref{eq:new02.10a}) and inserting
the orbital Fokker-Planck equation (\ref{eq:new02.010})
and the full Bloch equation (\ref{eq:new02.010d}) into the rhs of
(\ref{eq:new02.010new31}) we get 
\begin{eqnarray}                                                            
&&   \frac{\partial W}{\partial t} 
= \frac{1}{2} \Biggl( I_{2\times 2}\;l_{orb}\rho[W]
+ \sigma_i \biggl( ( l_{orb} + l_{hom})\vec{\cal P}[W]
+l_{inhom}\rho[W]\biggr)_i \Biggr)
\nonumber\\
&& = l_{orb}\frac{1}{2}\biggl( I_{2\times 2}\rho[W]
+\sigma_i {\cal P}_i[W]\biggr)  +\frac{1}{2}\sigma_i\biggl( l_{hom}\vec{\cal P}[W]
+l_{inhom}\rho[W]\biggr)_i
\nonumber\\
&&   =l_{orb}W+\frac{1}{2}\sigma_i\biggl( l_{hom}
\left( \begin{array}{c} 
Tr_{2\times 2}[\sigma_1 W]\\
Tr_{2\times 2}[\sigma_2 W]\\
Tr_{2\times 2}[\sigma_3 W] 
\end{array}\right)
+l_{inhom}Tr_{2\times 2}[W] \biggr)_i \; ,
\nonumber
\end{eqnarray}
in short, we got the linear PDE system:
\begin{eqnarray}                                                            
&&  \hspace{-10mm} 
 \frac{\partial W}{\partial t} 
=l_{orb}W+\frac{1}{2}\sigma_i\biggl( l_{hom}
\left( \begin{array}{c} 
Tr_{2\times 2}[\sigma_1 W]\\
Tr_{2\times 2}[\sigma_2 W]\\
Tr_{2\times 2}[\sigma_3 W] 
\end{array}\right)
+l_{inhom}Tr_{2\times 2}[W] \biggr)_i \; ,
\label{eq:new002.100}                                                          
\end{eqnarray}
which is the evolution equation for $W$.

Having thus shown that (\ref{eq:new002.100}) follows from
(\ref{eq:new02.010}), (\ref{eq:new02.010d}) we now show that the converse is also true.
We first note, by (\ref{eq:new02.10a}),
that (\ref{eq:new002.100}) implies
\begin{eqnarray}                                                            
&&  \hspace{-15mm}  \frac{\partial W}{\partial t} 
=l_{orb}W
+\frac{1}{2}\sigma_i\biggl( l_{hom} \vec{\cal P}[W]
+l_{inhom}\rho[W] \biggr)_i \; .
\label{eq:new02.010new32}   
\end{eqnarray}
We
compute, by (\ref{eq:new02.10000a}), (\ref{eq:new02.10a}), (\ref{eq:new02.010new32}), 
\begin{eqnarray}                                                            
&&  \hspace{-15mm}  
\frac{\partial \rho[W]}{\partial t} 
= \frac{\partial}{\partial t} Tr_{2\times 2}[W] 
=Tr_{2\times 2}[  \frac{\partial W}{\partial t}] 
= Tr_{2\times 2}\Biggl[  l_{orb} W
+\frac{1}{2}\sigma_i \biggl( l_{hom} \vec{\cal P}[W]
+l_{inhom}\rho[W] \biggr)_i
\Biggr\rbrack
\nonumber\\
&& = Tr_{2\times 2}[  l_{orb} W]
= l_{orb} Tr_{2\times 2}[W]
= l_{orb}\rho[W] \; ,
\label{eq:new02.010new32a}   
\end{eqnarray}
so that (\ref{eq:new02.010}) holds. Thus
(\ref{eq:new02.010}) follows from (\ref{eq:new002.100}).
To show that (\ref{eq:new02.010d}) follows from (\ref{eq:new002.100}) we multiply
(\ref{eq:new02.010new32}) from the left by $\sigma_j$ resulting in
\begin{eqnarray}                                                            
&&  \hspace{-20mm} \sigma_j \frac{\partial W}{\partial t} 
= \sigma_j l_{orb} W
+\frac{1}{2}\sigma_j\Biggl(
\sigma_i\biggl( l_{hom} \vec{\cal P}[W]
+l_{inhom}\rho[W] \biggr)_i\Biggr) \; ,
\label{eq:new02.010new33}   
\end{eqnarray}
where $j=1,2,3$. Note that (\ref{eq:new02.010new33}) follows from (\ref{eq:new002.100}) because
(\ref{eq:new02.010new32}) follows from (\ref{eq:new002.100}).
It follows from (\ref{eq:new02.10000a}), (\ref{eq:new02.10a}), (\ref{eq:new02.010new33})
that, for $j=1,2,3$,
\begin{eqnarray}                                                            
&& \hspace{-15mm}
\frac{\partial {\cal P}_j[W]}{\partial t} = \frac{\partial}{\partial t} 
Tr_{2\times 2}[\sigma_j W] 
=Tr_{2\times 2}[\sigma_j\frac{\partial W}{\partial t}] 
\nonumber\\
&& 
= Tr_{2\times 2}\Biggl[\sigma_j l_{orb} W
+\frac{1}{2}\sigma_j\Biggl(
\sigma_i\biggl( l_{hom} \vec{\cal P}[W]
+l_{inhom}\rho[W] \biggr)_i\Biggr)\Biggr\rbrack
\nonumber\\
&& = l_{orb} Tr_{2\times 2}[\sigma_j W] 
+\frac{1}{2}Tr_{2\times 2}\Biggl[  
\sigma_j\sigma_i
\biggl(  l_{hom} \vec{\cal P}
+   l_{inhom}\rho\biggr)_i\Biggr\rbrack
\nonumber\\
&& =  l_{orb} {\cal P}_j[W]
+\biggl(  l_{hom}\vec{\cal P}[W]
+ l_{inhom}\rho[W]\biggr)_j  \; ,
\nonumber
\end{eqnarray}
so that $\frac{\partial \vec{\cal P}[W]}{\partial t} 
= ( l_{orb}+l_{hom})\vec{\cal P}[W]
+ l_{inhom}\rho[W]$, i.e., (\ref{eq:new02.010d}) holds. Thus (\ref{eq:new02.010d}) follows from (\ref{eq:new002.100}).

Therefore (\ref{eq:new002.100}) implies (\ref{eq:new02.010}), (\ref{eq:new02.010d}) 
which completes the proof that the linear PDE system, 
(\ref{eq:new002.100}), is equivalent to the linear PDE system,
(\ref{eq:new02.010}), (\ref{eq:new02.010d}). 
While (\ref{eq:new02.010}), (\ref{eq:new02.010d}) are the central piece of \cite{DK75}
we note that
(\ref{eq:new002.100}) does not occur in \cite{DK75}
(where $W$ is merely mentioned).
However $W$ plays a key role in \cite{Kon82}
(see some related remarks in Section 5.5).

\subsection{The statistical conditions. The spin-$1/2$ Wigner function of a bunch}

To arrive at the notion of
the spin-$1/2$ Wigner function of a bunch we need statistical conditions on $W$ (which will lead us
to statistical conditions on $\rho[W]$ and $\vec{\cal P}[W]$ as well).
We require
the following three conditions on $W$ (which for lack of a better word we call
statistical conditions)
to be valid for $t\in\R,q\in\R^6$,
\begin{eqnarray}                                                            
&&  \hspace{-15mm} \biggl( W(t,q)\biggr)^\dagger = W(t,q) \; , 
\label{eq:new02.10c}  \\
&&  \hspace{-15mm}
\int_{\R^6} \; Tr_{2\times2}\biggl[W(t,\tilde{q})\biggr] d^6 \tilde{q} = 1 \; ,
\label{eq:new02.10d}\\
&&  \hspace{-15mm}
\Big{|} \int_{\R^6} \; 
\left( \begin{array}{c} 
Tr_{2\times 2}[\sigma_1 W(t,\tilde{q})]\\
Tr_{2\times 2}[\sigma_2 W(t,\tilde{q})]\\
Tr_{2\times 2}[\sigma_3 W(t,\tilde{q})] 
\end{array}\right) 
d^6\tilde{q} \Big{|} \leq 1 \; ,
\label{eq:new02.10k}                                                           
\end{eqnarray}
where $\dagger$ denotes the complex conjugate of the transpose of a matrix
and where (\ref{eq:new02.10c}) is the common hermiticity condition on Wigner functions
while (\ref{eq:new02.10d}) reflects the well known statistical interpretation of Wigner functions and
(\ref{eq:new02.10k}) is the normalization condition of the polarization.
For more details on (\ref{eq:new02.10c}), (\ref{eq:new02.10d}), (\ref{eq:new02.10k})
see the discussion after (\ref{eq:new02.14new}).

We thus arrive at the notion of
the spin-$1/2$ Wigner function of a bunch:
We call $W$ the spin-$1/2$ Wigner function of a bunch iff it satisfies (\ref{eq:new02.10c}), (\ref{eq:new02.10d}), (\ref{eq:new02.10k})
and the evolution equation, (\ref{eq:new002.100}).

The statistical conditions on $W$ can be expressed in terms of 
its building blocks, $\rho[W]$ and $\vec{\cal P}[W]$ as follows.
First of all,
since $I_{2\times 2}$ and $\sigma_1,\sigma_2,\sigma_3$ are Hermitian
$2\times 2$-matrices, it follows from (\ref{eq:new02.10a}) that 
(\ref{eq:new02.10c}) is equivalent to
\begin{eqnarray}                                                            
&&   \rho[W](t,q)\in\R \; , \quad \vec{\cal P}[W](t,q)\in\R^3 \; ,
\label{eq:new02.10e}  
\end{eqnarray}
where $t\in\R,q\in\R^6$.
Secondly, it follows from (\ref{eq:new02.10a}) that 
(\ref{eq:new02.10d}) is equivalent to
\begin{eqnarray}                                                            
&& \int_{\R^6} \; \rho[W](t,q)d^6q   = 1 \; ,
\label{eq:new02.10f}                                                           
\end{eqnarray}
where $t\in\R$.
Thirdly, we note, by (\ref{eq:new02.10a}) and for $t\in\R$, that 
\begin{eqnarray}                                                                
&& \hspace{-15mm}  \int_{\R^6} \; \left( \begin{array}{c} 
Tr_{2\times 2}[\sigma_1 W(t,q)]\\
Tr_{2\times 2}[\sigma_2 W(t,q)]\\
Tr_{2\times 2}[\sigma_3 W(t,q)] 
\end{array}\right) 
d^6q = \int_{\R^6} \; \vec{\cal P}[W](t,q)  d^6q\; ,
\label{eq:new002.14new}                                                                
\end{eqnarray}
which entails that (\ref{eq:new02.10k}) is equivalent to:
\begin{eqnarray}                                                                
&& \big{|}\int_{\R^6} \; \vec{\cal P}[W](t,q)  d^6q\big{|}  \leq  1\; .
\label{eq:new02.15}                                                                
\end{eqnarray}
Fourthly, we conclude that (\ref{eq:new02.10c}), (\ref{eq:new02.10d}) and (\ref{eq:new02.10k}) are equivalent to (\ref{eq:new02.10e}), (\ref{eq:new02.10f}) and (\ref{eq:new02.15}). 

The conditions (\ref{eq:new002.100}), (\ref{eq:new02.10c}), (\ref{eq:new02.10d}), (\ref{eq:new02.10k}) on  $W$ are equivalent to the conditions,
(\ref{eq:new02.010}), (\ref{eq:new02.010d}), 
(\ref{eq:new02.10e}), (\ref{eq:new02.10f}), (\ref{eq:new02.15}),
on $\rho[W],\vec{\cal P}[W]$. We thus call
$\rho[W]$ the orbital density and $\vec{\cal P}[W]$ the polarization density
of a bunch (or, shortly, we say: $\rho[W],\vec{\cal P}[W]$ are physically meaningful) iff 
$\rho[W],\vec{\cal P}[W]$ satisfy (\ref{eq:new02.010}), (\ref{eq:new02.010d}), 
(\ref{eq:new02.10e}), (\ref{eq:new02.10f}), (\ref{eq:new02.15}), i.e.,
iff $W$ is the spin-$1/2$ Wigner function of a bunch.

As an aside we make a remark on (\ref{eq:new02.10k}).
In fact we briefly discuss the option of strengthening
(\ref{eq:new02.10k}) to
\begin{eqnarray}                                                            
&&  \big{|}\vec{\cal P}[W]\big{|} \leq  \rho[W] \; ,
\label{eq:new02.12}                                                           
\end{eqnarray}
while sticking to (\ref{eq:new02.10c}) and (\ref{eq:new02.10d}).
For later reference we note that (\ref{eq:new02.12}) implies
\begin{eqnarray}                                                                
&& \rho[W] \geq 0 \; .
\label{eq:new2.013fa}                                                                
\end{eqnarray}
To show that (\ref{eq:new02.12}) 
is stronger than (\ref{eq:new02.10k}) we first compute, by (\ref{eq:new02.10f}), (\ref{eq:new002.14new}) and (\ref{eq:new02.12})
and for $t\in\R$,
\begin{eqnarray}                                                                
&& 
\Big{|} \int_{\R^6} \;  \left( \begin{array}{c} 
Tr_{2\times 2}[\sigma_1 W(t,q)]\\
Tr_{2\times 2}[\sigma_2 W(t,q)]\\
Tr_{2\times 2}[\sigma_3 W(t,q)] 
\end{array}\right) d^6 q \Big{|}
=\big{|}\int_{\R^6} \; \vec{\cal P}[W](t,q)  d^6q\big{|}  
\nonumber\\
&& \hspace{-5mm} \leq  \int_{\R^6} \;\big{|} \vec{\cal P}[W](t,q) \big{|} d^6q
\leq  \int_{\R^6} \; \rho[W](t,q) d^6q = 1\; ,
\label{eq:new02.15new}                                                                
\end{eqnarray}
so that (\ref{eq:new02.12}) 
is at least as strong as (\ref{eq:new02.10k}).
Secondly, it is easy to choose $W(0,\cdot)$ such that, at $t=0$,
(\ref{eq:new02.10k}) holds and (\ref{eq:new02.12}) does not hold. The `trick'
here is to modify, if necessary, $\rho[W](0,\cdot)$, e.g., by a scaling transformation by which
$\rho[W](0,q)$ is transformed to $\eta^6\rho[W](0,\eta q)$, where $\eta$ is a positive constant
while $\vec{\cal P}[W](0,\cdot)$ is kept unchanged
(we leave the details to the reader).
Thus indeed (\ref{eq:new02.12}) is stronger than (\ref{eq:new02.10k}). 
The strengthening, (\ref{eq:new02.12}), 
of (\ref{eq:new02.10k}) can be utilized
to interprete certain features of the kinetic approach (see the discussion
after (\ref{eq:new02.16dnew})).  
However in the present work we do not use (\ref{eq:new02.12})
except when explicitly mentioned.
Even the condition, (\ref{eq:new2.013fa}) (which is an implication of 
(\ref{eq:new02.12})), is not assumed in the present work except when explicitly mentioned.
Nevertheless (\ref{eq:new2.013fa}) is expected to hold for a realistic bunch.

\subsection{The spin-$1/2$ observables and their expectation values.
The polarization vector and polarization}

An arbitrary (scalar) observable is, in the
spin-$1/2$ Wigner function formalism, a function $A:\R^7\rightarrow\C^{2\times 2}$ 
whose values are Hermitian matrices, i.e., satisfies, for $t\in\R,q\in\R^6$,
$\biggl( A(t,q)\biggr)^\dagger = A(t,q)$
%
and thus 
\begin{equation}                                                            
 A(t,q) = I_{2\times 2}A_{orb}(t,q) + \sigma_i A_{spin,i}(t,q) \; ,
\label{eq:new02.10g}                                                           
\end{equation}
where $A_{orb}(t,q)\in\R$ and $\vec{A}_{spin}(t,q)\in\R^3$.
If $W$ is the spin-$1/2$ Wigner function of a bunch, i.e., if
(\ref{eq:new002.100}), (\ref{eq:new02.10c}), (\ref{eq:new02.10d}) and (\ref{eq:new02.10k}) holds
we define the function $\langle{A}\rangle_W:\R\rightarrow\R$ by
\begin{eqnarray}
&& \hspace{-10mm}
\langle{A}\rangle_W(t):=\int_{\R^6} \; Tr_{2\times 2}\biggl[A(t,q)W(t,q)\biggr] d^6q
\nonumber\\
&&  \hspace{-20mm}
=\int_{\R^6} \; \biggl( A_{orb}(t,q)  \rho[W](t,q) + A_{spin,i}(t,q){\cal P}_i[W](t,q)\biggr)d^6q \; ,
\label{eq:new02.10h}
\end{eqnarray}
where we also used (\ref{eq:new02.10}), (\ref{eq:new02.10a}),
and (\ref{eq:new02.10g}).
We call $\langle{A}\rangle_W(t)$ the expectation value of $A$ w.r.t. $W$ at time $t$.
The expectation values of vector observables like $\vec{\sigma}$ are defined by using (\ref{eq:new02.10h}) componentwise.
For example
\begin{eqnarray}                                                                
&& \langle\vec{\sigma}\rangle_W(t):=\int_{\R^6} \; 
\left( \begin{array}{c} 
Tr_{2\times 2}[\sigma_1 W(t,q)]\\
Tr_{2\times 2}[\sigma_2 W(t,q)]\\
Tr_{2\times 2}[\sigma_3 W(t,q)] 
\end{array}\right) d^6q 
=\int_{\R^6} \; \vec{\cal P}[W](t,q)  d^6q \; ,
\label{eq:new002.14}                                                                
\end{eqnarray}
where in the second equation of (\ref{eq:new002.14}) we used 
(\ref{eq:new002.14new}).
For later reference we note, by (\ref{eq:new002.14}), 
that the statistical condition, (\ref{eq:new02.10k}), is equivalent to
\begin{eqnarray}                                                                
&& |\langle\vec{\sigma}\rangle_W(t)| \leq 1 \; .
\label{eq:new0002.14}                                                                
\end{eqnarray}
If $W$ is the spin-$1/2$ Wigner function of a bunch then,
motivated by (\ref{eq:new002.14}), we define the  function
$\vec{P}[W]:\R\rightarrow\R^3$, by
\begin{eqnarray}                                                                
&& \vec{P}[W](t):=\langle\vec{\sigma}\rangle_W(t)
=\int_{\R^6} \; \vec{\cal P}[W](t,q)  d^6q \; ,
\label{eq:new02.14}                                                                
\end{eqnarray}
where we also used (\ref{eq:new002.14}).   

By (\ref{eq:new02.14}),
$\vec{P}[W](t)$ is the expectation value of $\vec{\sigma}$ w.r.t. $W$ at time, $t$,
and this underscores the importance of expectation values.
Note that the expectation values of 
observables will also play a role in Section 12.
If $W$ is the spin-$1/2$ Wigner function of a bunch
then by the above $\vec{\cal P}[W]$ is the polarization density
of a bunch and thus, by (\ref{eq:new02.14}),
$\vec{P}[W](t)$ is the so-called polarization vector of a bunch (at time, $t$) 
and therefore $|\vec{P}[W](t)|$ is the so-called polarization of a bunch (at time, $t$).
Note, by (\ref{eq:new0002.14})  and (\ref{eq:new02.14}), that the statistical condition, (\ref{eq:new02.10k}),
is equivalent to
\begin{eqnarray}                                                                
&& |\vec{P}[W](t)|\leq 1  \; .
\label{eq:new02.14new}                                                                
\end{eqnarray}

With the above we can now justify  our statistical conditions,
(\ref{eq:new02.10c}), (\ref{eq:new02.10d}) and (\ref{eq:new02.10k}),
in more detail.
First of all, orbital densities are $\R$-valued
and polarization densities are $\R^3$-valued, i.e., 
satisfy (\ref{eq:new02.10e}). Also, from the discussion before (\ref{eq:new02.10e}) we know
that (\ref{eq:new02.10e}) is equivalent to
(\ref{eq:new02.10c}).
Secondly, orbital densities are normalized by (\ref{eq:new02.10f}).         
Also, from the discussion before (\ref{eq:new02.10f}) we know that
(\ref{eq:new02.10f}) is equivalent to
(\ref{eq:new02.10d}). For another justification of (\ref{eq:new02.10c}), (\ref{eq:new02.10d}), see
the discussion before (\ref{eq:new2.100b}).
Thirdly, polarization vectors satisfy (\ref{eq:new02.14new}) which, by the discussion before
(\ref{eq:new02.14new}), is equivalent to (\ref{eq:new02.10k}).

\subsection{Relating the statistical conditions with the dynamics}

We now discuss how the statistical conditions, 
(\ref{eq:new02.10c}), (\ref{eq:new02.10d}) and (\ref{eq:new02.10k}),
on $W$ fit to (\ref{eq:new002.100}), i.e., to the dynamics of $W$.
First of all, we recall that (\ref{eq:new02.10c}) is equivalent to (\ref{eq:new02.10e}).
Moreover since the orbital Fokker-Planck equation is a Fokker-Planck equation
we observe that $\rho[W]$ is $\R$-valued if $\rho[W](0,\cdot)$ is $\R$-valued.
Furthermore since
the functions ${\cal D}_{orb}, {\cal D}_{spin,+}, \vec{\cal D}_{spin,0,ST}, {\cal B}_{orb}, \vec{\cal B}_{spin}$
are $\R^l$-valued (for positive integer $l$)
and since (\ref{eq:new02.010a}) holds we observe that
$\vec{\cal P}[W]$ is $\R^3$-valued if $\vec{\cal P}[W](0,\cdot)$ is $\R^3$-valued.
This guarantees
that (\ref{eq:new02.10e}) holds for all $t$ if it holds for $t=0$ and if the
coefficient functions of (\ref{eq:new002.100}) are sufficiently regular.
Thus, by the equivalence of (\ref{eq:new02.10c}) and (\ref{eq:new02.10e}) we indeed observe that
(\ref{eq:new02.10c}) holds for all $t$ if it holds for $t=0$.
Secondly, it follows from (\ref{eq:new02.010new32a}) that
\begin{eqnarray}                                                            
&&  
\frac{\partial}{\partial t} Tr_{2\times 2}[W] 
= l_{orb} Tr_{2\times 2}[W]\; ,
\nonumber
\end{eqnarray}
so that, for $t\in\R$ and by (\ref{eq:new02.010a}),
\begin{eqnarray}                                                            
&&  \hspace{-10mm} \frac{d}{dt} \int_{\R^6} \; Tr_{2\times 2}[W(t,q)]d^6 q
 =\int_{\R^6} \; \frac{\partial}{\partial t} Tr_{2\times 2}[W(t,q)] d^6 q 
=\int_{\R^6} \; l_{orb} Tr_{2\times 2}[W(t,q)]d^6 q 
\nonumber\\
&&\hspace{-10mm} 
=\int_{\R^6} \;\Biggl( \biggl(- \frac{\partial}{\partial q_k} {\cal D}_{orb,k}(t,q)
+\frac{1}{2}\frac{\partial}{\partial q_k} \frac{\partial}{\partial q_l} 
{\cal B}_{orb,k}(t,q)  {\cal B}_{orb,l}(t,q)\biggr)Tr_{2\times 2}[W(t,q)]\Biggr)d^6 q 
= 0 \; .
\label{eq:new002.101}                                                           
\end{eqnarray}
It follows from 
(\ref{eq:new002.101}) that the normalization condition, (\ref{eq:new02.10d}), obeys the following conservation law:
(\ref{eq:new02.10d}) holds for all $t$ if it holds for $t=0$ and if the
coefficient functions of (\ref{eq:new002.100}) are sufficiently regular.
Thirdly, in contrast to the constraints of
(\ref{eq:new02.10c}) and (\ref{eq:new02.10d}) we are not aware 
if (\ref{eq:new02.10k}) fits to the dynamics of $W$ 
(thus we are not aware if
(\ref{eq:new02.15}), (\ref{eq:new0002.14}) or (\ref{eq:new02.14new}) 
fit to the dynamics of $W$ either).
Therefore we leave the following question open: Does (\ref{eq:new02.10k}) hold 
for all $t$ if it holds for $t=0$? 
Note that for the unknown `true'
dynamics of the bunch the answer would be: yes.
Also we believe that (\ref{eq:new002.100}) is a good approximation of the `true' 
dynamics of the bunch and thus we believe
that in general the answer to the above question
is: yes (if the coefficient functions of (\ref{eq:new002.100}) are sufficiently regular).
Nevertheless the present work 
does not rely on the positive answer to this question, except when explicitly mentioned 
(namely in Section 15.1).
Fourthly, we
will reformulate, in Section 12, the statistical conditions on $W$, as statistical
conditions on the spin-orbit density $f$, i.e., as statistical conditions for the extension of
the kinetic approach.

\subsection{Miscellanea}

We now wrap up this section with final comments on $W$.
First of all, the evolution equation for $W$ was derived 
from the QED of the Dirac equation 
in \cite{Kon82} (recall from Section 1.2 that Kondratenko's thesis underlies \cite{DK75}).
Kondratenko's derivation rests on the fact that
$W$ is the so-called Wigner-Weyl transform of the density operator of the bunch
and it relies on various approximations, e.g.,
a semiclassical treatment and 
a Markov approximation (the latter necessitated after having traced over unwelcome degrees of
freedom, e.g., photon degrees of freedom).
Also, in order to separate the electron from the positron
degrees of freedom in the Dirac equation, a (fully relativistic) Foldy-Wouthuysen transformation is employed
(not the semirelativistic Foldy-Wouthuysen transformation
known from Atomic Physics!).
In short,
the spin-$1/2$ Wigner function of a bunch is a Wigner-Weyl transform
of an approximation
of the density operator of the true bunch (this fact can also be used as another justification of
(\ref{eq:new02.10c}), (\ref{eq:new02.10d})). By \cite{Kon82}
the time evolution takes into account the effect of the external electromagnetic field and
the associated (incoherent) synchrotron radiation on the bunch.
Other effects like coherent synchrotron radiation effects,
intrabeam scattering, the weak-strong beam-beam effect and radiative 
$g-2$ terms \cite{Man87-2}
may be included as well (but this is not pursued in the present work). 
Secondly, in the reduced setup, (\ref{eq:new02.010new30a}), the evolution   
equation, (\ref{eq:new002.100}), simplifies to:
\begin{eqnarray} 
&& \hspace{-10mm}
 \frac{\partial W}{\partial t} 
=l_{orb}W+\frac{1}{2}\sigma_i\biggl( 
\vec{\Omega}\times
\left( \begin{array}{c} 
Tr_{2\times 2}[\sigma_1 W]\\
Tr_{2\times 2}[\sigma_2 W]\\
Tr_{2\times 2}[\sigma_3 W] 
\end{array}\right)
\biggr)_i \; ,
\label{eq:new2.100b}                                                           
\end{eqnarray}
where we also used (\ref{eq:new2.99a}), (\ref{eq:new2.100d}).
We recall from Section 1.2 that (\ref{eq:new2.100b}),
contains all the information needed to study the
radiative depolarization effect
(by the discussion before (\ref{eq:new02.010new11})
this is even true if one neglects $\vec{\cal Q}$ in (\ref{eq:new2.100b})).
Even in the reduced setup we stick to our language used in this section.
In particular in analogy to the discussion after (\ref{eq:new02.15}) 
we call, in the reduced setup, $W$ the spin-$1/2$ Wigner function of a bunch iff $W$ satisfies
(\ref{eq:new02.10c}), (\ref{eq:new02.10d}), (\ref{eq:new02.10k}) and (\ref{eq:new2.100b}).
Moreover in analogy to the discussion after (\ref{eq:new02.15}) we call, in the reduced setup, 
$\rho[W]$ the orbital density and $\vec{\cal P}[W]$ the polarization density
of a bunch iff $W$ is the spin-$1/2$ Wigner function of a bunch, i.e., iff
$\rho[W],\vec{\cal P}[W]$ satisfy (\ref{eq:new02.10e}), (\ref{eq:new02.10f}), (\ref{eq:new02.15})  
and the orbital Fokker-Planck equation as well as the reduced Bloch 
equation.
Furthermore we stick to the definition of $\vec{P}[W](t)$: 
If in the reduced setup $W$ is the spin-$1/2$ Wigner function of a bunch
then by the above $\vec{\cal P}[W]$ is the polarization density
of a bunch and thus, by (\ref{eq:new02.14}),
$\vec{P}[W](t)$ is the so-called polarization vector of a bunch (at time, $t$) 
and therefore $|\vec{P}[W](t)|$ is the so-called polarization of a bunch (at time, $t$).
Thirdly, we note that there is an extensive literature on spin-$1/2$ Wigner functions.
For example the reader may consult
\cite{OCW84}. Note also that spin-$1/2$ Wigner functions are sometimes called Stratonovich functions.

\section{The local polarization vector field}

To keep the discussion concise
we assume in this section that
\begin{eqnarray}                                                                
&& \rho[W]>0 \; .
\label{eq:new02.16b}                                                                
\end{eqnarray}
Following \cite{BH01}, we define
\begin{eqnarray}                                                                
&& \hspace{-10mm} \vec{\cal P}_{loc}[W](t,q):=
\frac{\vec{\cal P}[W](t,q)}{\rho[W](t,q)} \; ,
\label{eq:new02.16}                                                                
\end{eqnarray}
so that
\begin{eqnarray}                                                                
&& \hspace{-10mm} \vec{\cal P}[W](t,q)=
\rho[W](t,q)  \vec{\cal P}_{loc}[W](t,q) \; ,
\label{eq:new02.16c}                                                                
\end{eqnarray}
which entails, by (\ref{eq:new02.14}), 
\begin{eqnarray}                                                                
&& \hspace{-10mm} \vec{P}[W](t)=
\int_{\R^6} \; \rho[W](t,q)  \vec{\cal P}_{loc}[W](t,q) d^6q \; .
\label{eq:new02.16a}                                                                
\end{eqnarray}
If $W$ is the spin-$1/2$ Wigner function of a bunch, i.e., if it
satisfies (\ref{eq:new002.100}), (\ref{eq:new02.10c}), (\ref{eq:new02.10d}) and (\ref{eq:new02.10k}) 
that is, if $\rho[W]$ is the orbital density and $\vec{\cal P}[W]$ is the polarization density of a
bunch (recall the discussion after (\ref{eq:new02.15}))
then, by following \cite{BH01}, we call $\vec{\cal P}_{loc}[W]$ the local polarization vector field.
The fact that we define the local polarization vector field only in a special case, namely (\ref{eq:new02.16b}), indicates that
this field only plays a side role in this work.

In the remaining parts of this section we comment on the relation between 
(\ref{eq:new02.12}) and $\vec{\cal P}_{loc}[W]$.
First of all, by (\ref{eq:new02.16b}),
\begin{eqnarray}                                                                
&& \hspace{-10mm} |\vec{\cal P}[W]| =
|\rho[W]|\;|\vec{\cal P}_{loc}[W]| 
=\rho[W] |\vec{\cal P}_{loc}[W]| \; .
\label{eq:new02.16cnew}                                                                
\end{eqnarray}
Secondly, if 
\begin{eqnarray}                                                            
&&  |\vec{\cal P}_{loc}[W]|  \leq 1  \; ,
\label{eq:new02.12a}                                                           
\end{eqnarray}
then, due to (\ref{eq:new02.16cnew}), we get
(\ref{eq:new02.12}), i.e., $|\vec{\cal P}[W]| \leq  \rho[W]$.
Thirdly, it follows from (\ref{eq:new02.12}), (\ref{eq:new02.16cnew}) that
\begin{eqnarray}                                                                
&& \hspace{-10mm} \rho[W]\;|\vec{\cal P}_{loc}[W]| \leq \rho[W] \; ,
\label{eq:new02.16dnew}                                                                
\end{eqnarray}
so that, by (\ref{eq:new02.16b}), we get (\ref{eq:new02.12a}).
Thus (\ref{eq:new02.12a}) is equivalent to the strengthening,
(\ref{eq:new02.12}), of (\ref{eq:new02.10k}).
Fourthly, if (\ref{eq:new02.12}) holds at a point, $(t,q)$, i.e., if
(\ref{eq:new02.12a}) holds at $(t,q)$
then, because of (\ref{eq:new02.16a}), one 
may interprete, $\vec{\cal P}_{loc}[W](t,q)$, as
the spin polarization vector at $(t,q)$ namely the ensemble average  of normalized 
single-particle spin vectors at that point. 
In the opposite case, where at $(t,q)$
(\ref{eq:new02.10k}) holds but where (\ref{eq:new02.12}) 
does not hold, one may interprete, $\vec{\cal P}_{loc}[W](t,q)$, as
the weighted spin polarization vector at $(t,q)$ where the possibly $t$-dependent
weight may be defined in the vein of the discussion after
(\ref{eq:new02.15new}), i.e., via a scaling transformation of $\rho[W]$. 
These interpretations of $\vec{\cal P}_{loc}[W]$
will be used in Section 10 (see the discussion after
(\ref{eq:new2.44vcdm})).

\section{The dynamical condition used to derive the full spin-orbit Fokker-Planck equation}

Sections 7-10 are focused on the following task: Translate, if possible, the dynamical information of the kinetic approach, i.e., the information contained in 
the evolution equation, (\ref{eq:new002.100}), of $W$
into a Fokker-Planck equation on the spin-orbit phase space, $\R^9$!
Such an equation we call full spin-orbit Fokker-Planck equation (and, in the reduced setup,
the reduced spin-orbit Fokker-Planck equation).
In the full setup, i.e., when (\ref{eq:new02.010}), (\ref{eq:new02.010d}), 
(\ref{eq:new002.100}) hold
the aforementioned task was fulfilled in \cite{HABBE19} resulting in
the full spin-orbit Fokker-Planck equation, (\ref{eq:new2.44vanew}), which is presented in
Section 10 together with an associated Ito SDE system. In fact,
being a Fokker-Planck equation, (\ref{eq:new2.44vanew}) has the same dynamical
content as each of its associated Ito SDE systems.
In \cite{BH01} a derivation of the reduced spin-orbit Fokker-Planck equation was given
for the (common) case where $\vec{Q}=\vec{0}$ and this derivation was partly
based on physical intuition, whereas so far we did not publish a derivation of the
full spin-orbit Fokker-Planck equation (not even in \cite{HABBE19}).
Thus the derivation of the full spin-orbit Fokker-Planck equation, which will be accomplished in
Section 9, is an important piece of the present work.
The key element of our derivation is the so-called dynamical condition
which is the topic of this section.

Since (\ref{eq:new2.100b}) is a special case of
(\ref{eq:new002.100}) it is no surprise that the reduced spin-orbit Fokker-Planck equation,
(\ref{eq:new2.44vanewnewred}), is a special case of (\ref{eq:new2.44vanew}).
Thus in our work of Sections 7-10 we focus on the full setup and do not have to care
about the reduced setup (the reduced setup is covered in Section 11).

Before we define the dynamical condition we need to make some preparations and we first
note that the
full spin-orbit Fokker-Planck equation, being a Fokker-Planck equation, 
can be formally written, for $f=f(t,q,\vec{s})$, as
\begin{eqnarray}                                                            
&&  \frac{\partial f}{\partial t} = -\frac{\partial}{\partial y_j} (d_{orb,spin,j}f)
+\frac{1}{2}\frac{\partial}{\partial y_j} \frac{\partial}{\partial y_k} 
\biggl( (b_{orb,spin}b_{orb,spin}^T)_{k,j}   f\biggr) \; ,
\label{eq:new2.44v}                                                           
\end{eqnarray}
where $d_{orb,spin}$ is $\R^9$-valued and
$b_{orb,spin}$ is $\R^{9\times K}$-valued with $K$ being a positive integer and
\begin{eqnarray}                                                            
&& y \equiv \left( \begin{array}{c}  q \\
 \vec{s} \end{array}\right) \; .
\label{eq:new2.44ube}                                                           
\end{eqnarray}
Thus the aforementioned task of Sections 7-10 boils down to find $d_{orb,spin}$ and
$b_{orb,spin}$. 
One calls the $\R^9$-valued function, $d_{orb,spin}$, the drift vector field and 
the $\R^{9\times 9}$-valued function, $b_{orb,spin}b_{orb,spin}^T$, 
the diffusion matrix field of (\ref{eq:new2.44v}).
For the notions of Fokker-Planck equation, drift vector field and diffusion matrix field 
see again \cite{Arn74,Gar04}.
We call the $\R^{9\times K}$-valued function, $b_{orb,spin}$, a
noise matrix field of (\ref{eq:new2.44v}).

Since $b_{orb,spin}$ is a
noise matrix field of (\ref{eq:new2.44v}), the following is an
Ito SDE system associated with
the formal Fokker-Planck equation, (\ref{eq:new2.44v}):
\begin{eqnarray}                                                            
&&   \hspace{-10mm}  \left( \begin{array}{c}  
Q' \\ 
\vec{S}' \end{array}\right)
= d_{orb,spin}(t,Q,\vec{S}) + b_{orb,spin}(t,Q,\vec{S})\omega(t) \; ,
\label{eq:new2.44uc}                                                           
\end{eqnarray}
where $\omega$ is the $K$-dimensional white-noise process.

How do we arrive at a dynamical condition which will allow us to translate the
information in (\ref{eq:new002.100}) into, (\ref{eq:new2.44v})?
Answer: Following \cite{BH01} there is a natural way to translate $f$ into a 
spin-$1/2$ Wigner function (which we denote by $W_f$).
In fact it is natural to mimick the trace operation in (\ref{eq:new02.10a}) by the operation
of $\vec{s}$-integration over $\R^3$ giving us the following ansatz 
for $\rho[W_f],\vec{\cal P}[W_f]$:
\begin{eqnarray}                                                            
&&   
\hspace{-20mm}
\rho[W_f](t,q) =\int_{\R^3} \; f(t,q,\vec{s})d^3s  \; , \quad
\vec{\cal P}[W_f](t,q) =\int_{\R^3} \; 
\left( \begin{array}{c} 
s_1 f(t,q,\vec{s}) \\
s_2 f(t,q,\vec{s}) \\
s_3 f(t,q,\vec{s}) 
\end{array}\right) d^3s
\nonumber\\
&&
=\int_{\R^3} \; \vec{s} f(t,q,\vec{s})d^3s \; ,
\nonumber
\end{eqnarray}
and thus, by using (\ref{eq:new02.10}), 
we arrive at the following ansatz for $W_f$:
\begin{eqnarray}                                                            
&&  \hspace{-15mm} W_f(t,q):= \frac{1}{2}\biggl(  I_{2\times 2}\int_{\R^3}f(t,q,\vec{s})d^3s 
+ \sigma_i \int_{\R^3} \; s_i f(t,q,\vec{s})d^3s\biggr) \; .
\label{eq:new2.44r}                                                       
\end{eqnarray}
For later reference we note, by (\ref{eq:new02.10a}), (\ref{eq:new2.44r}), that
\begin{eqnarray}                                                            
&&   
\hspace{-20mm}
\rho[W_f](t,q) =Tr_{2\times 2}[W_f(t,q)]
=\int_{\R^3} \; f(t,q,\vec{s})d^3s  \; ,
\label{eq:new2.44p} \\ 
&& 
\hspace{-20mm} 
\vec{\cal P}[W_f](t,q)= 
\left( \begin{array}{c} 
Tr_{2\times 2}[\sigma_1 W_f(t,q)]\\
Tr_{2\times 2}[\sigma_2 W_f(t,q)]\\
Tr_{2\times 2}[\sigma_3 W_f(t,q)] 
\end{array}\right) 
=\int_{\R^3} \; \vec{s} f(t,q,\vec{s})d^3s \; ,
\label{eq:new2.44q}                                                           
\end{eqnarray}
so that, by (\ref{eq:new2.44r}),   
\begin{eqnarray}                                                            
&&  \hspace{-22mm} W_f(t,q) = \frac{1}{2}\biggl(  I_{2\times 2}\rho[W_f](t,q)
+ \sigma_i {\cal P}_i[W_f](t,q)\biggr)  \; .
\end{eqnarray}
Note that (\ref{eq:new2.44p}), (\ref{eq:new2.44q}) explicitly display the mimicking 
of $Tr_{2\times 2}[\cdots]$ by $\int_{\R^3} \; d^3s\cdots$.

With the correspondence, (\ref{eq:new2.44r}),
between $f$ and $W_f$ 
we can now define the dynamical condition which reads as follows:
If $f$ satisfies, (\ref{eq:new2.44v}), and if $W_f$ in (\ref{eq:new2.44r}) exists and is sufficiently
regular then
$W_f$ satisfies, (\ref{eq:new002.100}), i.e.,
\begin{eqnarray}                                                            
&&  \hspace{-10mm} 
 \frac{\partial W_f}{\partial t} 
=l_{orb}W_f+\frac{1}{2}\sigma_i\biggl( l_{hom}
\left( \begin{array}{c} 
Tr_{2\times 2}[\sigma_1 W_f]\\
Tr_{2\times 2}[\sigma_2 W_f]\\
Tr_{2\times 2}[\sigma_3 W_f] 
\end{array}\right)
+l_{inhom}Tr_{2\times 2}[W_f] \biggr)_i \; .
\label{eq:new002.100f}                                                           
\end{eqnarray}
Recalling from Section 5.1
that (\ref{eq:new002.100}) is equivalent to the
system (\ref{eq:new02.010new0final}), (\ref{eq:new02.010d}) we note that
(\ref{eq:new002.100f}) is equivalent to
\begin{eqnarray}                                                            
&&                                                                          
\hspace{-15mm} 
\frac{\partial \rho[W_f]}{\partial t}  
= - \frac{\partial}{\partial q_j} ({\cal D}_{orb,j}\rho[W_f])
+\frac{1}{2}\frac{\partial}{\partial q_k} \frac{\partial}{\partial q_l} 
({\cal B}_{orb,k}  {\cal B}_{orb,l} \rho[W_f]) \; ,
\label{eq:new02.010f} \\
&&                                                                          
\hspace{-15mm} 
\frac{\partial \vec{\cal P}[W_f]}{\partial t} =(l_{orb} +l_{hom})\vec{\cal P}[W_f]
+ l_{inhom}\rho[W_f] \; .
\label{eq:new02.010dnew}                                                                                                             
\end{eqnarray}
Thus the dynamical condition 
is equivalent to the following criterion: If $f$ satisfies (\ref{eq:new2.44v}) 
and if $\rho[W_f],\vec{\cal P}[W_f]$ exist
and are sufficiently regular then
(\ref{eq:new02.010f}), (\ref{eq:new02.010dnew}) have to hold.
This criterion will be used in Section 9 and will
give us
the full spin-orbit Fokker-Planck equation, (\ref{eq:new2.44vanew}), in Section 10.

In the remaining parts of this section we make some comments on the dynamical condition.
First of all, the correspondence, (\ref{eq:new2.44r}),
between $f$ and $W_f$ is independent of the dynamics of $W_f$, i.e., it would also be used
if one would modify, (\ref{eq:new002.100f}), e.g., by adding further physical effects like the weak-strong beam-beam effect.
Secondly, for an arbitrary solution $f$ of (\ref{eq:new2.44v}),
$W_f$ does not necessarily exist 
(pick for example $f(0,\cdot,\cdot)$ as a constant function!).
Thirdly, if $W_f$ exists then it is uniquely determined by $f$, but
$f$ is not uniquely determined by $W_f$.
Thus $W_f$ and $f$ play similar roles as the electromagnetic
fields and their potentials in classical electrodynamics.
In other words it can happen that, under the dynamical condition, 
solutions $f,g$ of (\ref{eq:new2.44v}) exist such that $W_f=W_g$ and $f\neq g$. 
Explicit examples of $f,g$ with $f\neq g$ 
are presented, for the reduced setup, in Section 14 below, see
(\ref{eq:new2.44vcdn81newnew}), (\ref{eq:new2.44vcdn81newnewnew}).
Fourthly, conditions on $f$ which guarantee that $W_f$ 
is the spin-$1/2$ Wigner function of a bunch will be stated in Section 12 (if these 
conditions hold
then we will call $f$, from Section 12 onwards, a spin-orbit density).
Fifthly, it follows from (\ref{eq:new02.14}), (\ref{eq:new2.44q}) that 
if $W_f$ is the spin-$1/2$ Wigner function of a bunch, e.g., if $f$ is a spin-orbit density 
then the polarization vector $\vec{P}[W_f](t)$ at time $t$
is well defined and reads as
\begin{eqnarray}                                                            
&& \hspace{-10mm} \vec{P}[W_f](t) =\langle\vec{\sigma}\rangle_{W_f}(t)
=\int_{\R^6} \; \vec{\cal P}[W_f](t,q)d^6q 
= \int_{\R^9} \; \vec{s} f(t,q,\vec{s})d^3s d^6q \; .
\label{eq:new2.44vcdn90orbnew}                                                           
\end{eqnarray}
\section{The kinematical conditions used to derive the full spin-orbit Fokker-Planck equation}

We mentioned in Section 7 the nonuniqueness of $f$ in $W_f$ and there is even more to say about
nonuniqueness as follows.
Every choice of the coefficient functions
of the formal Fokker-Planck equation, (\ref{eq:new2.44v}),
which satisfies the dynamical 
condition, reproduces the 
evolution equation, (\ref{eq:new002.100f}), of $W_f$.
Thus the uniqueness of the full spin-orbit
Fokker-Planck equation
is neither required nor known to us (same for the reduced spin-orbit
Fokker-Planck equation).
In other words we are not interested in the question,
whether (\ref{eq:new2.44vanew}) is the only possible choice of (\ref{eq:new2.44v})
which satisfies the dynamical condition. 
Even this can be understood in terms of 
our analogy (from Section 7) with classical electrodynamics: While the electromagnetic fields of 
classical electrodynamics
obey the same PDE system, the potentials 
obey PDE systems which vary (since they may belong to different gauges).

Nevertheless in order to facilitate the specification of the
coefficient functions of (\ref{eq:new2.44v}) (not in order to make the coefficient functions unique!)
we add two kinematical conditions to the dynamical condition (kinematical means non-dynamical).
These two kinematical conditions read as follows:
\begin{itemize}
\item[(i)]  Choose the positive integer, $K$, in $b_{orb,spin}$ as small as possible 
(recall that $b_{orb,spin}$ is $\R^{9\times K}$-valued!).
In other words, first try $K=1$ then try $K=2$ and so on.
\item[(ii)] Let all coefficient functions of the 
formal Fokker-Planck equation, (\ref{eq:new2.44v}),
depend on $t,q,\vec{s}$ such that the dependence on $\vec{s}$ is of at most first order.
\end{itemize}
In the remaining parts of this section we make some comments on (i),(ii).
First of all, (i) is Occam's razor while (ii) is motivated by dealing with a spin-$1/2$ particle
(in fact (ii) would be modified in the case of a particle of spin different from $1/2$).
Secondly, (i) facilitates our work in an obvious way.
In fact we will see in Section 9 that, $K=1$, will suffice.
Thirdly, (ii) facilitates our work since it will ensure that
only few nonphysical entries occur in
(\ref{eq:new2.44vcd}), (\ref{eq:new2.44vcc}) (which are then
eliminated in Section 9.2 via the dynamical condition).
Nevertheless without (ii) we would arrive at the same full spin-orbit Fokker-Planck equation, (\ref{eq:new2.44vanew}).
In contrast, we leave open if we would arrive at (\ref{eq:new2.44vanew}) if we would drop (i).
In other words we leave unanswered the question if a non-scalar white-noise process $\omega$
in (\ref{eq:new2.44uc}) is
consistent with the dynamical condition, i.e., if the condition, $K>1$, is
consistent with the dynamical condition.
However this is not a drawback since,
as mentioned above, the uniqueness of the full spin-orbit Fokker-Planck equation
is not required. Nevertheless, thanks to the naturalness of the two kinematical conditions, (i), (ii),
we believe that the explicit form of the 
full spin-orbit Fokker-Planck equation, (\ref{eq:new2.44vanew}), is the simplest possible one.
Fourthly, as mentioned above, in \cite{HABBE19} we introduced (\ref{eq:new2.44vanew}) without
derivation, i.e., we did not explain in \cite{HABBE19} how we got (\ref{eq:new2.44vanew}).
In fact the unpublished derivation relied on the dynamical condition and parts of the two kinematical conditions
as well as some educated guesses.
Fifthly, in contrast to \cite{HABBE19},
in this work we explicitly derive (\ref{eq:new2.44vanew}) and our derivation here
is completely based on the dynamical condition and the two kinematical conditions and this
puts (\ref{eq:new2.44vanew}) on a firm ground.
It is however
not surprising that we get the same full spin-orbit Fokker-Planck equation 
as in \cite{HABBE19}, namely (\ref{eq:new2.44vanew}),
since our conditions here are very similar to the ones underlying \cite{HABBE19}.
Sixthly, the following question arises:
In which order should the conditions be enforced?
Answer: First enforce the kinematical conditions, (i), (ii), and then enforce 
the dynamical condition! 

\section{Deriving the full spin-orbit Fokker-Planck equation}

As announced in Section 8
we impose here, on the formal Fokker-Planck equation (\ref{eq:new2.44v}), 
the two kinematical conditions and then the 
dynamical condition (thus Section 9 is divided into Sections 9.1 and 9.2).
In particular we will show that the simplest case, $K=1$, 
of the kinematical conditions is sufficient.

\subsection{Enforcing the two kinematical conditions}

We begin with the kinematical condition, (i), whereby in the case, $K=1$, 
$b_{orb,spin}$ is $\R^9$-valued. Because $d_{orb,spin}(t,q,\vec{s})\in\R^9$ and
$b_{orb,spin}(t,q,\vec{s})\in\R^9$ we can write
\begin{eqnarray}                                                            
&&   d_{orb,spin}(t,q,\vec{s}) = \left( \begin{array}{c}  d_{orb}(t,q,\vec{s}) \\
\vec{d}_{spin}(t,q,\vec{s})
\end{array}\right) \; ,
\label{eq:new2.44td}  \\    
\nonumber\\
&&  
b_{orb,spin}(t,q,\vec{s}) = \left( \begin{array}{c}  b_{orb}(t,q,\vec{s}) \\
\vec{b}_{spin}(t,q,\vec{s})
\end{array}\right) \; ,
\label{eq:new2.44tc}                                                           
\end{eqnarray}
where $d_{orb}(t,q,\vec{s})\in\R^6,b_{orb}(t,q,\vec{s})\in\R^6$
and $\vec{d}_{spin}(t,q,\vec{s})\in\R^3,\vec{b}_{spin}(t,q,\vec{s})\in\R^3$.

For any solution, $f$, of  (\ref{eq:new2.44v}) we now compute, by (\ref{eq:new2.44ube}), (\ref{eq:new2.44td}), (\ref{eq:new2.44tc}),
\begin{eqnarray}                                                            
&&   -\frac{\partial}{\partial y_j}(b_{orb,spin,j}f) =
-\frac{\partial}{\partial q_j} (d_{orb,j}f)
-\frac{\partial}{\partial s_j} (d_{spin,j}f) \; ,
\nonumber\\
&&  \frac{\partial}{\partial y_k}\frac{\partial}{\partial y_l}(b_{orb,spin,l} b_{orb,spin,k}f)
= \frac{\partial}{\partial y_k}\biggl(  
\frac{\partial}{\partial q_l}( b_{orb,l} b_{orb,spin,k}f)
+ \frac{\partial}{\partial s_l}(b_{spin,l} b_{orb,spin,k}f)\biggr)
\nonumber\\
&&
= \frac{\partial}{\partial q_k}\frac{\partial}{\partial q_l}( b_{orb,l} b_{orb,k}f)
+ \frac{\partial}{\partial s_k} \frac{\partial}{\partial q_l}(b_{orb,l} b_{spin,k}f)
\nonumber\\
&&
+\frac{\partial}{\partial q_k}\frac{\partial}{\partial s_l}( b_{spin,l} b_{orb,k}f)
+ \frac{\partial}{\partial s_k}\frac{\partial}{\partial s_l}(b_{spin,l} b_{spin,k}f)
\nonumber\\
&&
= \frac{\partial}{\partial q_k}\frac{\partial}{\partial q_l}( b_{orb,l} b_{orb,k}f)
+ 2\frac{\partial}{\partial s_k} \frac{\partial}{\partial q_l}(b_{orb,l} b_{spin,k}f)
+ \frac{\partial}{\partial s_k}\frac{\partial}{\partial s_l}(b_{spin,l} b_{spin,k}f) \; ,
\nonumber
\end{eqnarray}
so that, for the case, $K=1$, 
the formal Fokker-Planck equation, (\ref{eq:new2.44v}), can be written 
under the kinematical condition, (i), as 
\begin{eqnarray}                                                            
&&  \hspace{-10mm} 
 \frac{\partial f}{\partial t} = -\frac{\partial}{\partial q_j} (d_{orb,j}f)
-\frac{\partial}{\partial s_j} (d_{spin,j}f) 
+\frac{1}{2}\frac{\partial}{\partial q_k}\frac{\partial}{\partial q_l}(b_{orb,l}b_{orb,k}f)
\nonumber\\
&&
+ \frac{\partial}{\partial s_k}\frac{\partial}{\partial q_l}(b_{orb,l} b_{spin,k}f)
+\frac{1}{2}\frac{\partial}{\partial s_k}\frac{\partial}{\partial s_l}(b_{spin,l} b_{spin,k}f)\; .
\label{eq:new2.44vnew}  
\end{eqnarray}
Recalling from Section 7 that $b_{orb,spin}$ is a
noise matrix field of (\ref{eq:new2.44v}) it follows from (\ref{eq:new2.44tc}) that 
$\left( \begin{array}{c}  b_{orb} \\
\vec{b}_{spin}\end{array}\right)$ is a noise matrix field of (\ref{eq:new2.44vnew}).
Thus in the case, $K=1$, the following is an Ito SDE system associated with
the formal Fokker-Planck equation, under the kinematical condition, (i):
\begin{eqnarray}                                                            
&&   Q' =  d_{orb}(t,Q,\vec{S}) +b_{orb}(t,Q,\vec{S}) \nu(t) \; ,
\label{eq:new2.44uca}\\   
&& \vec{S}'  = \vec{d}_{spin}(t,Q,\vec{S}) + \vec{b}_{spin}(t,Q,\vec{S})\nu(t) \; ,
\label{eq:new2.44ucb}                                                           
\end{eqnarray}
where $\nu$ is the one-dimensional white-noise process. Note that
(\ref{eq:new2.44uca}), (\ref{eq:new2.44ucb}) is obtained by
inserting, (\ref{eq:new2.44td}), (\ref{eq:new2.44tc}), into (\ref{eq:new2.44uc}).

Having enforced the kinematical condition, (i), for the case, $K=1$, we have arrived at 
(\ref{eq:new2.44vnew}). To enforce both kinematical conditions we now impose the kinematical
condition, (ii), onto (\ref{eq:new2.44vnew}). 
Recalling the definition of (ii) in
Section 8 we get from (\ref{eq:new2.44td}), (\ref{eq:new2.44tc}) 
\begin{eqnarray}                                                            
&&   d_{orb}(t,q,\vec{s}) = d_{orb,0}(t,q) + d_{orb,+}(t,q)\vec{s} \; ,
\nonumber\\
&&\vec{d}_{spin}(t,q,\vec{s})  =  \vec{d}_{spin,0}(t,q)  + d_{spin,+}(t,q)\vec{s} \; ,
\nonumber\\
&& b_{orb}(t,q,\vec{s}) = b_{orb,0}(t,q) + b_{orb,+}(t,q)\vec{s} \; ,
\nonumber\\
&&\vec{b}_{spin}(t,q,\vec{s})  =  \vec{b}_{spin,0}(t,q)  + b_{spin,+}(t,q)\vec{s} \; ,
\nonumber\\
\label{eq:new2.44vca} 
\end{eqnarray}
where $\vec{d}_{spin,0}(t,q),\vec{b}_{spin,0}(t,q)\in\R^3;
d_{orb,0}(t,q),b_{orb,0}(t,q)\in\R^6$ and
$d_{spin,+}(t,q),b_{spin,+}(t,q)\in\R^{3\times 3}$ as well 
as $d_{orb,+}(t,q),b_{orb,+}(t,q)\in\R^{6\times 3}$.
Note, by (\ref{eq:new2.44td}), (\ref{eq:new2.44tc}),
(\ref{eq:new2.44vca}), that
\begin{eqnarray}                                                            
&& \hspace{-10mm}   d_{orb,spin}(t,q,\vec{s}) = \left( \begin{array}{c}  
d_{orb,0}(t,q) + d_{orb,+}(t,q)\vec{s}  \\
 \vec{d}_{spin,0}(t,q)  + d_{spin,+}(t,q)\vec{s} 
\end{array}\right) \; ,
\label{eq:new2.44tdnew}  \\
&& \hspace{-10mm}  b_{orb,spin}(t,q,\vec{s}) = \left( \begin{array}{c}  
b_{orb,0}(t,q) + b_{orb,+}(t,q)\vec{s}  \\
 \vec{b}_{spin,0}(t,q)  + b_{spin,+}(t,q)\vec{s} 
\end{array}\right) \; .
\label{eq:new2.44tcnew}                                                           
\end{eqnarray}

With (\ref{eq:new2.44tdnew}), (\ref{eq:new2.44tcnew})              
our task, mentioned after (\ref{eq:new2.44ube})          
and to be fulfilled in this section, which is to specify 
$d_{orb,spin}, b_{orb,spin}$, boils down to the specification of the four vector fields,
$d_{orb,0},\vec{d}_{spin,0},b_{orb,0},\vec{b}_{spin,0}$, and of the 
four matrix fields,
$d_{orb,+},d_{spin,+},b_{orb,+},b_{spin,+}$.
Note that under the kinematical conditions and for the case, $K=1$, 
the formal Fokker-Planck equation can be written in terms of these eight functions as
\begin{eqnarray}                                                            
&&  \hspace{-10mm} 
 \frac{\partial f}{\partial t} 
= -\frac{\partial}{\partial q_j} \biggl( (d_{orb,0,j} + d_{orb,+,j,k}s_k)f\biggr)
-\frac{\partial}{\partial s_j} \biggl( (d_{spin,0,j} + d_{spin,+,j,k}s_k)f\biggr)
\nonumber\\
&&+\frac{1}{2}\frac{\partial}{\partial q_k}\frac{\partial}{\partial q_l}
\biggl( (b_{orb,0,l} + b_{orb,+,l,i}s_i)(b_{orb,0,k} + b_{orb,+,k,j}s_j)f\biggr)
\nonumber\\
&&+\frac{\partial}{\partial s_k}\frac{\partial}{\partial q_l}
\biggl( (b_{orb,0,l} + b_{orb,+,l,i}s_i)(b_{spin,0,k} + b_{spin,+,k,j}s_j)f\biggr)
\nonumber\\
&&+\frac{1}{2}
\frac{\partial}{\partial s_k}\frac{\partial}{\partial s_l}
\biggl( (b_{spin,0,l} + b_{spin,+,l,i}s_i)(b_{spin,0,k} + b_{spin,+,k,j}s_j)f\biggr)\; .
\label{eq:new2.44vbnew}  
\end{eqnarray}
Note that (\ref{eq:new2.44vbnew}) is obtained by inserting, (\ref{eq:new2.44vca}),
into (\ref{eq:new2.44vnew}).

As an aside we note that
under the kinematical conditions and for the case, $K=1$, 
the following is an Ito SDE system associated with the formal Fokker-Planck equation:
\begin{eqnarray}                                                            
&&    Q' = d_{orb,0}(t,Q) + d_{orb,+}(t,Q)\vec{S}
+\biggl( b_{orb,0}(t,Q) + b_{orb,+}(t,Q)\vec{S} \biggr)\nu(t) \; ,
\label{eq:new2.44vcnew}\\
&& \vec{S}' =  \vec{d}_{spin,0}(t,Q) + d_{spin,+}(t,Q)\vec{S}
+\biggl( \vec{b}_{spin,0}(t,Q) + b_{spin,+}(t,Q)\vec{S} \biggr)\nu(t) \; ,
\label{eq:new2.44vdnew}                                                           
\end{eqnarray}
where $\nu$ is the one-dimensional white-noise process.
Note that (\ref{eq:new2.44vcnew}), (\ref{eq:new2.44vdnew}) are obtained by
inserting, (\ref{eq:new2.44vca}), into 
(\ref{eq:new2.44uca}), (\ref{eq:new2.44ucb}).
This completes our enforcement of the kinematical conditions
for the case, $K=1$ (as we will see at the end of Section 9.2, we will not have to go beyond, $K=1$).

Having enforced both kinematical conditions (for the case, $K=1$) we have arrived
at (\ref{eq:new2.44vbnew}). Thus we will now
prepare ourselves for enforcing the dynamical condition
by computing $\frac{\partial \rho[W_f]}{\partial t}$ 
and $\frac{\partial \vec{\cal P}[W_f]}{\partial t}$ 
where $f$ is an arbitrary solution of the formal
Fokker-Planck equation, (\ref{eq:new2.44vbnew}).

We begin with $\frac{\partial \rho[W_f]}{\partial t}$ and
compute, by (\ref{eq:new2.44p}), (\ref{eq:new2.44q}), (\ref{eq:new2.44vnew}),
\begin{eqnarray}                                                            
&& 
\frac{\partial \rho[W_f]}{\partial t}(t,q) = \int_{\R^3} \;\frac{\partial f}{\partial t}(t,q,\vec{s}) d^3s
\nonumber\\
&&
= -\frac{\partial}{\partial q_j} 
\int_{\R^3} \; \biggl( 
d_{orb,0,j}(t,q) + d_{orb,+,j,k}(t,q)s_k\biggr)f(t,q,\vec{s}) d^3s  
\nonumber\\
&& 
+\frac{1}{2}\frac{\partial}{\partial q_k}\frac{\partial}{\partial q_l}
\int_{\R^3} \;
\biggl( (b_{orb,0,l}(t,q) + b_{orb,+,l,i}(t,q)s_i)
\nonumber\\
&&
\cdot (b_{orb,0,k}(t,q) + b_{orb,+,k,j}(t,q)s_j)\biggr)
f(t,q,\vec{s}) d^3s
\nonumber\\
&&
= -\frac{\partial}{\partial q_j} 
\biggl( d_{orb,0,j}(t,q) \int_{\R^3} \; f(t,q,\vec{s}) d^3s\biggr)
-\frac{\partial}{\partial q_j} 
\biggl( d_{orb,+,j,k}(t,q)  \int_{\R^3} \; s_k f(t,q,\vec{s}) d^3s\biggr)
\nonumber\\
&& 
+\frac{1}{2}
\frac{\partial}{\partial q_k}\frac{\partial}{\partial q_l}
\int_{\R^3} \; \biggl( b_{orb,0,l}(t,q)b_{orb,0,k}(t,q)
\nonumber\\
&&
+ b_{orb,+,l,i}(t,q)s_i b_{orb,0,k}(t,q)
+b_{orb,0,l}(t,q)b_{orb,+,k,j}(t,q)s_j  
\nonumber\\
&& +b_{orb,+,l,i}(t,q)s_i b_{orb,+,k,j}(t,q)s_j  
\biggr)f(t,q,\vec{s}) d^3s
\nonumber\\
&&
= -\frac{\partial}{\partial q_j} 
\biggl( d_{orb,0,j}(t,q)\rho[W_f](t,q)\biggr)
-\frac{\partial}{\partial q_j} 
\biggl(d_{orb,+,j,k}(t,q){\cal P}_k[W_f](t,q)\biggr)
\nonumber\\
&& +\frac{1}{2}\frac{\partial}{\partial q_k}\frac{\partial}{\partial q_l}
\biggl(b_{orb,0,l}(t,q)b_{orb,0,k}(t,q)\rho[W_f](t,q)\biggr)
\nonumber\\
&& + \frac{1}{2}
\frac{\partial}{\partial q_k}\frac{\partial}{\partial q_l}
\biggl( b_{orb,+,l,i}(t,q){\cal P}_i[W_f](t,q)b_{orb,0,k}(t,q)\biggr)
\nonumber\\
&& + \frac{1}{2}\frac{\partial}{\partial q_k}\frac{\partial}{\partial q_l}
\biggl( b_{orb,0,l}(t,q)b_{orb,+,k,j}(t,q){\cal P}_j[W_f](t,q)\biggr)
\nonumber\\
&& 
+\frac{1}{2}\frac{\partial}{\partial q_k}\frac{\partial}{\partial q_l}\biggl(
\int_{\R^3} \; 
b_{orb,+,l,i}(t,q)b_{orb,+,k,j}(t,q)s_i s_j f(t,q,\vec{s})
 d^3s\biggr)
\nonumber\\
&& = -\frac{\partial}{\partial q_j} 
\biggl( d_{orb,0,j}(t,q)\rho[W_f](t,q)\biggr)
-\frac{\partial}{\partial q_j} 
\biggl(d_{orb,+,j,k}(t,q){\cal P}_k[W_f](t,q)\biggr)
\nonumber\\
&& +\frac{1}{2}\frac{\partial}{\partial q_k}\frac{\partial}{\partial q_l}
\biggl( b_{orb,0,l}(t,q)b_{orb,0,k}(t,q)\rho[W_f](t,q)\biggr)
\nonumber\\
&& +\frac{\partial}{\partial q_k}\frac{\partial}{\partial q_l}
\biggl( b_{orb,0,l}(t,q)b_{orb,+,k,j}(t,q){\cal P}_j[W_f](t,q)\biggr)
\nonumber\\
&& 
+\frac{1}{2}\frac{\partial}{\partial q_k}\frac{\partial}{\partial q_l}
\biggl(
b_{orb,+,l,i}(t,q)b_{orb,+,k,j}(t,q)
\int_{\R^3} \; s_i s_j f(t,q,\vec{s})
 d^3s\biggr) \; ,
\nonumber
\end{eqnarray}
in short
\begin{eqnarray}                                                            
&& 
\frac{\partial \rho[W_f]}{\partial t}(t,q) 
= -\frac{\partial}{\partial q_j} 
\biggl( d_{orb,0,j}(t,q)\rho[W_f](t,q)\biggr)
\nonumber\\
&& -\frac{\partial}{\partial q_j} 
\biggl(d_{orb,+,j,k}(t,q){\cal P}_k[W_f](t,q)\biggr)
\nonumber\\
&& +\frac{1}{2}\frac{\partial}{\partial q_k}\frac{\partial}{\partial q_l}
\biggl( b_{orb,0,l}(t,q)b_{orb,0,k}(t,q)\rho[W_f](t,q)\biggr)
\nonumber\\
&& +\frac{\partial}{\partial q_k}\frac{\partial}{\partial q_l}
\biggl( b_{orb,0,l}(t,q)b_{orb,+,k,j}(t,q){\cal P}_j[W_f](t,q)\biggr)
\nonumber\\
&& \hspace{-20mm}
+\frac{1}{2}\frac{\partial}{\partial q_k}\frac{\partial}{\partial q_l}\biggl(b_{orb,+,l,i}(t,q)b_{orb,+,k,j}(t,q)
\int_{\R^3} \; s_i s_j f(t,q,\vec{s})
 d^3s \biggr)\; ,
\label{eq:new2.44vcd}                                                           
\end{eqnarray}
where we also used that (\ref{eq:new2.44vnew}) implies (\ref{eq:new2.44vbnew}).
The second and fourth entries on the rhs
of (\ref{eq:new2.44vcd}) are Stern-Gerlach-like terms and also 
the fifth entry is nonphysical (in the sense that $\int_{\R^3} \; s_i s_j f(t,q,\vec{s})d^3s$ 
cannot be expressed in terms of $\rho[W_f],\vec{\cal P}[W_f]$).
In fact these nonphysical entries will be eliminated in Section 9.2
when we enforce the
dynamical condition.

Having computed $\frac{\partial \rho[W_f]}{\partial t}$ 
we now compute $\frac{\partial \vec{\cal P}[W_f]}{\partial t}$ 
where again $f$ is an arbitrary solution of the formal
Fokker-Planck equation, (\ref{eq:new2.44vbnew}).
We first compute, by (\ref{eq:new2.44q}), (\ref{eq:new2.44vnew}) and via integration by parts,
\begin{eqnarray}                                                            
&& \hspace{-10mm} \frac{\partial {\cal P}_m[W_f]}{\partial t}(t,q) = \int_{\R^3} \; s_m  \frac{\partial f}{\partial t}(t,q,\vec{s}) d^3s  
= \int_{\R^3} \; s_m\biggl(
-\frac{\partial}{\partial q_j} d_{orb,j}(t,q,\vec{s}) 
-\frac{\partial}{\partial s_j} d_{spin,j}(t,q,\vec{s}) 
\nonumber\\
&&
+\frac{1}{2}\frac{\partial}{\partial q_k}\frac{\partial}{\partial q_l} 
b_{orb,l}(t,q,\vec{s}) b_{orb,k}(t,q,\vec{s}) 
+ \frac{\partial}{\partial s_k}\frac{\partial}{\partial q_l}b_{orb,l}(t,q,\vec{s})  b_{spin,k}(t,q,\vec{s}) 
\nonumber\\
&&+\frac{1}{2}\frac{\partial}{\partial s_k}\frac{\partial}{\partial s_l}b_{spin,l}(t,q,\vec{s})  
b_{spin,k}(t,q,\vec{s}) \biggr) f(t,q,\vec{s}) d^3s
\nonumber\\
&&  = -\frac{\partial}{\partial q_j} 
 \int_{\R^3} \; d_{orb,j}(t,q,\vec{s}) s_m  f(t,q,\vec{s}) d^3s
-\int_{\R^3} \; s_m \frac{\partial}{\partial s_j} d_{spin,j}(t,q,\vec{s}) f(t,q,\vec{s}) d^3s
\nonumber\\
&&
+\frac{1}{2}\frac{\partial}{\partial q_k}\frac{\partial}{\partial q_l} 
\int_{\R^3} \;b_{orb,l}(t,q,\vec{s}) b_{orb,k}(t,q,\vec{s})s_m f(t,q,\vec{s}) d^3s
\nonumber\\
&& + \int_{\R^3} \; s_m
\frac{\partial}{\partial s_k}\frac{\partial}{\partial q_l}b_{orb,l}(t,q,\vec{s})  b_{spin,k}(t,q,\vec{s}) 
f(t,q,\vec{s}) d^3s
\nonumber\\
&& + \frac{1}{2}\int_{\R^3} \; s_m\frac{\partial}{\partial s_k}\frac{\partial}{\partial s_l}b_{spin,l}(t,q,\vec{s})  
b_{spin,k}(t,q,\vec{s}) f(t,q,\vec{s}) d^3s
\nonumber\\
&& = -\frac{\partial}{\partial q_j} 
 \int_{\R^3} \; d_{orb,j}(t,q,\vec{s}) s_m  f(t,q,\vec{s}) d^3s
+ \int_{\R^3} \; d_{spin,m}(t,q,\vec{s})f(t,q,\vec{s}) d^3s
\nonumber\\
&&
+\frac{1}{2}\frac{\partial}{\partial q_k}\frac{\partial}{\partial q_l} 
\int_{\R^3} \;b_{orb,l}(t,q,\vec{s}) b_{orb,k}(t,q,\vec{s})s_m f(t,q,\vec{s}) d^3s
\nonumber\\
&& -\frac{\partial}{\partial q_l} 
\int_{\R^3} \; b_{orb,l}(t,q,\vec{s})  b_{spin,m}(t,q,\vec{s}) f(t,q,\vec{s}) d^3s \; ,
\nonumber
\end{eqnarray}
where $m=1,2,3$ which implies, by (\ref{eq:new2.44p}), (\ref{eq:new2.44q}), (\ref{eq:new2.44vca}),
\begin{eqnarray}                                                            
&&  \hspace{-10mm} \frac{\partial {\cal P}_m[W_f]}{\partial t}(t,q) 
= -\frac{\partial}{\partial q_j} 
 \int_{\R^3} \; \biggl(  d_{orb,0,j}(t,q) 
+d_{orb,+,j,i}(t,q)s_i\biggr) s_m  f(t,q,\vec{s}) d^3s
\nonumber\\
&& + \int_{\R^3} \; \biggl(  d_{spin,0,m}(t,q) 
+d_{spin,+,m,i}(t,q)s_i\biggr) f(t,q,\vec{s}) d^3s
\nonumber\\
&&
+\frac{1}{2}\frac{\partial}{\partial q_k}\frac{\partial}{\partial q_l} 
\int_{\R^3} \;\biggl(  b_{orb,0,l}(t,q) 
+b_{orb,+,l,i}(t,q)s_i\biggr) 
\nonumber\\
&& \cdot \biggl(  b_{orb,0,k}(t,q) 
+b_{orb,+,k,j}(t,q)s_j\biggr) s_m f(t,q,\vec{s}) d^3s
\nonumber\\
&&  \hspace{-10mm} 
-\frac{\partial}{\partial q_l} 
\int_{\R^3} \; 
\biggl(  b_{orb,0,l}(t,q) 
+b_{orb,+,l,i}(t,q)s_i\biggr) 
\biggl(  b_{spin,0,m}(t,q) 
+b_{spin,+,m,j}(t,q)s_j\biggr) f(t,q,\vec{s}) d^3s
\nonumber\\
&& \hspace{-10mm} = -\frac{\partial}{\partial q_j} d_{orb,0,j}(t,q) 
 \int_{\R^3} \; s_m  f(t,q,\vec{s}) d^3s
-\frac{\partial}{\partial q_j} d_{orb,+,j,i}(t,q) 
 \int_{\R^3} \; s_i s_m  f(t,q,\vec{s}) d^3s
\nonumber\\
&& + d_{spin,0,m}(t,q) \int_{\R^3} \; f(t,q,\vec{s}) d^3s
+ d_{spin,+,m,i}(t,q) \int_{\R^3} \; s_i f(t,q,\vec{s}) d^3s
\nonumber\\
&&  
+\frac{1}{2}\frac{\partial}{\partial q_k}\frac{\partial}{\partial q_l} 
b_{orb,0,l}(t,q) b_{orb,0,k}(t,q) 
\int_{\R^3} \;  s_m f(t,q,\vec{s}) d^3s
\nonumber\\
&& 
+\frac{1}{2}\frac{\partial}{\partial q_k}\frac{\partial}{\partial q_l} b_{orb,0,l}(t,q) 
b_{orb,+,k,j}(t,q)\int_{\R^3} \; s_j s_m f(t,q,\vec{s}) d^3s
\nonumber\\
&&
+\frac{1}{2}\frac{\partial}{\partial q_k}\frac{\partial}{\partial q_l} 
b_{orb,+,l,i}(t,q) b_{orb,0,k}(t,q) 
\int_{\R^3} \; s_i s_m f(t,q,\vec{s}) d^3s
\nonumber\\
&& 
+\frac{1}{2}\frac{\partial}{\partial q_k}\frac{\partial}{\partial q_l} 
b_{orb,+,l,i}(t,q)b_{orb,+,k,j}(t,q)
\int_{\R^3} \; s_i s_j s_m f(t,q,\vec{s}) d^3s
\nonumber\\
&&  
-\frac{\partial}{\partial q_l} 
b_{orb,0,l}(t,q) b_{spin,0,m}(t,q) 
\int_{\R^3} \;  f(t,q,\vec{s}) d^3s
\nonumber\\
&& -\frac{\partial}{\partial q_l} b_{orb,0,l}(t,q) 
b_{spin,+,m,j}(t,q)\int_{\R^3} \; s_j  f(t,q,\vec{s}) d^3s
\nonumber\\
&&
-\frac{\partial}{\partial q_l} 
b_{orb,+,l,i}(t,q) b_{spin,0,m}(t,q) 
\int_{\R^3} \; s_i  f(t,q,\vec{s}) d^3s
\nonumber\\
&& -\frac{\partial}{\partial q_l} 
b_{orb,+,l,i}(t,q)b_{spin,+,m,j}(t,q)
\int_{\R^3} \; s_i s_j  f(t,q,\vec{s}) d^3s
\nonumber\\
&&  \hspace{-10mm} 
= -\frac{\partial}{\partial q_j} d_{orb,0,j}(t,q){\cal P}_m[W_f](t,q) 
-\frac{\partial}{\partial q_j} d_{orb,+,j,i}(t,q) 
 \int_{\R^3} \; s_i s_m  f(t,q,\vec{s}) d^3s
\nonumber\\
&& + d_{spin,0,m}(t,q) \rho[W_f](t,q) + d_{spin,+,m,i}(t,q) {\cal P}_i[W_f](t,q) 
\nonumber\\
&&   
+\frac{1}{2}\frac{\partial}{\partial q_k}\frac{\partial}{\partial q_l} 
b_{orb,0,l}(t,q) b_{orb,0,k}(t,q) {\cal P}_m[W_f](t,q) 
\nonumber\\
&& 
+\frac{1}{2}\frac{\partial}{\partial q_k}\frac{\partial}{\partial q_l} b_{orb,0,l}(t,q) 
b_{orb,+,k,j}(t,q)\int_{\R^3} \; s_j s_m f(t,q,\vec{s}) d^3s
\nonumber
\end{eqnarray}
\begin{eqnarray}                                                            
&&
+\frac{1}{2}\frac{\partial}{\partial q_k}\frac{\partial}{\partial q_l} 
b_{orb,+,l,i}(t,q) b_{orb,0,k}(t,q) 
\int_{\R^3} \; s_i s_m f(t,q,\vec{s}) d^3s
\nonumber\\
&& 
+\frac{1}{2}\frac{\partial}{\partial q_k}\frac{\partial}{\partial q_l} 
b_{orb,+,l,i}(t,q)b_{orb,+,k,j}(t,q)
\int_{\R^3} \; s_i s_j s_m f(t,q,\vec{s}) d^3s
\nonumber\\
&&  -\frac{\partial}{\partial q_l} 
b_{orb,0,l}(t,q) b_{spin,0,m}(t,q) \rho[W_f](t,q) 
-\frac{\partial}{\partial q_l} b_{orb,0,l}(t,q) 
b_{spin,+,m,j}(t,q){\cal P}_j[W_f](t,q) 
\nonumber\\
&&
-\frac{\partial}{\partial q_l} 
b_{orb,+,l,i}(t,q) b_{spin,0,m}(t,q) {\cal P}_i[W_f](t,q) 
\nonumber\\
&& -\frac{\partial}{\partial q_l} 
b_{orb,+,l,i}(t,q)b_{spin,+,m,j}(t,q)
\int_{\R^3} \; s_i s_j  f(t,q,\vec{s}) d^3s \; ,
\nonumber
\end{eqnarray}
where $m=1,2,3$ and thus we got
\begin{eqnarray}                                                            
&&  \hspace{-10mm} \frac{\partial \vec{\cal P}[W_f]}{\partial t}(t,q) 
= -\frac{\partial}{\partial q_j} \biggl( d_{orb,0,j}(t,q)\vec{\cal P}[W_f](t,q) \biggr)
\nonumber\\
&& -\frac{\partial}{\partial q_j} \biggl( d_{orb,+,j,i}(t,q) 
 \int_{\R^3} \; \vec{s}s_i  f(t,q,\vec{s}) d^3s\biggr)
\nonumber\\
&& + \vec{d}_{spin,0}(t,q) \rho[W_f](t,q) + d_{spin,+}(t,q) \vec{\cal P}[W_f](t,q) 
\nonumber\\
&&   
+\frac{1}{2}\frac{\partial}{\partial q_k}\frac{\partial}{\partial q_l} 
\biggl( b_{orb,0,l}(t,q) b_{orb,0,k}(t,q) \vec{\cal P}[W_f](t,q) \biggr)
\nonumber\\
&& 
+\frac{1}{2}\frac{\partial}{\partial q_k}\frac{\partial}{\partial q_l} \biggl( b_{orb,0,l}(t,q) 
b_{orb,+,k,j}(t,q)\int_{\R^3} \; \vec{s} s_j f(t,q,\vec{s}) d^3s\biggr)
\nonumber\\
&&
+\frac{1}{2}\frac{\partial}{\partial q_k}\frac{\partial}{\partial q_l} 
\biggl( b_{orb,+,l,i}(t,q) b_{orb,0,k}(t,q) 
\int_{\R^3} \; \vec{s} s_i f(t,q,\vec{s}) d^3s\biggr)
\nonumber\\
&& 
+\frac{1}{2}\frac{\partial}{\partial q_k}\frac{\partial}{\partial q_l} 
\biggl( b_{orb,+,l,i}(t,q)b_{orb,+,k,j}(t,q)
\int_{\R^3} \; \vec{s} s_i s_j f(t,q,\vec{s}) d^3s\biggr)
\nonumber\\
&&  -\frac{\partial}{\partial q_l} 
\biggl( b_{orb,0,l}(t,q) \vec{b}_{spin,0}(t,q) \rho[W_f](t,q) \biggr)
\nonumber\\
&& 
-\frac{\partial}{\partial q_l} \biggl( b_{orb,0,l}(t,q) 
b_{spin,+}(t,q)\vec{\cal P}[W_f](t,q) \biggr)
\nonumber\\
&&
-\frac{\partial}{\partial q_l} \biggl(
b_{orb,+,l,i}(t,q) {\cal P}_i[W_f](t,q) \vec{b}_{spin,0}(t,q) \biggr)
\nonumber\\
&& \hspace{-10mm}
-\frac{\partial}{\partial q_l} \biggl(
b_{orb,+,l,i}(t,q)b_{spin,+}(t,q)
\int_{\R^3} \; s_i \vec{s}  f(t,q,\vec{s}) d^3s\biggr) \; .
\label{eq:new2.44vcc}  
\end{eqnarray}
Note that the second, sixth, seventh, eighth and twelfth entries
on the rhs of (\ref{eq:new2.44vcc}) are nonphysical in the sense 
that terms of the form: $\int_{\R^3} \; \vec{s} s_n f(t,q,\vec{s})d^3s$ and of the form:
$\int_{\R^3} \;  \vec{s} s_n s_p f(t,q,\vec{s})d^3s$ 
cannot be expressed via $\rho[W_f],\vec{\cal P}[W_f]$.
We will see in Section 9.2 that also the tenth entry
on the rhs of (\ref{eq:new2.44vcc}) is nonphysical.
All these nonphysical entries will be eliminated in Section 9.2 when we enforce 
the dynamical condition.

\subsection{Enforcing the dynamical condition}

With (\ref{eq:new2.44vcd}), (\ref{eq:new2.44vcc})  
we have computed $\frac{\partial \rho[W_f]}{\partial t}$ 
and $\frac{\partial \vec{\cal P}[W_f]}{\partial t}$ 
where $f$ is a solution of the formal
Fokker-Planck equation, (\ref{eq:new2.44vnew}) (and thus is a solution of the formal
Fokker-Planck equation, (\ref{eq:new2.44vbnew})). Therefore
(\ref{eq:new2.44vcd}), (\ref{eq:new2.44vcc}) hold
under the kinematical conditions for the case, $K=1$).
Thus we are ready to enforce the dynamical condition, i.e., we will try (and in fact we will succeed)
to choose the eight coefficient functions,
$d_{orb,0},...,b_{spin,+}$, in (\ref{eq:new2.44vbnew})
such that the dynamical condition holds, i.e., such that (\ref{eq:new2.44vcd}), (\ref{eq:new2.44vcc})  
are the same as (\ref{eq:new02.010f}), (\ref{eq:new02.010dnew}).

We begin with the first part of the dynamical condition whereby we have to ensure
that (\ref{eq:new2.44vcd}) is the same as (\ref{eq:new02.010f}).
Thus in particular we have to get rid of the nonphysical entries in (\ref{eq:new2.44vcd}) which
were specified in the discussion after (\ref{eq:new2.44vcd}).
This we accomplish by choosing $d_{orb,+},b_{spin,+}$ as zero matrix fields:
\begin{eqnarray}                                                            
&&  d_{orb,+}:= 0_{6\times 3} \; , \quad  b_{orb,+}:= 0_{6\times 3} \; ,
\label{eq:new2.44vcda}  
\end{eqnarray}
whereby (\ref{eq:new2.44vcd}) simplifies to
\begin{eqnarray}                                                            
&& 
\hspace{-15mm} 
\frac{\partial \rho[W_f]}{\partial t}
= -\frac{\partial}{\partial q_j} 
\biggl( d_{orb,0,j}\rho[W_f]\biggr)
+\frac{1}{2}\frac{\partial}{\partial q_k}\frac{\partial}{\partial q_l}
\biggl( b_{orb,0,l}b_{orb,0,k}\rho[W_f]\biggr)  \; .
\label{eq:new2.44vcdc}  
\end{eqnarray}

Thus the task of enforcing the first part of the dynamical condition 
has been boiled down to the task of
ensuring that (\ref{eq:new2.44vcdc}) is the same as (\ref{eq:new02.010f}).
This we accomplish by choosing:    
\begin{eqnarray}                                                            
&&  d_{orb,0}:= {\cal D}_{orb} \; , \quad b_{orb,0}:= {\cal B}_{orb} \; .
\label{eq:new2.44vcdb}  
\end{eqnarray}
With the choices, (\ref{eq:new2.44vcda}), (\ref{eq:new2.44vcdb}), of 
$d_{orb,0},b_{orb,0},d_{orb,+},b_{orb,+}$, we observe that (\ref{eq:new2.44vcdc}) is the same as 
(\ref{eq:new02.010f})
and so the first part of the dynamical condition has been successfully enforced.

Thus we now enforce the second part of the dynamical condition.
We will accomplish this by using 
(\ref{eq:new2.44vcda}), (\ref{eq:new2.44vcdb}), and by using the
freedom of properly choosing the still arbitrary coefficient functions,
$\vec{d}_{spin,0},\vec{b}_{spin,0},d_{spin,+},b_{spin,+}$.
Enforcing the second part of the dynamical condition therefore means that we have 
to ensure that (\ref{eq:new2.44vcc}) is the same as  (\ref{eq:new02.010dnew}).
We recall from Section 4 that (\ref{eq:new02.010d}) 
is the same as (\ref{eq:new2.44vcdc0new}) and thus (\ref{eq:new02.010dnew})
is the same as
\begin{eqnarray}                                                            
&&        \hspace{-20mm}                                                                
\frac{\partial \vec{\cal P}[W_f]}{\partial t} 
= l_{orb}\vec{\cal P}[W_f]
+{\cal D}_{spin,+}\vec{\cal P}[W_f]
+\vec{\cal D}_{spin,0,ST}\rho[W_f]
- \frac{\partial}{\partial q_k} ({\cal B}_{orb,k}
\vec{\cal B}_{spin}\rho[W_f]) \; ,
\label{eq:new2.44vcdc0}                                                           
\end{eqnarray}
which will be used in this subsection and in Section 16.

As an aside we note, by (\ref{eq:new02.010new27}),
that (\ref{eq:new2.44vcdc0}) can be written as
\begin{eqnarray}                                                            
&&        \hspace{-15mm}                                                                
\frac{\partial \vec{\cal P}[W_f]}{\partial t} 
= l_{orb}\vec{\cal P}[W_f]
+{\cal D}_{spin,+,TBMT}\vec{\cal P}[W_f]
+{\cal D}_{spin,+,ST}\vec{\cal P}[W_f]
\nonumber\\
&&  \hspace{-10mm}   
+{\cal D}_{spin,+,BK}\vec{\cal P}[W_f]
+\vec{\cal D}_{spin,0,ST}\rho[W_f]
- \frac{\partial}{\partial q_k} ({\cal B}_{orb,k}
\vec{\cal B}_{spin}\rho[W_f]) \; ,
\label{eq:new2.44vcdc0newnew}                                                           
\end{eqnarray}
which will be used in Section 16.

Enforcing the second part of the dynamical condition 
now means that we have 
to ensure that (\ref{eq:new2.44vcc}) is the same as (\ref{eq:new2.44vcdc0}).
In particular we have to get rid of the nonphysical entries in (\ref{eq:new2.44vcc}) which
are specified in the discussion after (\ref{eq:new2.44vcc}).
Thanks to our choice,
(\ref{eq:new2.44vcda}), (\ref{eq:new2.44vcdb}),  
of $d_{orb,0},b_{orb,0},d_{orb,+},b_{orb,+}$, (\ref{eq:new2.44vcc}) simplifies to
\begin{eqnarray}                                                            
&&  \hspace{-15mm} \frac{\partial \vec{\cal P}[W_f]}{\partial t}
= -\frac{\partial}{\partial q_j} ( {\cal D}_{orb,j}\vec{\cal P}[W_f] )
+ \vec{d}_{spin,0}\rho[W_f] + d_{spin,+}\vec{\cal P}[W_f]
\nonumber\\
&&   
+\frac{1}{2}\frac{\partial}{\partial q_k}\frac{\partial}{\partial q_l} 
\biggl( {\cal B}_{orb,l} {\cal B}_{orb,k}\vec{\cal P}[W_f]\biggr)
\nonumber\\
&&  \hspace{-10mm} 
-\frac{\partial}{\partial q_l} 
\biggl( {\cal B}_{orb,l} \vec{b}_{spin,0}\rho[W_f] \biggr)
-\frac{\partial}{\partial q_l} \biggl( {\cal B}_{orb,l}
b_{spin,+}\vec{\cal P}[W_f] \biggr) \; ,
\label{eq:new2.44vccnew}  
\end{eqnarray}
so that most of the nonphysical entries of (\ref{eq:new2.44vcc})
have disappeared without having yet enforced the second part of the dynamical condition.
We next note, by (\ref{eq:new02.010a}),
that (\ref{eq:new2.44vccnew}) can be written as
\begin{eqnarray}                                                            
&&  \hspace{-10mm} \frac{\partial \vec{\cal P}[W_f]}{\partial t}
=l_{orb}\vec{\cal P}[W_f]
+ \vec{d}_{spin,0} \rho[W_f] + d_{spin,+} \vec{\cal P}[W_f]
\nonumber\\
&&  -\frac{\partial}{\partial q_l} 
\biggl( {\cal B}_{orb,l} \vec{b}_{spin,0} \rho[W_f] \biggr)
-\frac{\partial}{\partial q_l} \biggl( {\cal B}_{orb,l}
b_{spin,+}\vec{\cal P}[W_f]\biggr) \; .
\label{eq:new2.44vcf}    
\end{eqnarray}

Thus the task of enforcing the second part of the dynamical 
condition has been boiled down to the task of
ensuring that (\ref{eq:new2.44vcf}) is the same as (\ref{eq:new2.44vcdc0}).
In other words we have to choose $\vec{d}_{spin,0},\vec{b}_{spin,0},d_{spin,+},b_{spin,+}$ 
such that (\ref{eq:new2.44vcf}) is the same as (\ref{eq:new2.44vcdc0}).
To accomplish this we next note that phase space derivatives of $\vec{\cal P}[W_f]$ occur in
(\ref{eq:new2.44vcdc0}) only via the entry, $l_{orb}\vec{\cal P}[W_f]$, 
whereas phase space 
derivatives of $\vec{\cal P}[W_f]$ occur in (\ref{eq:new2.44vcf}) not only via
the entry, $l_{orb}\vec{\cal P}[W_f]$, but also via
the fifth entry on the rhs of (\ref{eq:new2.44vcf}).
Thus the fifth entry on the rhs of (\ref{eq:new2.44vcf}) is nonphysical (and therefore
the tenth entry on the rhs of (\ref{eq:new2.44vcc}) is nonphysical!)
and so we have to eliminate it.
This we accomplish by choosing $b_{spin,+}$ as the zero matrix field:
\begin{eqnarray}                                                            
&& b_{spin,+}:= 0_{3\times 3} \; ,
\label{eq:new2.44vcdh0}  
\end{eqnarray}
whereby (\ref{eq:new2.44vcf}) simplifies to
\begin{eqnarray}                                                            
&&  \hspace{-10mm} \frac{\partial \vec{\cal P}[W_f]}{\partial t}
=l_{orb}\vec{\cal P}[W_f]
+ \vec{d}_{spin,0} \rho[W_f] + d_{spin,+} \vec{\cal P}[W_f]
-\frac{\partial}{\partial q_l} 
\biggl( {\cal B}_{orb,l} \vec{b}_{spin,0} \rho[W_f] \biggr) \; .
\label{eq:new2.44vcfnew}    
\end{eqnarray}

Thus the task of enforcing the second part of the dynamical 
condition has been simplified to the task of
ensuring that (\ref{eq:new2.44vcfnew}) is the same as (\ref{eq:new2.44vcdc0}).
This we accomplish by choosing:
\begin{eqnarray}                                                            
&&  \hspace{-15mm} \vec{d}_{spin,0}:=\vec{\cal D}_{spin,0,ST} \; ,
\label{eq:new2.44vcdh}  \\
&&\hspace{-15mm}  d_{spin,+}:={\cal D}_{spin,+} \; ,
\label{eq:new2.44vcdi} \\
&& \hspace{-15mm}
\vec{b}_{spin,0}:= \vec{\cal B}_{spin}\; ,
\label{eq:new2.44vcdha}  
\end{eqnarray}
where $\vec{\cal B}_{spin},\vec{\cal D}_{spin,0,ST},{\cal D}_{spin,+}$ are defined by
(\ref{eq:new2.99d}), (\ref{eq:new02.010new26}), (\ref{eq:new02.010new27}).

With the choice, (\ref{eq:new2.44vcdh}), (\ref{eq:new2.44vcdi}), (\ref{eq:new2.44vcdha}),
we note that (\ref{eq:new2.44vcfnew}) is the same as (\ref{eq:new2.44vcdc0})
and so the second part of the dynamical condition has been
successfully enforced as well. Thus both parts of the dynamical condition have been
successfully enforced and so the aim of this section is fulfilled. In particular we  see that
the case, $K=1$, is consistent with the kinematical 
and dynamical conditions.

\section{The full spin-orbit Fokker-Planck equation and the full spin-orbit stochastic ODE system}

With Section 9 at hand we can now write down the full spin-orbit Fokker-Planck equation and an
associated Ito SDE systems.

Choosing the coefficient functions, 
$d_{orb,0},\vec{d}_{spin,0},b_{orb,0},\vec{b}_{spin,0}$, and\\ 
$d_{orb,+},d_{spin,+},b_{orb,+},b_{spin,+}$
via (\ref{eq:new2.44vcda}), (\ref{eq:new2.44vcdb}), (\ref{eq:new2.44vcdh0}),
(\ref{eq:new2.44vcdh}), (\ref{eq:new2.44vcdi}), (\ref{eq:new2.44vcdha}),
the formal Fokker-Planck equation,
(\ref{eq:new2.44vbnew}), becomes
\begin{eqnarray}                                                            
&&  \hspace{-10mm} 
\frac{\partial f}{\partial t} 
= -\frac{\partial}{\partial q_j} ({\cal D}_{orb,j}f)
-\frac{\partial}{\partial s_j} \biggl( ({\cal D}_{spin,0,ST,j} + {\cal D}_{spin,+,j,k}s_k)f\biggr)
+\frac{1}{2}\frac{\partial}{\partial q_k}\frac{\partial}{\partial q_l}
({\cal B}_{orb,l} {\cal B}_{orb,k}f)
\nonumber\\
&&
+\frac{\partial}{\partial s_k}\frac{\partial}{\partial q_l}
({\cal B}_{orb,l} {\cal B}_{spin,k}f)
+\frac{1}{2}
\frac{\partial}{\partial s_k}\frac{\partial}{\partial s_l}
({\cal B}_{spin,l} {\cal B}_{spin,k}f)\; .
\label{eq:new2.44vanew}  
\end{eqnarray}
By its derivation in Section 9
it is clear that  (\ref{eq:new2.44vanew}) 
satisfies the dynamical condition.
Thus, by Section 7, (\ref{eq:new2.44vanew}) is a Fokker-Planck equation and it is
called the full spin-orbit Fokker-Planck equation.
Also, by Section 7, for every solution $f$ of (\ref{eq:new2.44vanew}), $W_f$ satisfies (\ref{eq:new002.100f})
(if $W_f$ is sufficiently regular).

The expression (\ref{eq:new2.44vanew}) of the full spin-orbit Fokker-Planck equation
is not
always the most convenient one and so we will present in this section and in Section 13
five more expressions, namely (\ref{eq:new2.44vanewz}),  
(\ref{eq:new2.44vanewnew}), (\ref{eq:new2.44vvanew}),  
(\ref{eq:new2.44vvanewfinal}) and (\ref{eq:109new}).

Using (\ref{eq:new2.44vanewnewnew}) we can rewrite (\ref{eq:new2.44vanew}) 
into the following simpler form
\begin{eqnarray}                                                            
&&  \hspace{-10mm} 
\frac{\partial f}{\partial t} 
= -\frac{\partial}{\partial q_j} ({\cal D}_{orb,j}f)
-\frac{\partial}{\partial s_j} ( {\cal D}_{spin,j} f)
+\frac{1}{2}\frac{\partial}{\partial q_k}\frac{\partial}{\partial q_l}
({\cal B}_{orb,l} {\cal B}_{orb,k}f)
\nonumber\\
&&
+\frac{\partial}{\partial s_k}\frac{\partial}{\partial q_l}
({\cal B}_{orb,l} {\cal B}_{spin,k}f)
+\frac{1}{2}
\frac{\partial}{\partial s_k}\frac{\partial}{\partial s_l}
({\cal B}_{spin,l} {\cal B}_{spin,k}f)\; ,
\label{eq:new2.44vanewz}  
\end{eqnarray}
where ${\cal D}_{orb},{\cal B}_{orb},\vec{\cal B}_{spin},\vec{\cal D}_{spin}$ are defined by
(\ref{eq:new2.44a}), (\ref{eq:new2.44h}), (\ref{eq:new2.99d}), (\ref{eq:new2.44vanewnewnew}).
By its derivation it is clear that (\ref{eq:new2.44vanewz}) 
is the same as (\ref{eq:new2.44vanew}) and so it is the full spin-orbit Fokker-Planck equation.

For the casual reader, neither (\ref{eq:new2.44vanew}) nor
(\ref{eq:new2.44vanewz})
may look like a Fokker-Planck equation. In order to make this property clearly visible
we now write (\ref{eq:new2.44vanewz}) as
\begin{eqnarray}                                                            
&&  \hspace{-10mm} 
\frac{\partial f}{\partial t}
= -\frac{\partial}{\partial y_j}
\Biggl( \left( \begin{array}{c}  {\cal D}_{orb} \\
\vec{\cal D}_{spin}\end{array}\right)_j f\Biggr)
+\frac{1}{2}
\frac{\partial}{\partial y_k}\frac{\partial}{\partial y_l}
\Biggl( 
\left( \begin{array}{c}  {\cal B}_{orb} \\
\vec{\cal B}_{spin}\end{array}\right)_l
\left( \begin{array}{c}  {\cal B}_{orb} \\
\vec{\cal B}_{spin}\end{array}\right)_k f\Biggr) \; .
\label{eq:new2.44vanewnew}  
\end{eqnarray}
Note, by (\ref{eq:new2.44ube}), that $y=\left( \begin{array}{c}  q \\ \vec{s} \end{array}\right)$.
Clearly (\ref{eq:new2.44vanewnew}) is the same as (\ref{eq:new2.44vanewz}) 
(and so the same as (\ref{eq:new2.44vanew})).
The key point about 
(\ref{eq:new2.44vanewnew}) is being an expression in terms of the 
so-called drift vector field and of the
so-called diffusion matrix field. Thus (\ref{eq:new2.44vanewnew}) is that 
expression of the full spin-orbit Fokker-Planck equation by which it is immediately 
clear that it is a Fokker-Planck equation.
In fact $\left( \begin{array}{c}  {\cal D}_{orb} \\
\vec{\cal D}_{spin}\end{array}\right)$
is the so-called drift vector field and 
$\left( \begin{array}{c}  {\cal B}_{orb} \\
\vec{\cal B}_{spin}\end{array}\right)
\left( \begin{array}{c}  {\cal B}_{orb} \\
\vec{\cal B}_{spin}\end{array}\right)^T$ is the so-called diffusion matrix field of
(\ref{eq:new2.44vanewnew})
(for the physical interpretation of both fields, see Section 16 below).
Since $\left( \begin{array}{c}  {\cal B}_{orb} \\
\vec{\cal B}_{spin}\end{array}\right)$ 
is a vector field, which we call the noise vector field,
every value of  the diffusion matrix field is an outer product matrix.
The noise vector field of (\ref{eq:new2.44vanewnew})
is not uniquely determined
since each of its values can be multiplied by either $1$ or $-1$ without
affecting the diffusion matrix field (our choice of the noise vector field is given by
(\ref{eq:new2.44h}), (\ref{eq:new2.99d})).
For example $\left( \begin{array}{c}  -{\cal B}_{orb} \\
-\vec{\cal B}_{spin}\end{array}\right)$ 
is a noise vector field of (\ref{eq:new2.44vanewnew}), too.
For the notions of Fokker-Planck equation, drift vector field and diffusion matrix field 
see, e.g., \cite{Arn74,Gar04}.

Let us briefly continue the discussion of a fine point which began after
(\ref{eq:new2.44vcdc0new}) and which is based on the observation that 
${\cal B}_{orb}$ occurs twice on the rhs of (\ref{eq:new2.44vcdc0new}). 
The point to be made here is the following:
It follows from (\ref{eq:new2.44vanewnew}) 
that ${\cal B}_{orb}$ occurs twice on the rhs of (\ref{eq:new2.44vanewz}) 
(and thus twice on the rhs of (\ref{eq:new2.44vanew})).
The fact that ${\cal B}_{orb}$ occurs twice on the rhs of (\ref{eq:new2.44vanew}) we owe, by Section 9, to the fact that
${\cal B}_{orb}$ occurs twice on the rhs of (\ref{eq:new2.44vcdc0}). Thus the fact that  
${\cal B}_{orb}$ occurs twice on the rhs of the full Bloch equation (\ref{eq:new2.44vcdc0}) is 
crucial for obtaining the extension of the kinetic approach.
Note also that ${\cal B}_{orb}$ occurs twice on the rhs of (\ref{eq:new2.44vcdc0})
because it occurs twice on the rhs (\ref{eq:new2.44vcdc0new}). 
Thus, hypothetically speaking,
the extension of the kinetic approach 
would not be possible if ${\cal B}_{orb}$ would occur only once on the rhs of (\ref{eq:new2.44vcdc0new}). 

Since $\left( \begin{array}{c}  {\cal B}_{orb} \\
\vec{\cal B}_{spin}\end{array}\right)$ 
is a noise vector field of (\ref{eq:new2.44vanewnew}) the following is an
Ito SDE system
associated with the full spin-orbit Fokker-Planck equation, (\ref{eq:new2.44vanewnew}):
\begin{eqnarray}                                                            
&&  Q'  = {\cal D}_{orb}(t,Q)  +  {\cal B}_{orb}(t,Q) \nu(t)\; ,
\nonumber\\
&& \vec{S}'
=  \vec{\cal D}_{spin}(t,Q,\vec{S})  + \vec{\cal B}_{spin}(t,Q)\nu(t)\; ,
\nonumber\\
\label{eq:new2.44vbnewnew}  
\end{eqnarray}
where $\nu$ is the one-dimensional white-noise process.
We call, (\ref{eq:new2.44vbnewnew}), the  
full spin-orbit SDE system. Note that the first equality in (\ref{eq:new2.44vbnewnew}) is the same as
(\ref{eq:new2.44uanew}).
For the notion of associated Ito SDE system we recommend again, \cite{Arn74,Gar04}.

It is sometimes convenient to express the full spin-orbit Fokker-Planck equation
in terms of $l_{orb}$. Thus using (\ref{eq:new02.010a}) we write for later reference 
the full spin-orbit Fokker-Planck equation, (\ref{eq:new2.44vanewz}), as:
\begin{eqnarray}                                                            
&&  \hspace{-10mm} 
\frac{\partial f}{\partial t} 
= l_{orb}f
-\frac{\partial}{\partial s_j} ({\cal D}_{spin,j}f)
+\frac{\partial}{\partial s_k}\frac{\partial}{\partial q_l}
({\cal B}_{orb,l} {\cal B}_{spin,k}f)
+\frac{1}{2}
\frac{\partial}{\partial s_k}\frac{\partial}{\partial s_l}
({\cal B}_{spin,l} {\cal B}_{spin,k}f)\; .
\label{eq:new2.44vvanew}  
\end{eqnarray}
This expression of the full spin-orbit Fokker-Planck equation will be used in Section 11.

It is sometimes also convenient to express the full spin-orbit Fokker-Planck equation
in a more explicit form than in the expressions above. Thus using (\ref{eq:new02.010new27}), (\ref{eq:new2.44vanewnewnew})
we write the full spin-orbit Fokker-Planck equation, (\ref{eq:new2.44vanewz}), as
\begin{eqnarray}                                                            
&&  \hspace{-10mm} 
\frac{\partial f}{\partial t} 
= -\frac{\partial}{\partial q_j} ({\cal D}_{orb,j}f)
-\frac{\partial}{\partial s_j} \Biggl( \biggl( 
\vec{\cal D}_{spin,0,ST}(t,q)
+{\cal D}_{spin,+,TBMT}(t,q)\vec{s}
\nonumber\\
&& + {\cal D}_{spin,+,ST}(t,q)\vec{s}
+{\cal D}_{spin,+,BK}(t,q)\vec{s}\biggr)_j f\Biggr)
+\frac{1}{2}\frac{\partial}{\partial q_k}\frac{\partial}{\partial q_l}
({\cal B}_{orb,l} {\cal B}_{orb,k}f) 
\nonumber\\
&&
+\frac{\partial}{\partial s_k}\frac{\partial}{\partial q_l}
({\cal B}_{orb,l}(t,q) {\cal B}_{spin,k}(t,q)f)
+\frac{1}{2}
\frac{\partial}{\partial s_k}\frac{\partial}{\partial s_l}
({\cal B}_{spin,l}(t,q) {\cal B}_{spin,k}(t,q)f)\; .
\label{eq:new2.44vvanewfinal}  
\end{eqnarray}
Note that (\ref{eq:new2.44vvanewfinal}) will be used in Section 16
(in particular in Section 16 we interprete the terms on the rhs of
(\ref{eq:new2.44vvanewfinal})).
By its derivation it is clear that (\ref{eq:new2.44vvanewfinal}) 
is the same as (\ref{eq:new2.44vanewz}) and thus (\ref{eq:new2.44vvanewfinal}) 
is the same as (\ref{eq:new2.44vanew}), (\ref{eq:new2.44vanewnew}),
(\ref{eq:new2.44vvanew}) 
and so it is the full spin-orbit Fokker-Planck equation.

The following is an Ito SDE system
associated with the full spin-orbit Fokker-Planck equation, (\ref{eq:new2.44vvanewfinal}):
\begin{eqnarray}                                                            
&&   \hspace{-10mm} 
Q' = {\cal D}_{orb}(t,Q) +{\cal B}_{orb}(t,Q)\nu(t) \; ,
\nonumber\\
&& \hspace{-10mm} \vec{S}' =  
\vec{\cal D}_{spin,0,ST}(t,Q)+{\cal D}_{spin,+,TBMT}(t,Q)\vec{S}
\nonumber\\
&&\hspace{-5mm} 
+ {\cal D}_{spin,+,ST}(t,Q)\vec{S}
+{\cal D}_{spin,+,BK}(t,Q)\vec{S}
+\vec{\cal B}_{spin}(t,Q)  \nu(t) \; .
\nonumber\\
\label{eq:new2.44vcdmfinal}                                                           
\end{eqnarray}
Note that (\ref{eq:new2.44vcdmfinal})  is obtained by inserting
(\ref{eq:new02.010new27}), (\ref{eq:new2.44vanewnewnew}) into (\ref{eq:new2.44vbnewnew})
and that (\ref{eq:new2.44vcdmfinal}) will be used in Section 16
(in particular in Section 16 we interprete the terms on the rhs of
(\ref{eq:new2.44vcdmfinal})).
In fact (\ref{eq:new2.44vcdmfinal}) is the same as
(\ref{eq:new2.44vbnewnew}) and so (\ref{eq:new2.44vcdmfinal}) is the
full spin-orbit SDE system.

Most importantly, the
full spin-orbit SDE system is identical to eq. 19 in \cite{HABBE19}.
Thus having derived the SDE system of \cite{HABBE19} we have achieved an important
aim of the present work (see Section 1 for the aims of the present work).
Note that the full spin-orbit SDE system uniquely determines the full
spin-orbit Fokker-Planck equation since the former is an Ito SDE system associated
with the latter. Thus it is no shortcoming of \cite{HABBE19} that it does not
explicitly display the full spin-orbit Fokker-Planck equation.

As an aside we use (\ref{eq:new2.44anew}), (\ref{eq:new2.44anewnew}), 
(\ref{eq:new2.44a}), (\ref{eq:new02.010new26}), (\ref{eq:new02.010new26TBMT}), 
(\ref{eq:new02.010new26ST}), (\ref{eq:new02.010new26BK})
to write (\ref{eq:new2.44vcdmfinal}) more explicitly as
\begin{eqnarray}                                                            
&&   \hspace{-10mm} 
Q'  =
\left( \begin{array}{c}  \vec{v}(Q) \\
\vec{\cal F}(t,Q)+\vec{\cal C}(t,Q)+\vec{\cal Q}(t,Q)\end{array}\right) 
+  {\cal B}_{orb}(t,Q)\nu(t)\; ,
\nonumber\\
&& \hspace{-10mm} \vec{S}'   
=-\frac{1}{m\gamma(Q)}\lambda(t,Q)
\frac{\vec{\Pi}\times\vec{a}_{\cal F}(t,Q)}
{|\vec{a}_{\cal F}(t,Q)|} 
+\vec{\Omega}_{TBMT}(t,Q)\times \vec{S}
\nonumber\\
&&  \hspace{-5mm}
-\frac{5\sqrt{3}}{8}\lambda(t,Q)\vec{S}
+\frac{5\sqrt{3}}{36 m^2\gamma^2(Q)}\lambda(t,Q)
\vec{\Pi}\vec{\Pi}^T\vec{S}  +\vec{\cal B}_{spin}(t,Q)  \nu(t) \; .
\nonumber\\
\label{eq:new2.44vcdm}                                        
\end{eqnarray}
By its derivation (\ref{eq:new2.44vcdm}) is equal to (\ref{eq:new2.44vcdmfinal}) and thus equal to
the full spin-orbit SDE system.
We will use (\ref{eq:new2.44vcdm}) further below in this section.

In the remaining parts of this section we make
remarks on the full spin-orbit SDE system.
First of all, let us now give some practical instructions on numerical solution methods
of the full spin-orbit SDE system, e.g., standard SDE solvers 
or particle tracking codes (the latter perhaps obtained
by extending an existing Monte-Carlo 
spin tracking code developed for computing the radiative
depolarization time). 
As pointed out in Section 1.2
in order to perform such computations one transforms, from cartesian coordinates to machine
coordinates, all equations involved, that is,
the full spin-orbit SDE system, the definition, (\ref{eq:new2.44vcdn90orbnew}),
of the polarization vector etc.
To be specific let us sketch the typical computational procedure to be performed by a numerical
solution method: Compute the initial polarization vector and the polarization vector
at a later time (here `time' in the sense of machine coordinates)!
To fulfill this task one may pick $N$ spin-orbit vectors in machine coordinates, 
$(Q_{0,n},\vec{S}_{0,n})$, where $n=1,...N$
and $Q_{0,n}\in\R^6,\vec{S}_{0,n}\in\R^3$ with $|\vec{S}_{0,n}|\leq 1$.
One now defines 
$p_1,...,p_N$ as the values of the initial orbital density  at the 
initial points $(0,Q_{0,1}),..., (0,Q_{0,N})$. The initial
polarization vector is now defined as
$\frac{p_1\vec{S}_{0,1}+\cdots+p_N\vec{S}_{0,N}}{p_1+\cdots + p_N}$. 
Thus $|\frac{p_1\vec{S}_{0,1}+\cdots+p_N\vec{S}_{0,N}}{p_1+\cdots + p_N}|\leq 1$. One 
then uses the chosen 
numerical solution method to evolve the $N$
spin-orbit vectors to a final 
`time' resulting in $(Q_{n},\vec{S}_{n})$ where
$n=1,...N$.  
The final polarization 
vector is thus defined as
$\frac{p_1\vec{S}_{1}+\cdots+p_N\vec{S}_{N}}{p_1+\cdots + p_N}$.
Note that this sketch of 
computing the initial and final
polarization vector is schematic and
may even have to be modified. But we believe it reflects the main idea which is
familiar from Monte-Carlo spin
tracking codes developed for
computing the radiative
depolarization time.

Secondly, in order to further clarify these practical instructions we note that, 
in contrast to the reduced spin-orbit SDE system (see 
(\ref{eq:new2.44vbnewnewred})), in the full spin-orbit SDE system
$|\vec{S}(t)|$ is not conserved in time. Therefore one cannot interprete  $\vec{S}(t)$ in the
full spin-orbit SDE system as the spin of a single particle.
To find an interpretation we assume, for simplicity,
in this paragraph, that (\ref{eq:new02.16b})            
holds, i.e., 
$\rho[W_f]>0$ so that we can apply Section 6 and in particular the remarks after
(\ref{eq:new02.16dnew}).   
If (\ref{eq:new02.12}) holds at $(t,Q(t))$, i.e., 
$|\vec{\cal P}[W_f](t,Q(t))| \leq  \rho[W_f](t,Q(t))$, 
then, by the remarks after
(\ref{eq:new02.16dnew}), we can interprete $\vec{S}(t)$ in 
the full spin-orbit SDE system  as the spin polarization vector at $(t,Q(t))$.
When (\ref{eq:new02.12}) does not hold at $(t,Q(t))$
then, by the remarks after
(\ref{eq:new02.16dnew}), 
we can interprete $\vec{S}(t)$ in 
the full spin-orbit SDE system  as the weighted spin polarization vector at $(t,Q(t))$.
Most importantly we believe that the distinction between the
two cases (where  (\ref{eq:new02.12}) does or does not hold at a point)
has no have detrimental effect on practical computations.
In fact in the above sketch of computing
the initial and final polarization vector the two cases are treated in the same way.
This comes as a relief since determining, which case holds at a point,
would be computationally expensive. Note also that from now on in this work 
the condition, (\ref{eq:new02.16b}), will not be used anymore.

Thirdly, we recall from Sections 3 and 4 that  
when neglecting all radiative effects, the functions
$\vec{\cal C},\vec{\cal Q},\lambda,{\cal B}_{orb},\vec{\cal D}_{spin,0,ST},
{\cal D}_{spin,+,ST},{\cal D}_{spin,+,BK},\vec{\cal B}_{spin}$ vanish and
${\cal D}_{orb}$ becomes ${\cal D}_{orb,nrad}$.
Thus we note that when neglecting all radiative effects in
(\ref{eq:new2.44vcdm}) we get
\begin{eqnarray}                                                            
&&   \hspace{-10mm} \vec{R}'=\vec{v}(Q) \; , \quad
\vec{\Pi}'  = \vec{\cal F}(t,Q)  \; , \quad
\vec{S}' =  \vec{\Omega}_{TBMT}(t,Q)\times \vec{S} \; ,
\label{eq:new2.44vcdmnew}                                                           
\end{eqnarray}
and, when neglecting all radiative effects in (\ref{eq:new2.44vvanewfinal}), we get
\begin{eqnarray}                                                            
&&  \hspace{-10mm} 
\frac{\partial f}{\partial t} 
= -\frac{\partial}{\partial r_j} (v_j f)-\frac{\partial}{\partial \pi_j} ({\cal F}_j f)
-\frac{\partial}{\partial s_j} \Biggl( \biggl( \vec{\Omega}_{TBMT}\times \vec{s}\biggr)_j f\Biggr) \; ,
\label{eq:new2.44vanradnew}  
\end{eqnarray}
where we also used 
(\ref{eq:new2.44anew}), (\ref{eq:new02.010new26TBMT}).
Note that the first two equations in (\ref{eq:new2.44vcdmnew})   
are the same as (\ref{eq:new2.44uaanewnew}). 
The point to be made now is the following:
The solutions $\left( \begin{array}{c}  \vec{R} \\   \vec{\Pi} \\  \vec{S} \end{array}\right)$ of
(\ref{eq:new2.44vcdmnew}) are characteristics of (\ref{eq:new2.44vanradnew})  
in the sense that they satisfy:
$\frac{d}{dt} \Biggl( f\biggl(t,\vec{R}(t),\vec{\Pi}(t),\vec{S}(t)\biggr)\Biggr) = 0$, 
%
%
where $f$ is any solution of (\ref{eq:new2.44vanradnew}).
As an aside we note, by (\ref{eq:new2.44vcdmnew}), that the radiative terms on the rhs of
(\ref{eq:new2.44vcdm}) are those terms in (\ref{eq:new2.44vcdm}) which depend nonlinearly
on $\vec{a}_{\cal F}$ reflecting that every radiative effect in this work 
is a (magnetic) bremsstrahlung effect.

Fourthly, since 
(\ref{eq:new2.44vbnewnew})
is an Ito SDE system 
associated with (\ref{eq:new2.44vanewnew})
it is natural to ask for a
Stratonovich SDE system associated with 
(\ref{eq:new2.44vanewnew}). 
In fact, using for example
Section 10.2 in \cite{Arn74}, the following is such an SDE system:
\begin{eqnarray}                                               
&&  Q'  = {\cal D}_{orb}(t,Q)  -\frac{1}{2} 
{\cal B}_{orb,k}(t,Q) 
\frac{\partial {\cal B}_{orb}}{\partial q_k}(t,Q)  
+  {\cal B}_{orb}(t,Q) \nu(t)\; ,
\nonumber\\
&& \vec{S}'
=  \vec{\cal D}_{spin}(t,Q,\vec{S})  + \vec{\cal B}_{spin}(t,Q)\nu(t)\; .
\nonumber\\
\label{eq:new2.44vbnewOL}  
\end{eqnarray}
As an aside we note that the white-noise part of the
Stratonovich SDE system is nonunique in the
same way as for the Ito SDE system.
To get insight into ${\cal B}_{orb,k}
\frac{\partial {\cal B}_{orb}}{\partial q_k}$
we compute, by (\ref{eq:new2.44g}), (\ref{eq:new2.44e}),(\ref{eq:new2.44h}),
\begin{eqnarray}     
&&  {\cal B}_{orb,k}(t,q) 
\frac{\partial {\cal B}_{orb}}{\partial q_k}(t,q) 
= \frac{55}{48\sqrt{3}}
\biggl(  2\lambda(t,q)  + \pi_i \frac{\partial \lambda}{\partial \pi_i}(t,q) \biggr)
\left( \begin{array}{c}  \vec{0} \\   
\vec{\pi} \end{array}\right) \; ,
\nonumber\\
&& \pi_i \frac{\partial \lambda}{\partial \pi_i}(t,q) 
=\biggl( 2+ \frac{1}{\gamma^2(\vec{\pi})}\biggr)\lambda(t,q) \; ,
\nonumber\\
\label{eq:1.28} 
\end{eqnarray}
which implies, by (\ref{eq:new2.44f}),
\begin{eqnarray}                                               
&&  -\frac{1}{2} 
{\cal B}_{orb,k}(t,q) 
\frac{\partial {\cal B}_{orb}}{\partial q_k}(t,q)  
=-\frac{55}{96\sqrt{3}}
\biggl(  4 +  \frac{1}{\gamma^2(q)}\biggr)\lambda(t,q) 
\left( \begin{array}{c}  \vec{0} \\   
\vec{\pi} \end{array}\right) 
\nonumber\\
&& \hspace{5mm} 
=  - \frac{4\gamma^2(\vec{\pi})+1}{12\gamma^2(\vec{\pi})+2}
\left( \begin{array}{c}  \vec{0} \\  
\vec{\cal Q}(t,q)\end{array}\right) 
\; ,
\label{eq:1.30}  
\end{eqnarray}
and so (\ref{eq:new2.44vbnewOL}) can be written 
more explicitly as
\begin{eqnarray}                                              
&&  Q'  = {\cal D}_{orb}(t,Q)  
+\left( \begin{array}{c}  \vec{0} \\   
- \frac{4\gamma^2(\vec{\pi})+1}{12\gamma^2(\vec{\pi})+2}\vec{\cal Q}(t,q)
\end{array}\right) +  {\cal B}_{orb}(t,Q) \nu(t)\; ,
\nonumber\\
&& \vec{S}'
=  \vec{\cal D}_{spin}(t,Q,\vec{S})  + \vec{\cal B}_{spin}(t,Q)\nu(t)\; .
\nonumber\\
\label{eq:new2.44vbnewnewOL}  
\end{eqnarray}
Comparing (\ref{eq:new2.44vbnewnewOL}) with  
(\ref{eq:new2.44vbnewnew})
we see that the only difference between Ito and Stratonovich is the term
$\left( \begin{array}{c}  \vec{0} \\   
- \frac{4\gamma^2(\vec{\pi})+1}{12\gamma^2(\vec{\pi})+2}\vec{\cal Q}(t,q)
\end{array}\right)$ in (\ref{eq:new2.44vbnewnewOL}). 
Thus in those applications, where one neglects 
$\vec{\cal Q}$, i.e.,
where $\vec{\cal C}+\vec{\cal Q}$
is approximated by $\vec{\cal C}$, Ito
and Stratonovich are the same (recall from the discussion before (\ref{eq:new02.010new11})
that neglecting
$\vec{\cal Q}$ is very common).
That the second equation in (\ref{eq:new2.44vbnewnewOL}) 
is the same as the second equation in 
(\ref{eq:new2.44vbnewnew})
is due to the fact that $\vec{\cal B}_{spin}(t,Q)$ is
independent of $\vec{S}$.
The main interest in the Stratonovich form of
the SDE system arises from its convenience when it 
comes to transformations, e.g., the transformation from
cartesian to machine coordinates (of course, 
in applications, where one neglects $\vec{\cal Q}$, the Ito form is as convenient for performing transformations as is the 
Stratonovich form).

Fifthly, let us ask how the full 
spin-orbit SDE system, (\ref{eq:new2.44vcdm}),
would look like in 
Gaussian cgs units (=Gaussian cgs units
without $c=1$). Writing (\ref{eq:new2.44vcdm}) in
Gaussian cgs units as
\begin{eqnarray}                                              
&&   \hspace{-10mm} 
Q'  =
\left( \begin{array}{c}  \vec{v}_{cgs}(Q_{cgs}) \\
\vec{\cal F}_{cgs}(t_{cgs},Q_{cgs})
+\vec{\cal C}_{cgs}(t_{cgs},Q_{cgs})
+\vec{\cal Q}_{cgs}(t_{cgs},Q_{cgs})\end{array}\right) 
+  {\cal B}_{orb}^{cgs}(t_{cgs},Q_{cgs})
\nu(t_{cgs}) \; ,
\nonumber\\
&& \hspace{-10mm} \vec{S}'   
=-\frac{1}{m_{cgs}c\gamma_{cgs}(Q_{cgs})}
\lambda_{cgs}(t_{cgs},Q_{cgs})
\frac{\vec{\Pi}_{cgs}\times
\vec{a}_{\cal F}^{cgs}(t_{cgs},Q_{cgs})}
{|\vec{a}_{\cal F}^{cgs}(t_{cgs},Q_{cgs}|} 
+\vec{\Omega}_{TBMT}^{cgs}(t_{cgs},Q_{cgs})\times \vec{S}
\nonumber\\
&&  \hspace{-5mm}
-\frac{5\sqrt{3}}{8}\lambda_{cgs}(t_{cgs},Q_{cgs})\vec{S}
+\frac{5\sqrt{3}}{36 m_{cgs}^2 c^2\gamma^2_{cgs}(Q_{cgs})}
\lambda_{cgs}(t_{cgs},Q_{cgs})
\vec{\Pi}_{cgs}\vec{\Pi}^T_{cgs}\vec{S}  
\nonumber\\
&&  \hspace{-5mm}
+\vec{\cal B}_{spin}^{cgs}(t_{cgs},Q_{cgs})  \nu(t_{cgs}) \; ,
\nonumber\\
\label{eq:new2.44vcdmnewcgs}                                        
\end{eqnarray}
we thus have to identify 
$\vec{v}_{cgs},...,\vec{\cal B}_{spin}^{cgs}$.
To accomplish this we use the following
three facts: In Gaussian cgs units 
$e\vec{E}$ and $e\vec{B}$ have the dimension of force and the classical electron
radius is given by $e^2/mc^2$.
With these facts in mind one arrives, by using 
(\ref{eq:new2.44b}), (\ref{eq:new2.44c}),
(\ref{eq:new2.44d}), (\ref{eq:new2.44g}),
(\ref{eq:new2.44e}), (\ref{eq:new2.44h}),
(\ref{eq:new2.99anew}), (\ref{eq:new2.99d}), at:
\begin{eqnarray}     
&& 
\gamma_{cgs}(q_{cgs}):=
\sqrt{ \frac{|\vec{\pi}_{cgs}|^2+m_{cgs}^2 c^2}{m_{cgs}^2 c^2}} \; ,
\quad  
\vec{v}_{cgs}(q_{cgs}):=
\frac{\vec{\pi}_{cgs}}{m_{cgs}\gamma_{cgs}(q_{cgs})} \; , 
\nonumber\\
&& 
\vec{a}_{\cal F}^{cgs}(t_{cgs},q_{cgs}):=
c^{-1}\cdot\Biggl(
\frac{e_{cgs}}{m_{cgs}\gamma_{cgs}(q_{cgs})}
\biggl(\vec{v}_{cgs}(q_{cgs})\times
\vec{B}_{cgs}(t_{cgs},\vec{r}_{cgs})\biggr)\Biggr) \; ,
\nonumber\\
&&  
\lambda_{cgs}(t_{cgs},q_{cgs}):=c^{-8}\cdot\biggl(
\hbar\frac{e_{cgs}^2}{m_{cgs}^2}\gamma^5_{cgs}(q_{cgs})
|\vec{a}_{\cal F}^{cgs}(t_{cgs},q_{cgs})|^3\biggr) \; ,
\nonumber\\
&&
{\cal F}_{cgs}(t_{cgs},q_{cgs}):= e_{cgs}
\biggl(  \vec{E}_{cgs}(t_{cgs},\vec{r}_{cgs}) +\frac{\vec{v}_{cgs}(q_{cgs})}{c}\times
\vec{B}_{cgs}(t_{cgs},\vec{r}_{cgs})\biggr) \; ,
\nonumber\\
&& 
\vec{\cal C}_{cgs}(t_{cgs},q_{cgs}):=c^{-5}\cdot\biggl(
-\frac{2}{3}\frac{e_{cgs}^2}{m_{cgs}}
\gamma^3_{cgs}(q_{cgs})
|\vec{a}_{\cal F}^{cgs}(t_{cgs},q_{cgs})|^2
\vec{\pi}_{cgs}\biggr) \; ,
\nonumber\\
&&  
\vec{\cal Q}_{cgs}(t_{cgs},q_{cgs}):=  \frac{55}{48\sqrt{3}}
\biggl( 6+ \frac{1}{\gamma^2_{cgs}(q_{cgs})}\biggr)\lambda_{cgs}(t_{cgs},q_{cgs})\vec{\pi}_{cgs} \; ,
\nonumber\\
&&  
{\cal B}_{orb}^{cgs}(t_{cgs},q_{cgs}):=
\sqrt{ \frac{55}{24\sqrt{3}}}
\sqrt{\lambda_{cgs}(t_{cgs},q_{cgs})}
\left( \begin{array}{c}  \vec{0} \\
\vec{\pi}_{cgs}\end{array}\right) \; ,
\nonumber\\
&&  \vec{\Omega}_{TBMT}^{cgs}(t_{cgs},q_{cgs}):=
c^{-1}\cdot\Biggl(
-\frac{e_{cgs}}{m_{cgs}\gamma_{cgs}(q_{cgs})}\biggl( 1+
\frac{g-2}{2}\gamma_{cgs}(q_{cgs})\biggr)
\vec{B}_{cgs}(t_{cgs},\vec{r}_{cgs})\Biggr)
\nonumber\\
&& \hspace{5mm}
+ c^{-3}\cdot\biggl(
\frac{e_{cgs}}{m_{cgs}} \frac{g-2}{2}
\frac{\gamma_{cgs}(q_{cgs})}{1+\gamma_{cgs}(q_{cgs})}
\vec{v}_{cgs}(q_{cgs})v_{cgs,i}(q_{cgs}) 
B_{cgs,i}(t_{cgs},\vec{r}_{cgs})\biggr)
\nonumber\\
&& \hspace{5mm} + c^{-2}\cdot\Biggl(
\frac{e_{cgs}}{m_{cgs}}
\biggl( \frac{g-2}{2} +\frac{1}{1+\gamma_{cgs}(q_{cgs})}\biggr)
\biggl(\vec{v}_{cgs}(q_{cgs})\times
\vec{E}_{cgs}(t_{cgs},\vec{r}_{cgs})\biggr) \Biggr)\; ,
\nonumber
\end{eqnarray}
\begin{eqnarray}     
&&     
\vec{\cal B}_{spin}^{cgs}(t_{cgs},q_{cgs}):=   
c^{-1}\cdot\biggl(
\sqrt{ \frac{24\sqrt{3}}{55}}
\sqrt{\lambda_{cgs}(t_{cgs},q_{cgs})}
\frac{1}{m_{cgs}\gamma_{cgs}(q_{cgs})}
\frac{\vec{\pi}_{cgs}\times
\vec{a}_{cgs}(t_{cgs},q_{cgs})}
{|\vec{a}_{cgs}(t_{cgs},q_{cgs})|}\biggr) \; ,
\nonumber\\
\label{eq:new2.44gcgsOL} 
\end{eqnarray}
where we also used the fact that, in Gaussian cgs units with $c=1$, every power of $c$
is equal to $1$. Note that, by the above method,
every equation of this work could be easily 
converted to Gaussian cgs
units.

\section{The reduced spin-orbit Fokker-Planck equation and the reduced spin-orbit 
stochastic ODE system}

In this section we revisit Section 10 by focusing on the reduced setup.

In the reduced setup, defined by (\ref{eq:new02.010new30a}) resp.
(\ref{eq:new02.010new30b}), (\ref{eq:new02.010new30c}),
the full spin-orbit Fokker-Planck
equation, (\ref{eq:new2.44vvanew}),
simplifies to
the so-called reduced spin-orbit Fokker-Planck equation:
\begin{eqnarray}                                                            
&&  \hspace{-10mm} 
\frac{\partial f}{\partial t} 
= l_{orb}f
-\frac{\partial}{\partial s_j} \Biggl( \biggl( \vec{\Omega}_{TBMT}\times \vec{s}\biggr)_j f\Biggr)\; .
\label{eq:new2.44vanewnewred}  
\end{eqnarray}
Note that (\ref{eq:new2.44vanewnewred})  
is obtained by inserting (\ref{eq:new02.010new30b}), (\ref{eq:new02.010new30c})
into (\ref{eq:new2.44vvanew}).
By (\ref{eq:new02.010a})
we can write (\ref{eq:new2.44vanewnewred}) also in the form:
\begin{eqnarray}                                                            
&&  \hspace{-10mm} 
\frac{\partial f}{\partial t} 
= -\frac{\partial}{\partial q_j} ({\cal D}_{orb,j}f)
+\frac{1}{2}\frac{\partial}{\partial q_k}\frac{\partial}{\partial q_l}
({\cal B}_{orb,l} {\cal B}_{orb,k}f)
-\frac{\partial}{\partial s_j} \Biggl( \biggl( \vec{\Omega}_{TBMT}\times \vec{s}\biggr)_j f\Biggr)\; .
\label{eq:new2.44vanewnewredfinal}  
\end{eqnarray}
Note that $\left( \begin{array}{c}  {\cal B}_{orb} \\
\vec{0}\end{array}\right)$ 
is a noise vector field of (\ref{eq:new2.44vanewnewredfinal}) and thus the following is an
Ito SDE system
associated with the reduced spin-orbit Fokker-Planck equation, (\ref{eq:new2.44vanewnewredfinal}):
\begin{eqnarray}                                                            
&&  \hspace{-10mm}
Q'  =  {\cal D}_{orb}(t,Q) +  {\cal B}_{orb}(t,Q)\nu(t)\; ,
\nonumber\\
&&  \hspace{-10mm}
\vec{S}'
=\vec{\Omega}_{TBMT}(t,Q)\times \vec{S}  \; ,
\nonumber\\
\label{eq:new2.44vbnewnewred}  
\end{eqnarray}
where $\nu$ is the one-dimensional white-noise process.
We call, (\ref{eq:new2.44vbnewnewred}), the  
reduced spin-orbit SDE system. 
With (\ref{eq:new2.44anew}), (\ref{eq:new2.44anewnew}), 
(\ref{eq:new2.44a}) we can write (\ref{eq:new2.44vbnewnewred}) more explicitly as
\begin{eqnarray}                                                            
&&   \hspace{-10mm}
Q'  =
\left( \begin{array}{c}  \vec{v}(Q) \\
\vec{\cal F}(t,Q)+\vec{\cal C}(t,Q)+\vec{\cal Q}(t,Q)\end{array}\right) 
+  {\cal B}_{orb}(t,Q)\nu(t) \; ,
\nonumber\\
&& \hspace{-10mm}
 \vec{S}' =  
\vec{\Omega}_{TBMT}(t,Q)\times \vec{S} \; .
\nonumber\\
\label{eq:new2.44vcdmred}                                                           
\end{eqnarray}
Clearly (\ref{eq:new2.44vbnewnewred}) is the same as
(\ref{eq:new2.44vcdmred}) and so (\ref{eq:new2.44vcdmred}) is the
reduced spin-orbit SDE system. 
If one neglects $\vec{\cal Q}$ then
(\ref{eq:new2.44vcdmred}) becomes the SDE system which was
already derived in \cite{BH01} and (\ref{eq:new2.44vanewnewred}) 
becomes the Fokker-Planck equation which was
already derived in \cite{BH01} (recall from Section 3 that $\vec{\cal Q}$ is neglible).

Note that if $f$ satisfies the reduced spin-orbit Fokker-Planck equation, 
(\ref{eq:new2.44vanewnewred}), 
then (if $W_f$ is sufficiently regular), by the dynamical
condition stated in Section 7, $W_f$ satisfies (\ref{eq:new2.100b}), i.e.,
\begin{eqnarray} 
&& \hspace{-10mm}
 \frac{\partial W_f}{\partial t} 
=l_{orb}W_f
+\frac{1}{2}\sigma_i\biggl( 
\vec{\Omega}\times
\left( \begin{array}{c} 
Tr_{2\times 2}[\sigma_1 W_f]\\
Tr_{2\times 2}[\sigma_2 W_f]\\
Tr_{2\times 2}[\sigma_3 W_f] 
\end{array}\right)
\biggr)_i \; .
\label{eq:new2.100bf}                                                           
\end{eqnarray}
By our remarks made after (\ref{eq:new2.100b})
we also note that (\ref{eq:new2.100bf}) 
contains all the information to study the
radiative depolarization effect (same for the reduced spin-orbit
Fokker-Planck equation and for the reduced spin-orbit SDE system).
Thus by the dynamical condition the reduced spin-orbit
Fokker-Planck equation (as well as the reduced spin-orbit SDE system)
contains all the information to study the
radiative depolarization effect.

\section{Spin-orbit densities: The physically meaningful solutions of the full and the reduced spin-orbit 
Fokker-Planck equations}

In this section we define and discuss those 
solutions of the full spin-orbit Fokker-Planck equation which we call spin-orbit densities or, shortly, physically meaningful solutions.
Since the reduced spin-orbit Fokker-Planck equation is a special case of the
full spin-orbit Fokker-Planck equation, the results of the present section apply to the reduced setup
as well.

Using from Section 5.2 the notion 
of spin-$1/2$ Wigner function of a bunch we thus require for 
every spin-orbit density, $f$, that $W_f$ is the spin-$1/2$ Wigner function of a bunch.
In other words
we want $W_f$ to
satisfy the statistical conditions, 
(\ref{eq:new02.10c}), (\ref{eq:new02.10d}), (\ref{eq:new02.10k}),
that is, for $t\in\R,q\in\R^6$,
\begin{eqnarray}                                                            
&&  \hspace{-15mm} \biggl( W_f(t,q)\biggr)^\dagger = W_f(t,q) \; , 
\label{eq:new02.10cf}  \\
&&  \hspace{-15mm}
\int_{\R^6} \; Tr_{2\times 2}\biggl[W_f(t,\tilde{q})\biggr] d^6 \tilde{q} = 1 \; ,
\label{eq:new02.10df} \\
&&  \hspace{-15mm}
\Big{|} \int_{\R^6} \; Tr_{2\times 2}\biggl[\vec{\sigma}W_f(t,\tilde{q})\biggr] d^6\tilde{q} \Big{|} \leq 1 \; ,
\label{eq:new02.10kf}                                                           
\end{eqnarray}
and we want $W_f$ to satisfy the evolution equation (which in the full setup is
(\ref{eq:new002.100f}) and which in the reduced setup is (\ref{eq:new2.100bf})). 

We know from Section 5.2 that
the statistical conditions,
(\ref{eq:new02.10c}), (\ref{eq:new02.10d}), (\ref{eq:new02.10k}), are equivalent to the
conditions, (\ref{eq:new02.10e}), (\ref{eq:new02.10f}), (\ref{eq:new02.15}), 
on $\rho[W_f],\vec{\cal P}[W_f]$ which read for $W=W_f$ as
\begin{eqnarray}                                                            
&&   \rho[W_f](t,q)\in\R \; , \quad \vec{\cal P}[W_f](t,q)\in\R^3 \; ,
\label{eq:new02.10ef}  \\
&& \int_{\R^6} \; \rho[W_f](t,\tilde{q})d^6\tilde{q}   = 1 \; ,
\label{eq:new02.10gsof}  \\                                               
&& \big{|}\int_{\R^6} \; \vec{\cal P}[W_f](t,\tilde{q})  d^6\tilde{q}\big{|}  \leq  1\; ,
\label{eq:new02.10fsox}                                                                
\end{eqnarray}
where $t\in\R,q\in\R^6$.

Motivated by (\ref{eq:new02.10ef}), (\ref{eq:new02.10gsof}), (\ref{eq:new02.10fsox}) 
and the dynamical condition stated in Section 7
we call $f$ a spin-orbit density of the full spin-orbit Fokker-Planck equation iff it is a solution of the
full spin-orbit Fokker-Planck equation
and satisfies, for $t\in\R,q\in\R^6,\vec{s}\in\R^3$,
\begin{eqnarray}                                                            
&&   f(t,q,\vec{s})\in\R \; ,
\label{eq:new02.10eso} \\ 
&&   \int_{\R^9} \; f(t,\tilde{q},\vec{\tau})d^3\tau d^6\tilde{q}   = 1 \; , 
\label{eq:new02.10dso} \\                                                
&&  \big{|} \int_{\R^9} \; \vec{\tau} f(t,\tilde{q},\vec{\tau})d^3\tau
d^6\tilde{q} \big{|} \leq 1  \; .
\label{eq:new02.10fso}                                                                
\end{eqnarray}
Note, by (\ref{eq:new2.44p}),
that (\ref{eq:new02.10ef}), (\ref{eq:new02.10gsof}) 
are equivalent to (\ref{eq:new02.10eso}), (\ref{eq:new02.10dso}) 
except that the condition, (\ref{eq:new02.10eso}), 
is, for the sake of convenience,
stronger than (\ref{eq:new02.10ef}).
Note also that (\ref{eq:new02.10fso}) is equivalent to   
(\ref{eq:new02.10fsox}) (because of (\ref{eq:new2.44q})).

Clearly if $f$ is a spin-orbit density then, by the discussion before (\ref{eq:new02.10ef}), 
$W_f$ satisfies the statistical conditions, 
(\ref{eq:new02.10cf}), (\ref{eq:new02.10df}), (\ref{eq:new02.10kf}),
and, by the dynamical condition, $W_f$ satisfies the evolution equation
(\ref{eq:new002.100f}). Thus, by Section 5.2, if $f$ is a spin-orbit density then $W_f$ is the
spin-$1/2$ Wigner function of a bunch. 

Let us now, in analogy to Section 5.4, discuss how
the three statistical conditions, 
(\ref{eq:new02.10eso}), (\ref{eq:new02.10dso}), (\ref{eq:new02.10fso}), 
on the spin-orbit density, $f$,
fit to the dynamics of $f$.
In fact, since the full spin-orbit Fokker-Planck equation is a Fokker-Planck equation we observe that
(\ref{eq:new02.10eso}), (\ref{eq:new02.10dso}) hold for all $t$ if they hold for $t=0$ (and if the
coefficient functions of the full spin-orbit Fokker-Planck equation are sufficiently regular), i.e.,
for $q\in\R^6,\vec{s}\in\R^3$,
\begin{eqnarray}                                                            
&&   f(0,q,\vec{s})\in\R \; ,
\label{eq:new02.10eso0} \\
&&   \int_{\R^9} \; f(0,\tilde{q},\vec{\tau})d^3\tau d^6\tilde{q}   = 1 \; .
\label{eq:new02.10dso0}  
\end{eqnarray}
In analogy to Section 5.4
we leave the following question unanswered: Does (\ref{eq:new02.10fso}) hold 
for all $t$ if it holds for $t=0$? 
Nevertheless it follows from (\ref{eq:new2.44q}) that
(\ref{eq:new02.10fso}) is equivalent to:
\begin{eqnarray}                                                            
&&  \hspace{-15mm} 
\Big{|} \int_{\R^6} \; 
\left( \begin{array}{c} 
Tr_{2\times 2}[\sigma_1 W_f(t,q)]\\
Tr_{2\times 2}[\sigma_2 W_f(t,q)]\\
Tr_{2\times 2}[\sigma_3 W_f(t,q)] 
\end{array}\right) 
d^6q \Big{|} \leq 1 \; ,
\label{eq:new02.10fsoxx}                                                                
\end{eqnarray}
and so by the discussion after
(\ref{eq:new002.101})
we believe that in general the answer to the above question
is: yes (if the coefficient functions of the full spin-orbit Fokker-Planck equation 
are sufficiently regular). Thus we believe that in general  
(\ref{eq:new02.10fso}) holds if:
\begin{eqnarray}                                                            
&& 
\big{|} \int_{\R^9} \; \vec{\tau} f(0,\tilde{q},\vec{\tau})d^3\tau d^6\tilde{q} \big{|}  \leq  1\; .
\label{eq:new02.10fso0}                                                                
\end{eqnarray}
In summary, under the belief, (\ref{eq:new02.10fso0}), 
$f$ is a spin-orbit density (and thus
$W_f$ is the spin-$1/2$ 
Wigner function of a bunch and 
$\rho[W_f]$ is the orbital density and $\vec{\cal P}[W_f]$ is the polarization
density of that bunch) if $f$ satisfies the full spin-orbit Fokker-Planck equation.
Nevertheless the present work does not rely on this belief, 
except when explicitly mentioned (namely in Section 15.1).

For later reference we
now discuss expectation values w.r.t. $W_f$. To consider these expectation values
let $f$ be a spin-orbit density of the full
spin-orbit Fokker-Planck equation.
Thus $W_f$ is the spin-$1/2$ Wigner function of a bunch and so
the expectation values $\langle{A}\rangle_{W_f}$ are well defined 
and can be directly expressed in terms of $f$.
In fact, by (\ref{eq:new02.10h}),
(\ref{eq:new2.44p}), (\ref{eq:new2.44q}), we have
\begin{eqnarray}                                                            
&& \hspace{-15mm}
\langle{A}\rangle_{W_f}(t)=\int_{\R^6} \; \biggl( A_{orb}(t,q)  \rho[W_f](t,q) + A_{spin,j}(t,q){\cal P}_j[W_f](t,q)\biggr)d^6q 
\nonumber\\
&& =\int_{\R^6} \; \biggl( A_{orb}(t,q)  \rho[W_f](t,q) +A_{spin,j}(t,q){\cal P}_j[W_f](t,q)\biggr)d^6q 
\nonumber\\
&& =\int_{\R^9} \; 
\biggl( A_{orb}(t,q)  + s_j A_{spin,j}(t,q)\biggr)f(t,q,\vec{s})d^3s d^6q
\; ,
\nonumber
\end{eqnarray}
in short
\begin{eqnarray}                                                            
&& \hspace{-20mm}
\langle{A}\rangle_{W_f}(t)
=\int_{\R^9} \; 
\biggl( A_{orb}(t,q)  +  s_j A_{spin,j}(t,q)\biggr)f(t,q,\vec{s})d^3s d^6q
\; .
\label{eq:new02.10hf}                                                           
\end{eqnarray}
In the special case, $\vec{A}_{spin}=\vec{0}$, i.e., $A= I_{2\times 2}A_{orb}$ 
we conclude from (\ref{eq:new02.10hf})     
\begin{eqnarray}                                                            
&& \hspace{-5mm}
\langle I_{2\times 2} A_{orb} \rangle_{W_f}(t)
    = \int_{\R^6} \; A_{orb}(t,q) \rho[W_f](t,q) d^6q 
= \int_{\R^9} \; A_{orb}(t,q) f(t,q,\vec{s})d^3s d^6q \; .
\label{eq:new2.44vcdn88orb}
\end{eqnarray}
In the special case, $A_{orb}=0$, (\ref{eq:new02.10hf}) results in
\begin{eqnarray}                                                            
&& 
\langle \sigma_j A_{spin,j}\rangle_{W_f}(t) = \int_{\R^6} \;  A_{spin,j}(t,q) 
{\cal P}_j[W_f](t,q) d^6q 
\nonumber\\
&& = \int_{\R^9} \; s_j A_{spin,j}(t,q) f(t,q,\vec{s})d^3s d^6q \; ,
\label{eq:new2.44vcdn89orb}
\end{eqnarray}
in particular we get for the polarization vector
\begin{eqnarray}                                                            
&& \hspace{-15mm} \langle\vec{\sigma}\rangle_{W_f}(t) 
= \int_{\R^9} \; \vec{s} f(t,q,\vec{s})d^3s d^6q 
= \vec{P}[W_f](t) \; ,
\label{eq:new2.44vcdn90orb}                                                           
\end{eqnarray}
where in the second equation of (\ref{eq:new2.44vcdn90orb}) we used
(\ref{eq:new2.44vcdn90orbnew}).  

Note finally, by the above, that
we do not require that a spin-orbit density,
$f$, is a probability density, i.e., negative values of $f$ are not problematic
(for explicit examples with negative values, see Section 14).

\section{Some general properties of the full and the reduced spin-orbit Fokker-Planck equations}
We here discuss some 
features of the spin-orbit Fokker-Planck equation
(full or reduced). These features are related to (\ref{eq:new2.44vcdn77}), (\ref{eq:new2.44vcdn79c}), (\ref{eq:new2.44vcdn79d}).

A simple observation is the following:  
If $f=f(t,q,\vec{s})$ then we can write:
\begin{eqnarray}                                                            
&& f(t,q,\vec{s}) =  f_{even}(t,q,\vec{s}) + f_{odd}(t,q,\vec{s}) \; ,
\label{eq:new2.44vcdn77}                                                           
\end{eqnarray}
where
\begin{eqnarray}                                                            
&& f_{even}(t,q,\vec{s}):=\frac{f(t,q,\vec{s}) + f(t,q,-\vec{s})}{2} \; , 
\label{eq:new2.44vcdn79c} \\
&& f_{odd}(t,q,\vec{s}):=\frac{f(t,q,\vec{s}) - f(t,q,-\vec{s})}{2} \; .
\label{eq:new2.44vcdn79d}                                                           
\end{eqnarray}
Note, by (\ref{eq:new2.44vcdn79c}), (\ref{eq:new2.44vcdn79d}),     
that $f_{even}$ is even in $\vec{s}$ and $f_{odd}$ is odd in $\vec{s}$.

If $f$ is a spin-orbit density of the spin-orbit Fokker-Planck equation (either full or reduced)
we observe, by (\ref{eq:new2.44p}), (\ref{eq:new2.44q}),
(\ref{eq:new2.44vcdn79c}), (\ref{eq:new2.44vcdn79d}),     
that $\rho[W_{f_{even}}],\rho[W_{f_{odd}}]$ and 
$\vec{\cal P}[W_{f_{even}}],\vec{\cal P}[W_{f_{odd}}]$ exist and satisfy
\begin{eqnarray}                                                            
&&\hspace{-18mm} \rho[W_{f_{even}}] = \rho[W_f] \; , \; \rho[W_{f_{odd}}] = 0 \; , \;
\vec{\cal P}[W_{f_{even}}] = \vec{0}\; , \;  \vec{\cal P}[W_{f_{odd}}]= \vec{\cal P}[W_f] \; .
\label{eq:new2.44vcdn78}                                                           
\end{eqnarray}
With (\ref{eq:new2.44vcdn78}) 
we arrive at the observation that the information about $\rho[W_f]$ resides entirely in $f_{even}$
and that the information about $\vec{\cal P}[W_f]$ resides entirely in $f_{odd}$.
Note also by the second equality of (\ref{eq:new2.44vcdn78}) that  
$f_{odd}$ is never a spin-orbit density (for the notion of spin-orbit density, see Section 12).

With (\ref{eq:new2.44vcdn77}), (\ref{eq:new2.44vcdn79c}), (\ref{eq:new2.44vcdn79d})
we are led to write the
full spin-orbit Fokker-Planck equation, (\ref{eq:new2.44vanewz}), as
\begin{eqnarray}                                                            
&&           \hspace{-15mm}                                                               
\frac{\partial f}{\partial t} = (l_\uparrow + l_\downarrow) f \; ,
\label{eq:109new} 
\end{eqnarray}
where the linear operators $l_\uparrow$ and $l_\downarrow$ are defined by
\begin{eqnarray}                                                            
&&  \hspace{-10mm} 
l_\uparrow g:=
-\frac{\partial}{\partial q_j} ({\cal D}_{orb,j}g)
+\frac{1}{2}\frac{\partial}{\partial q_k}\frac{\partial}{\partial q_l}
({\cal B}_{orb,l} {\cal B}_{orb,k}g)
-\frac{\partial}{\partial s_j} \biggl( ( {\cal D}_{spin,j} - {\cal D}_{spin,0,ST,j})g\biggr)
\nonumber\\
&&
+\frac{1}{2}
\frac{\partial}{\partial s_k}\frac{\partial}{\partial s_l}
({\cal B}_{spin,l} {\cal B}_{spin,k}g)\; ,
\label{eq:new2.44vanewzeo} \\ 
&&  \hspace{-10mm} 
l_\downarrow g:=
-\frac{\partial}{\partial s_j} ( {\cal D}_{spin,0,ST,j} g)
+\frac{\partial}{\partial s_k}\frac{\partial}{\partial q_l}
({\cal B}_{orb,l} {\cal B}_{spin,k}g) \; ,
\label{eq:new2.44vanewzeoeo}  
\end{eqnarray}
and where $g=g(t,q,\vec{s})$ is a two times differentiable $\R$-valued function.
It is clear, by (\ref{eq:new2.44vanewzeo}), (\ref{eq:new2.44vanewzeoeo}), that
(\ref{eq:109new}) is the same as (\ref{eq:new2.44vanewz}).
The key point about
(\ref{eq:new2.44vanewzeo}), (\ref{eq:new2.44vanewzeoeo})
is the following:
Taking the even part of (\ref{eq:109new})
we get from (\ref{eq:new2.44vcdn77}), (\ref{eq:new2.44vcdn79c}), (\ref{eq:new2.44vcdn79d}),  
(\ref{eq:new2.44vanewzeo}), (\ref{eq:new2.44vanewzeoeo}):
\begin{eqnarray}                                                            
&&           \hspace{-15mm}                                                               
\frac{\partial f_{even}}{\partial t} = l_\uparrow f_{even} + l_\downarrow f_{odd} \; ,
\label{eq:109} 
\end{eqnarray}
and taking the odd part of (\ref{eq:109new})
we get from (\ref{eq:new2.44vcdn77}), (\ref{eq:new2.44vcdn79c}), (\ref{eq:new2.44vcdn79d}),  
(\ref{eq:new2.44vanewzeo}), (\ref{eq:new2.44vanewzeoeo}):
\begin{eqnarray}                                                            
&&           \hspace{-15mm}                                                               
\frac{\partial f_{odd}}{\partial t} = l_\uparrow f_{odd} + l_\downarrow f_{even} \; .
\label{eq:110} 
\end{eqnarray}
Thus and by (\ref{eq:new2.44vcdn77}), (\ref{eq:new2.44vcdn79c}), (\ref{eq:new2.44vcdn79d})
the full spin-orbit Fokker-Planck equation, (\ref{eq:109new}),
is equivalent to the linear PDE system, (\ref{eq:109}), (\ref{eq:110}), for $f_{even}$ 
and $f_{odd}$.

In the remaining parts of this section we make remarks on 
(\ref{eq:109new})-(\ref{eq:110}). First of all, 
$l_\downarrow$ is responsible for the coupling between
$f_{even}$ and $f_{odd}$ in (\ref{eq:109}), (\ref{eq:110}).
Note also, by (\ref{eq:new2.44h}), (\ref{eq:new2.99d}), (\ref{eq:new02.010new26}), 
(\ref{eq:new2.44vanewzeoeo}), that $l_\downarrow$ is of order $\hbar$.
Secondly, the PDE system, (\ref{eq:109}), (\ref{eq:110}), opens 
up an alternative to (\ref{eq:new2.44vanew}), (\ref{eq:new2.44vanewz}),  
(\ref{eq:new2.44vanewnew}), (\ref{eq:new2.44vvanew}),  
(\ref{eq:new2.44vvanewfinal}),(\ref{eq:109new}) for studying 
the full spin-orbit Fokker-Planck equation
(this alternative will be pursued in future work), e.g., via
a perturbative treament of $f_{even},f_{odd}$ w.r.t. the perturbation parameter, $\hbar$.
Thirdly, if $f$ satisfies the full spin-orbit Fokker-Planck equation then, due to  
(\ref{eq:109}), (\ref{eq:110}), in general
neither $f_{even}$ nor $f_{odd}$ do.
Fourthly, in the reduced setup, (\ref{eq:new02.010new30a}) resp. (\ref{eq:new02.010new30b}), (\ref{eq:new02.010new30c}),
we observe, by (\ref{eq:new2.44vanewzeoeo}), that 
$l_\downarrow$ vanishes
and so, by (\ref{eq:109new}), (\ref{eq:109}), (\ref{eq:110})
and if $f$ satisfies the reduced spin-orbit Fokker-Planck equation,
then $f_{even},f_{odd}$ do, too
(this case will be discussed in Section 14). Fifthly, if $f$ is a spin-orbit density of the
reduced setup then $f_{even}$ is a spin-orbit density, too (we leave the proof to the reader).
In contrast, by the discussion after (\ref{eq:new2.44vcdn78}), $f_{odd}$ is never a spin-orbit density (even in the
reduced setup).

\section{Further general properties of the reduced spin-orbit Fokker-Planck equation}

The properties discussed Section 13 hold in particular in the reduced setup.
In this section we focus entirely on the reduced setup by studying
further properties valid for this setup (the extension of this study to the full setup is left to future work).
Thus the spin-orbit Fokker-Planck equation underlying this section is the
reduced spin-orbit Fokker-Planck equation.

We know from Section 12 that if $f$ is a spin-orbit density then $W_f$
is the spin-$1/2$ Wigner function of a bunch so one may wonder if the converse is also true.
In other words, the following question, which is the key question underlying this section,
will now be addressed: If $W$ is the spin-$1/2$ Wigner function of a bunch, does  
a spin-orbit density, $f$, exist such that:
\begin{eqnarray}                                                                
&& W = W_f \; .
\label{eq:new2.44vcdn80newnew}   
\end{eqnarray}
Perhaps surprisingly, the answer is: yes!
In fact, motivated by (\ref{eq:new2.44vcdn77}), we define the function, $f$, by
\begin{eqnarray}                                                                
&& \hspace{-10mm} f(t,q,\vec{s}):=(2\pi)^{-3/2}
\exp(-\frac{1}{2} s_i s_i) \biggl( \rho[W](t,q)
+ s_j {\cal P}_j[W](t,q)\biggr) \; ,
\label{eq:new2.44vcdn81new}   
\end{eqnarray}
and observe that if $W$ is the
spin-$1/2$ Wigner function of a bunch then
$f$ in (\ref{eq:new2.44vcdn81new}) satisfies
(\ref{eq:new2.44vcdn80newnew}) and is a spin-orbit density (in particular $f$ satisfies the reduced
spin-orbit Fokker-Planck equation, (\ref{eq:new2.44vanewnewredfinal})).
To prove that $f$ in (\ref{eq:new2.44vcdn81new}) is a spin-orbit density one may use, from Section 5, the
statistical conditions for $W$ and the
fact that $\rho[W]$ satisfies the orbital Fokker-Planck equation
and that $\vec{\cal P}[W]$ satisfies the reduced Bloch equation.
To prove that $f$ in (\ref{eq:new2.44vcdn81new}) satisfies 
(\ref{eq:new2.44vcdn80newnew})
one may insert (\ref{eq:new2.44vcdn81new}) into (\ref{eq:new2.44r})
and then show, by (\ref{eq:new02.10}), that 
$W=W_f$. The details we leave to the reader.

Having settled with (\ref{eq:new2.44vcdn81new})
the existence problem of $f$ we now discuss the nonuniqueness
of $f$ which was already alleged in Section 7. 
In other words the following question will now be addressed:
If $W$ is the spin-$1/2$ Wigner function of a bunch, does more than one
spin-orbit density, $f$, exists which satisfies (\ref{eq:new2.44vcdn80newnew})?
Unsurprisingly, the answer is: yes!
In fact, we define for an arbitrary positive number, $\eta$, the function, $f$, by
\begin{eqnarray}                                                                
&& \hspace{-10mm} f(t,q,\vec{s}):=\frac{1}{\eta^5}
(2\pi)^{-3/2}
\exp(-\frac{1}{2\eta^2} s_i s_i) \biggl( \eta^2\rho[W](t,q)
+ s_j {\cal P}_j[W](t,q)\biggr) \; ,
\label{eq:new2.44vcdn81newnew}   
\end{eqnarray}
and observe that
if $W$ is the
spin-$1/2$ Wigner function of a bunch then
each $f$ in (\ref{eq:new2.44vcdn81newnew}) satisfies, 
(\ref{eq:new2.44vcdn80newnew}), and is a spin-orbit density.
We leave the proof, which is almost the same as for (\ref{eq:new2.44vcdn81new}), to the reader.
Of course, the functions in
(\ref{eq:new2.44vcdn81newnew}) are infinitely many and pairwise distinct and so the nonuniqueness
problem has been solved.

In the remaining parts of this section we make some remarks on 
(\ref{eq:new2.44vcdn80newnew}), (\ref{eq:new2.44vcdn81new}),
(\ref{eq:new2.44vcdn81newnew}).
First of all, $f$ in (\ref{eq:new2.44vcdn81newnew})
is equal to $f$ in (\ref{eq:new2.44vcdn81new})
iff $\eta=1$.
Secondly, if $W$ is the spin-$1/2$ Wigner function of a bunch
then, since every spin-orbit density $f$ in (\ref{eq:new2.44vcdn81newnew}) satisfies
(\ref{eq:new2.44vcdn80newnew}), we can write every such function as
\begin{eqnarray}                                                                
&& \hspace{-10mm} f(t,q,\vec{s})=\frac{1}{\eta^5}
(2\pi)^{-3/2}
\exp(-\frac{1}{2\eta^2} s_i s_i) \biggl( \eta^2\rho[W_f](t,q)
+ s_j {\cal P}_j[W_f](t,q)\biggr) \; ,
\label{eq:new2.44vcdn83new}   
\end{eqnarray}
where, as in (\ref{eq:new2.44vcdn81newnew}),
$\eta$ is an arbitrary positive number.
Thirdly, by (\ref{eq:new2.44vcdn77}), (\ref{eq:new2.44vcdn79c}), (\ref{eq:new2.44vcdn79d}),  
(\ref{eq:new2.44vcdn80newnew}) and for $f$ in (\ref{eq:new2.44vcdn81newnew}), 
\begin{eqnarray}                                                                
&& \hspace{-10mm} f_{even}(t,q,\vec{s})
=\frac{1}{\eta^3}
(2\pi)^{-3/2}\exp(-\frac{1}{2\eta^2} s_i s_i) \rho[W](t,q) \; ,
\label{eq:new2.44vcdn82new}\\    
&& \hspace{-10mm} f_{odd}(t,q,\vec{s})
=\frac{1}{\eta^5}
(2\pi)^{-3/2}\exp(-\frac{1}{2\eta^2} s_i s_i) s_j{\cal P}_j[W](t,q) \; .
\label{eq:new2.44vcdn82newnew}   
\end{eqnarray}
Fourthly, it follows from the discussions after (\ref{eq:110})
that $f_{odd},f_{odd}$ in 
(\ref{eq:new2.44vcdn82new}), (\ref{eq:new2.44vcdn82newnew}) satisfy 
the reduced spin-orbit Fokker-Planck equation, (\ref{eq:new2.44vanewnewredfinal}),
and that $f_{even}$ is a spin-orbit density while $f_{odd}$ is not.
In fact all this can be checked directly in terms of
(\ref{eq:new2.44vcdn82new}), (\ref{eq:new2.44vcdn82newnew}) (the details we leave to the reader).
Fifthly, let $W$ be the spin-$1/2$ Wigner function of a bunch.
The point to be made is that $f$ in (\ref{eq:new2.44vcdn81newnew}) has some  negative values 
whenever $\vec{\cal P}[W]$ has some nonzero values. In other words, all values
of $f$ in (\ref{eq:new2.44vcdn81newnew})
are nonnegative only in the trivial case where $\vec{\cal P}[W]=\vec{0}$.
Sixthly, we point out that the spin-orbit densities of the form (\ref{eq:new2.44vcdn81newnew})
are not the only spin-orbit densities which satisfy (\ref{eq:new2.44vcdn80newnew}).
In fact, we define for an arbitrary positive number, $\eta$, the function, $f$, by
\begin{eqnarray}                                                                
&& \hspace{-25mm} f(t,q,\vec{s}):=\frac{1}{4\pi\eta^4}
\delta( |\vec{s}|-\eta)
\biggl( \eta^2\rho[W](t,q)
+ 3 s_j {\cal P}_j[W](t,q)\biggr) \; ,
\label{eq:new2.44vcdn81newnewnew}   
\end{eqnarray}
where $\delta$ is Dirac's delta function. We observe that
if $W$ is the
spin-$1/2$ Wigner function of a bunch then
each $f$ in (\ref{eq:new2.44vcdn81newnewnew}) satisfies
(\ref{eq:new2.44vcdn80newnew}) and is a spin-orbit density. 
We leave the proof, which is almost the same as for
(\ref{eq:new2.44vcdn81new}), to the reader.
Note that the factor $\delta( |\vec{s}|-\eta)$ in (\ref{eq:new2.44vcdn81newnewnew})
reflects the fact that, in the reduced setup, $|\vec{S}(t)|$ is
conserved in time (this follows from (\ref{eq:new2.44vbnewnewred})).
Seventhly, we point out, without proof,   
that even the spin-orbit densities of the forms, (\ref{eq:new2.44vcdn81newnew}),
and, (\ref{eq:new2.44vcdn81newnewnew}),   
are not the only spin-orbit densities which satisfy (\ref{eq:new2.44vcdn80newnew}).

\section{Deriving the Baier-Katkov-Strakhovenko equation 
from the full spin-orbit Fokker-Planck equation}

In this section we derive
from the full spin-orbit Fokker-Planck equation the BKS equation
(BKS=Baier-Katkov-Strakhovenko), the latter having been
introduced in \cite{Bai69} and \cite{BKS70}.

The motivation underlying this section is at least threefold. First of all, 
the BKS equation will help us, in Section 16, to 
identify the physical meaning of terms of the full spin-orbit Fokker-Planck equation, of
the full spin-orbit SDE system and of the full Bloch equation. Secondly, 
we obtain new insights into the BKS equation, e.g., by tying it with an SDE system,  
namely (\ref{eq:new2.44vcdq}), and with
a Fokker-Planck
equation, namely (\ref{eq:new2.44vcnewnew}). 
Thirdly, our derivation of the
BKS equation leads to a generalization of the BKS equation  
(see Section 15.2).

\subsection{The derivation}

To derive the BKS equation from the full spin-orbit Fokker-Planck equation
we first have to simplify, motivated by \cite{Bai69} and \cite{BKS70}, the full spin-orbit Fokker-Planck equation to 
(\ref{eq:new2.44vcnewnew}) since, as we will see, the BKS equation does not
hold in the full setup.
Thus in this subsection we simplify the full setup
to what we call the BKS setup
which is characterized by neglecting all orbital radiative terms, i.e., the BKS setup is defined by
the conditions
\begin{eqnarray}                                                            
&&  {\cal D}_{orb}={\cal D}_{orb,nrad} \; , \quad  {\cal B}_{orb}=0 \; ,
\label{eq:new02.010anew}                                                           
\end{eqnarray}
where ${\cal D}_{orb,nrad}$ is defined by (\ref{eq:new2.44anew}). 
We can now phrase the task of this section more precisely: Derive the BKS equation
from the spin-orbit Fokker-Planck equation in the BKS setup!

Before we begin with the derivation of the BKS equation we make some remarks on the
BKS setup. First of all,
in the BKS setup the orbital Fokker-Planck equation simplifies to 
(\ref{eq:new02.010new13}) in particular we have
$\vec{\cal C}=\vec{\cal Q}=\vec{0}$.
Secondly, in the BKS setup, the full spin-orbit SDE system, (\ref{eq:new2.44vbnewnew}), 
simplifies to:
\begin{eqnarray}                                                            
&& \hspace{-20mm}  \left( \begin{array}{c}  Q \\ \vec{S} \end{array}\right)'
=
\left( \begin{array}{c}  {\cal D}_{orb,nrad}(t,Q) \\
\vec{\cal D}_{spin}(t,Q,\vec{S})\end{array}\right) 
+  \left( \begin{array}{c}  0 \\
\vec{\cal B}_{spin}(t,Q)\end{array}\right) \nu(t)\; .
\label{eq:new2.44vcdq} 
\end{eqnarray}
Thirdly, in the BKS setup the radiative spin terms, unlike the
radiative orbital terms, have not disappeared
(these terms are carried
in (\ref{eq:new2.44vcdq}), due to Section 4, by
$\vec{\cal B}_{spin}$ and $\vec{\cal D}_{spin}$!).
Fourthly, in the BKS setup the full spin-orbit Fokker-Planck equation (\ref{eq:new2.44vanew}) 
simplifies, by (\ref{eq:new02.010anew}), to
\begin{eqnarray}                                                            
&&  \hspace{-10mm} 
\frac{\partial f}{\partial t} 
= -\frac{\partial}{\partial q_j} ({\cal D}_{orb,nonrad,j}f)
-\frac{\partial}{\partial s_j} \biggl( ({\cal D}_{spin,0,ST,j} + {\cal D}_{spin,+,j,k}s_k)f\biggr)
\nonumber\\
&& +\frac{1}{2}
\frac{\partial}{\partial s_k}\frac{\partial}{\partial s_l}
({\cal B}_{spin,l} {\cal B}_{spin,k}f)\; ,
\label{eq:new2.44vcnewnew}  
\end{eqnarray}
where we also used (\ref{eq:new2.44vanewnewnew}).  
Unsurprisingly, (\ref{eq:new2.44vcdq}) is an Ito SDE system associated with 
(\ref{eq:new2.44vcnewnew}).
Fifthly, no additional approximation beyond the BKS setup is required
in this subsection, i.e., we here deal with exact solutions of   
(\ref{eq:new2.44vcnewnew}).
We call $f$ a spin-orbit density of the BKS setup iff it satisfies, (\ref{eq:new2.44vcnewnew}),
and the statistical conditions, (\ref{eq:new02.10eso}),
(\ref{eq:new02.10dso}), (\ref{eq:new02.10fso}).
Sixthly, to keep things simple we stick in this subsection to the belief, stated in Section 12,
that the statistical condition, (\ref{eq:new02.10fso}), 
on $f$ fits to the dynamics of $f$ (see also the explanation, made after 
after (\ref{eq:new2.44vcdrnew}), why this belief is
not necessary for the derivation of the BKS equation).
We recall from Section 12 that this belief can be phrased as follows: 
If $f$ satisfies (\ref{eq:new02.10fso0}) then it satisfies (\ref{eq:new02.10fso}) for all $t$.
Seventhly, the orbital Fokker-Planck equation, (\ref{eq:new02.010f}), 
and the full Bloch equation, (\ref{eq:new2.44vcdc0}), simplify, by 
(\ref{eq:new2.44a}), (\ref{eq:new02.010a}), (\ref{eq:new02.010anew}),  
in the BKS setup to
\begin{eqnarray}                                                            
&& \hspace{-10mm} \frac{\partial \rho[W_f]}{\partial t}  
= - \frac{\partial}{\partial q_j} ({\cal D}_{orb,nrad,j}\rho[W_f])\; ,
\label{eq:new2.44vcdknewnew}  \\   
&&        \hspace{-10mm}                                                                
\frac{\partial \vec{\cal P}[W_f]}{\partial t} 
= - \frac{\partial}{\partial q_j} ({\cal D}_{orb,nrad,j}\vec{\cal P}[W_f])
+{\cal D}_{spin,+}\vec{\cal P}[W_f]
+\vec{\cal D}_{spin,0,ST}\rho[W_f]  \; .
\label{eq:new02.010cnew}  
\end{eqnarray}

After these remarks on the BKS setup
we now start the derivation of the BKS equation.
The key point is to show that the
BKS equation is an ODE for the polarization vector, 
$\vec{P}[W_f]$, if one properly restricts the generality of $f$ (in fact, $f$ will eventually 
be restricted via (\ref{eq:new2.44vcdn76}) where the function $h$ has to satisfy certain conditions).
To come to this we first 
compute, by (\ref{eq:new2.44vcdn90orbnew}), (\ref{eq:new02.010cnew}) and 
for an arbitrary spin-orbit density $f$ of the BKS setup,
\begin{eqnarray}                                                            
&& \hspace{-20mm}
\frac{d}{dt} \vec{P}[W_f](t) 
=\frac{d}{dt}\int_{\R^6} \; \vec{\cal P}[W_f](t,q) d^6q
= \int_{\R^6} \; \frac{\partial \vec{\cal P}[W_f]}{\partial t}(t,q)d^6q
\nonumber\\
&& \hspace{-15mm} = \int_{\R^6} \; \Biggl(
- \frac{\partial}{\partial q_j} \biggl({\cal D}_{orb,nrad,j}(t,q)\vec{\cal P}[W_f]\biggr)
\nonumber\\
&& +{\cal D}_{spin,+}(t,q)\vec{\cal P}[W_f](t,q) 
+\vec{\cal D}_{spin,0,ST}(t,q)\rho[W_f](t,q) \Biggr) d^6q
\nonumber\\
&& \hspace{-10mm} = \int_{\R^6} \; \Biggl(  {\cal D}_{spin,+}(t,q)\vec{\cal P}[W_f](t,q) 
+\vec{\cal D}_{spin,0,ST}(t,q)\rho[W_f](t,q) \Biggr) d^6q \; ,
\nonumber
\end{eqnarray}
in short,
\begin{eqnarray}                                                            
&& \hspace{-10mm}
\frac{d}{dt} \vec{P}[W_f](t) = \int_{\R^6} \; \Biggl(  {\cal D}_{spin,+}(t,q)\vec{\cal P}[W_f](t,q) 
+\vec{\cal D}_{spin,0,ST}(t,q)\rho[W_f](t,q) \Biggr) d^6q \; .
\label{eq:new2.44vcdrnewnew} 
\end{eqnarray}
Although $f$ 
in (\ref{eq:new2.44vcdrnewnew}) is a spin-orbit density of
the BKS setup (i.e., $f$ carries the `right' dynamics and the `right' statistics)
it is too general.
In fact for such a general $f$ the evolution equation
(\ref{eq:new2.44vcdrnewnew}) is not an ODE for $\vec{P}[W_f]$, let alone
the BKS equation for $\vec{P}[W_f]$.

To properly restrict the generality of 
$f$ in (\ref{eq:new2.44vcdrnewnew}) we have to recall
from \cite{Bai69} or \cite{BKS70} that the BKS equation is defined w.r.t. deterministic orbital motion.
Thus in order for (\ref{eq:new2.44vcdrnewnew}) to become an ODE for $\vec{P}[W_f]$,
not only $f$ has to be a spin-orbit density of the BKS setup but
$f$ also has to encapsulate deterministic orbital motion.
To define the latter notion we note that the orbital part, 
\begin{eqnarray}                                                            
&&   Q'  ={\cal D}_{orb,nrad}(t,Q) \; ,
\label{eq:new2.44vcdqnewnew} 
\end{eqnarray}
of the Ito SDE system, (\ref{eq:new2.44vcdq}),
is an ODE system and so it allows for deterministic orbital motion
(note, by (\ref{eq:new2.44vcdqnewnew}), that this orbital motion is nonradiative).
In fact an arbitrary deterministic solution,
$Q_0\equiv \left( \begin{array}{c} 
\vec{R}_0 \\ \vec{\Pi}_0 \end{array}\right)$, of 
(\ref{eq:new2.44vcdqnewnew}) satisfies
\begin{eqnarray}  
&&   Q_0'  ={\cal D}_{orb,nrad}(t,Q_0) \; ,
\label{eq:new2.44vcdqnewnewxx}
\end{eqnarray}
with $Q_0(0)\in\R^6$.
We can now state what it means for $f$ to encapsulate $Q_0$, i.e.,
to encapsulate deterministic orbital motion.
In fact it means that $f$ reads as
\begin{eqnarray}                                                            
&& f(t,q,\vec{s})
= \delta\biggl(q-Q_0(t)\biggr)h(t,\vec{s}) \; ,
\label{eq:new2.44vcdn76}                                                           
\end{eqnarray}
where the function $h$ will be discussed below.
To ensure that $f$ in (\ref{eq:new2.44vcdn76}) is a spin-orbit density we first have to ensure that it is
a solution of (\ref{eq:new2.44vcnewnew}). In fact
the ansatz, (\ref{eq:new2.44vcdn76}), is consistent with 
(\ref{eq:new2.44vcnewnew}) since, when inserting (\ref{eq:new2.44vcdn76}) into 
(\ref{eq:new2.44vcnewnew}), we get
\begin{eqnarray}                                                            
&&  \hspace{-10mm} 
\frac{\partial h}{\partial t} 
= -\frac{\partial}{\partial s_j} \Biggl( \biggl( {\cal D}_{spin,+,j,k}(t,Q_0(t))s_k
+{\cal D}_{spin,0,ST,j}(t,Q_0(t))\biggr)h\Biggr)
\nonumber\\
&& +\frac{1}{2}
\frac{\partial}{\partial s_k}\frac{\partial}{\partial s_l}\biggl(
{\cal B}_{spin,l}(t,Q_0(t)) {\cal B}_{spin,k}(t,Q_0(t))h\biggr)\; .
\label{eq:new2.44vddknew}  
\end{eqnarray}
We denote the initial condition by $h_0$, i.e.,
\begin{eqnarray}                                                            
&& h_0(\vec{s}):= h(0,\vec{s}) \; .
\label{eq:new2.44vcdn11orbnew}
\end{eqnarray}
Since (\ref{eq:new2.44vddknew}) is a Fokker-Planck equation
for $h$ we note that
$h$ exists and is uniquely determined
by $h_0$ via (\ref{eq:new2.44vddknew}), (\ref{eq:new2.44vcdn11orbnew}) (if the coefficient
functions of (\ref{eq:new2.44vddknew}) are sufficiently regular).
As an aside we note
that (\ref{eq:new2.44vddknew}) has solutions which are Gaussian in $\vec{s}$
(we leave the proof to the reader).
To make $f$ in (\ref{eq:new2.44vcdn76}) a spin-orbit density 
we have to impose the statistical conditions, (\ref{eq:new02.10eso}), (\ref{eq:new02.10dso}), (\ref{eq:new02.10fso}), which, by (\ref{eq:new2.44vcdn76}), are equivalent to
\begin{eqnarray}                                                            
&&   h(t,\vec{s})\in\R \; ,
\label{eq:new02.10esobks} \\ 
&&   \int_{\R^9} \; h(t,\vec{\tau})d^3\tau  = 1 \; , 
\label{eq:new02.10dsobks} \\                                                
&& \big{|} \int_{\R^9} \; \vec{\tau} h(t,\vec{\tau})d^3\tau\big{|} \leq 1  \; .
\label{eq:new02.10fsobks}                                                                
\end{eqnarray}
Note that since (\ref{eq:new2.44vddknew}) is a Fokker-Planck equation, we can fulfill
(\ref{eq:new02.10esobks}), (\ref{eq:new02.10dsobks}) by imposing on $h_0$:
\begin{eqnarray}                                                            
&&   h_0(\vec{s})\in\R \; ,
\label{eq:new02.10eso0bks} \\
&& \int_{\R^3} \; h_0(\vec{\tau}) d^3\tau = 1 \; .
\label{eq:new2.44vcdn11orb}
\end{eqnarray}
Compare this argumentation with the discussion made before (\ref{eq:new02.10eso0}) for the full setup!
To fulfill (\ref{eq:new02.10fsobks}) we follow
the discussion after (\ref{eq:new2.44vcnewnew}) and thus
we assume that the statistical conditions fit to the dynamics. In other words we assume
that (\ref{eq:new02.10fsobks}) holds whenever
\begin{eqnarray}                                                            
&& \big{|}\int_{\R^3} \; \vec{s}h_0(\vec{s}) d^3s \big{|} \leq 1 \; ,
\label{eq:new2.44vcdn13orb}
\end{eqnarray}
where we also used (\ref{eq:new2.44vcdn11orbnew}).
In summary we have found those 
spin-orbit densities $f$ of the BKS setup which encapsulate $Q_0$, i.e., which encapsulate
the deterministic orbital motion: They are those $f$ which are given by
(\ref{eq:new2.44vcdn76}) where $h$ satisfies (\ref{eq:new2.44vddknew})
and where $h_0$ satisfies 
(\ref{eq:new02.10eso0bks}), (\ref{eq:new2.44vcdn11orb}), (\ref{eq:new2.44vcdn13orb})
with $h_0$ being given by (\ref{eq:new2.44vcdn11orbnew}) (for comments how to avoid 
assuming that (\ref{eq:new2.44vcdn13orb}) implies (\ref{eq:new02.10fsobks}), see the discussion after
(\ref{eq:new2.44vcdrnew})).

Before we complete the derivation of the BKS equation we make some comments on 
$\rho[W_f]$ and $\vec{P}[W_f]$.
In fact it follows from (\ref{eq:new2.44p}), (\ref{eq:new2.44vcdn76}) that
\begin{eqnarray}                                                            
&& \rho[W_f](t,q) = \delta(q-Q_0(t)) \; ,
\label{eq:new2.44vcdn72000t}                                                       
\end{eqnarray}
which, as an aside, justifies the terminology `encapsulating
the deterministic orbital motion'.
Note, by (\ref{eq:new2.44vcdn72000t}), that
\begin{eqnarray}                                                            
&& \rho[W_f](0,q) = \delta(q-Q_0(0)) \; .
\label{eq:new2.44vcdn72000}                                                       
\end{eqnarray}
That $\rho[W_f]$ in (\ref{eq:new2.44vcdn72000t})
solves (\ref{eq:new2.44vcdknewnew}) can be shown either directly and
by using (\ref{eq:new2.44vcdn72000}) or by using (\ref{eq:new2.44vcdn72000}) 
and the method of characteristics (note that 
(\ref{eq:new2.44vcdknewnew}) is a first-order PDE!).
Moreover it follows from (\ref{eq:new2.44q}), (\ref{eq:new2.44vcdn76}), that
\begin{eqnarray}  
&& \hspace{-10mm}  
\vec{\cal P}[W_f](t,q) =\delta\biggl(q-Q_0(t)\biggr)
\int_{\R^3} \; \vec{s} h(t,\vec{s})d^3s \; ,
\label{eq:new2.44vcdn86} 
\end{eqnarray}
and so, by (\ref{eq:new2.44vcdn90orbnew}),
$\vec{P}[W_f](t) =\int_{\R^3} \; \vec{s} h(t,\vec{s})d^3s$
%
%
which, by (\ref{eq:new2.44vcdn86}), leads to
\begin{eqnarray}  
&& \hspace{-27mm}  
\vec{\cal P}[W_f](t,q) =\delta\biggl(q-Q_0(t)\biggr)\vec{P}[W_f](t) \; .
\label{eq:new2.44vcdrnewnewnew} 
\end{eqnarray}

To complete the derivation of the BKS equation we insert (\ref{eq:new2.44vcdn72000t}), 
(\ref{eq:new2.44vcdrnewnewnew}) 
into (\ref{eq:new2.44vcdrnewnew}) resulting in
\begin{eqnarray}                                                            
&& \frac{d}{dt}\vec{P}[W_f](t)  
= \vec{P}[W_f](t)  \int_{\R^6} \; 
{\cal D}_{spin,+}(t,q)\delta\biggl(q-Q_0(t)\biggr)d^6q
\nonumber\\
&& 
+\int_{\R^6} \; 
\vec{\cal D}_{spin,0,ST}(t,q)\delta\biggl(q-Q_0(t)\biggr)d^6q
\nonumber\\
&& =  {\cal D}_{spin,+}(t,Q_0(t)) \vec{P}[W_f](t)  
+\vec{\cal D}_{spin,0,ST}(t,Q_0(t)) \; ,
\nonumber
\end{eqnarray}
in short,
\begin{eqnarray}                                                            
&& \hspace{-28mm} \frac{d}{dt}\vec{P}[W_f](t)  
=  {\cal D}_{spin,+}(t,Q_0(t)) \vec{P}[W_f](t)  
+\vec{\cal D}_{spin,0,ST}(t,Q_0(t))  \; ,
\label{eq:new2.44vcen70new}                                                           
\end{eqnarray}
which indeed is an ODE for $\vec{P}[W_f]$.
Using (\ref{eq:new02.010new27}) 
we can write (\ref{eq:new2.44vcen70new}) more explicitly as
\begin{eqnarray}  
&& \hspace{-10mm}  \frac{d}{dt}\vec{P}[W_f](t) =  
{\cal D}_{spin,+,TBMT}\biggl(t,Q_0(t)\biggr)\vec{P}[W_f](t)
+{\cal D}_{spin,+,ST}\biggl(t,Q_0(t)\biggr)\vec{P}[W_f](t)
\nonumber\\
&&+{\cal D}_{spin,+,BK}\biggl(t,Q_0(t)\biggr)\vec{P}[W_f](t)
+\vec{\cal D}_{spin,0,ST}\biggl(t,Q_0(t)\biggr) 
\; ,
\label{eq:new2.44vcdrnewfinal} 
\end{eqnarray}
and using (\ref{eq:new2.99d}),  
(\ref{eq:new02.010new26}), (\ref{eq:new02.010new26TBMT}), (\ref{eq:new02.010new26ST}),
(\ref{eq:new02.010new26BK})
we can write (\ref{eq:new2.44vcdrnewfinal}) more explicitly as
\begin{eqnarray}  
&& \hspace{-10mm}  \frac{d}{dt}\vec{P}[W_f](t) =  
\vec{\Omega}_{TBMT}\biggl(t,Q_0(t)\biggr)\times \vec{P}[W_f](t)
\nonumber\\
&&  \hspace{-5mm} 
-\frac{5\sqrt{3}}{8}\lambda\biggl(t,Q_0(t)\biggr)\vec{P}[W_f](t)
+ \frac{5\sqrt{3}}{36}\lambda\biggl(t,Q_0(t)\biggr)
\vec{\Pi}_0(t)\vec{\Pi}_0^T(t)\vec{P}[W_f](t)
\nonumber\\
&&  \hspace{-5mm} 
-\frac{1}{m\gamma(Q_0(t))}\lambda(t,Q_0(t))
\frac{\vec{\Pi}_0(t)\times\vec{a}(t,Q_0(t))}
{|\vec{a}(t,Q_0(t))|}  \; .
\label{eq:new2.44vcdrnew} 
\end{eqnarray}
Comparing with \cite{BKS70} or with
\cite{Bai69} we observe that (\ref{eq:new2.44vcdrnew}) (and thus:
(\ref{eq:new2.44vcen70new}), (\ref{eq:new2.44vcdrnewfinal})) is the BKS equation.
It follows from \cite{Bai69,BKS70} that the first term on the rhs of 
(\ref{eq:new2.44vcdrnew}) and of (\ref{eq:new2.44vcdrnewfinal})
encapsulates the T-BMT precession effect and that
the second and fourth terms  on the rhs of 
(\ref{eq:new2.44vcdrnew}) and of (\ref{eq:new2.44vcdrnewfinal})
encapsulate the Sokolov-Ternov effect
while the third term  on the rhs of 
(\ref{eq:new2.44vcdrnew}) and of (\ref{eq:new2.44vcdrnewfinal})
encapsulates the Baier-Katkov correction.
These remarks will be used in Section 16 and they justify,
in combination with (\ref{eq:new02.010new26}), (\ref{eq:new02.010new26TBMT}), (\ref{eq:new02.010new26ST}),
(\ref{eq:new02.010new26BK}), the notations
${\cal D}_{spin,+,TBMT}, {\cal D}_{spin,+,ST},{\cal D}_{spin,+,BK},\vec{\cal D}_{spin,0,ST}$.
This completes our derivation of the BKS equation.

We now make some remarks on the above derivation of the BKS equation.
First of all, the only contentious point of this derivation is the assumption that
(\ref{eq:new2.44vcdn13orb}) implies (\ref{eq:new02.10fsobks}).
However close inspection of the derivation shows that this assumption is redundant since
(\ref{eq:new02.10fsobks}) has no effect on obtaining the BKS equation
(thus, as promised, we indeed derived the BKS equation from the full 
spin-orbit Fokker-Planck equation). Note that
the reason for assuming that
(\ref{eq:new2.44vcdn13orb}) implies (\ref{eq:new02.10fsobks}) was merely pedagogical, namely to
ensure that $f$ is a spin-orbit density.


Secondly, inspection of
our above derivation of the BKS equation, shows that the fourth term on the rhs of the Fokker-Planck
equation (\ref{eq:new2.44vcnewnew}) has no
impact on the derivation of the BKS equation. 
The fourth term on the rhs of 
(\ref{eq:new2.44vcnewnew}) corresponds to the spin white-noise term,
$\vec{\cal B}_{spin}(t,Q)\nu(t)$, on the rhs of (\ref{eq:new2.44vcdq}) (since (\ref{eq:new2.44vcdq})
is an Ito SDE system associated with (\ref{eq:new2.44vcnewnew})) and 
so this term has no
impact on the derivation of the BKS equation
(thus the BKS equation is independent of both
white-noise terms in (\ref{eq:new2.44vcdmfinal})). In other words, if one would have neglected 
the spin white-noise term in (\ref{eq:new2.44vcdq})
one would still have arrived at the BKS equation. For more details on the spin white-noise term
(in the full setup), see Section 16.

Thirdly, although both white-noise terms
in (\ref{eq:new2.44vcdmfinal}) have no effect on the
BKS equation we will see in Section 16 that these
terms are related to the radiative depolarization effect
and the kinetic polarization effect. Thus both of these
effects elude the BKS equation (and elude the
generalized BKS equation as well). 
In fact the inclusion of these two effects is the key achievement
of the non-kinetic and kinetic approaches. Nevertheless the
BKS equation is the most important piece in the prehistory
of both approaches.

\subsection{The generalized Baier-Katkov-Strakhovenko equation}

Inspection of the derivation of the BKS equation in Section 15.1 
reveals that the full spin-orbit Fokker-Planck equation
implies even a bit more than the BKS equation namely
an equation which for lack of a better word we call
the generalized BKS equation which, by definition,
is the same as each of (\ref{eq:new2.44vcen70new}), (\ref{eq:new2.44vcdrnewfinal}), (\ref{eq:new2.44vcdrnew}) 
except that $Q_0$ is now replaced by $Q_1$ which satisfies the ODE 
\begin{eqnarray}  
&&   Q_1'  ={\cal D}_{orb}(t,Q_1) \; ,
\label{eq:new2.44vcdqnewnewx} 
\end{eqnarray}
and which replaces the ODE, (\ref{eq:new2.44vcdqnewnewxx}) (note that $Q_1(0)\in\R^6$).
More precisely, in case of the generalized BKS equation, the BKS setup, (\ref{eq:new02.010anew}),
is replaced by the generalized BKS setup which is 
defined by: ${\cal B}_{orb}=0$.
Thus in the generalized BKS setup (\ref{eq:new2.44vcdq}) is replaced by 
\begin{eqnarray}                                                            
&& \hspace{-20mm}  \left( \begin{array}{c}  Q \\ \vec{S} \end{array}\right)'
=
\left( \begin{array}{c}  {\cal D}_{orb}(t,Q) \\
\vec{\cal D}_{spin}(t,Q,\vec{S})\end{array}\right) 
+  \left( \begin{array}{c}  0 \\
\vec{\cal B}_{spin}(t,Q)\end{array}\right) \nu(t)\; ,
\label{eq:new2.44vcdqbksg} 
\end{eqnarray}
which, in turn, leads to (\ref{eq:new2.44vcdqnewnewx}).  
Note that $\vec{\cal C},\vec{\cal Q}$ are not neglected
in the generalized BKS setup, i.e., the generalized BKS equation takes into account
the radiation-force field $\vec{\cal C} +\vec{\cal Q}$. 
Since $\vec{\cal C} +\vec{\cal Q}\neq\vec{0}$ we have, by 
(\ref{eq:new2.44anew}), (\ref{eq:new2.44anewnew}), (\ref{eq:new2.44a}), 
$\frac{\partial {\cal D}_{orb,j}}{\partial q_j}\neq 0$ (in contrast, in the situation of Section 15.1, we had $\frac{\partial {\cal D}_{orb,nrad,j}}{\partial q_j}= 0$).
Thus, at face value, the validity of the generalized BKS equation may seem questionable (note
that $\frac{\partial {\cal D}_{orb,j}}{\partial q_j}$ plays a key role if one applies the
method of characteristics to the orbital Fokker-Planck equation in the generalized BKS setup!).
Therefore it may come as a surprise that in the generalized BKS setup, (\ref{eq:new2.44vcdn72000t}),
is merely to be replaced by:                
\begin{eqnarray}                                                            
&& \rho[W_f](t,q) = \delta(q-Q_1(t)) \; ,
\label{eq:new2.44vcdn72000tg}                                                       
\end{eqnarray}
which can be shown for example by the method of characteristics (even in the
generalized BKS setup the orbital Fokker-Planck equation is a first-order PDE!).
We leave the details of deriving the generalized BKS equation from the
full spin-orbit Fokker-Planck equation to the reader (in fact this proof is, thanks to 
(\ref{eq:new2.44vcdn72000tg}), almost identical to 
the proof in Section 15.1 of the BKS equation!).
Thus many of our above remarks on the 
BKS equation also apply to the generalized BKS equation 
(for example the fact that both white-noise terms
in the  full spin-orbit SDE system, (\ref{eq:new2.44vcdmfinal}), have no effect on the
generalized BKS equation).

\section{Interpreting the full spin-orbit Fokker-Planck equation, the full spin-orbit stochastic
ODE system and the full Bloch equation}

This section wraps up our presentation of the kinetic approach. 
In fact in Sections 16.1 and 16.2 we show how each of the main effects on the polarization, mentioned
in Section 1, are tied
to specific terms of the full spin-orbit SDE system,
of the full spin-orbit Fokker-Planck equation, and of the full Bloch equation. In Section 16.3 
we draw conclusions.

To accomplish these tasks we
take advantage of the fact that the full spin-orbit SDE system
is an Ito SDE system associated
with the full spin-orbit Fokker-Planck equation whereby
the terms in the full spin-orbit SDE system
correspond in a well-defined way to 
the terms in the full spin-orbit Fokker-Planck equation.
Moreover we take advantage of how the full Bloch equation was used in
Section 9 to derive the full spin-orbit Fokker-Planck equation.
%
%
%
%

\subsection{Interpreting the full spin-orbit Fokker-Planck equation and the full spin-orbit stochastic
ODE system}

We first recall from the discussion after (\ref{eq:new2.100bf})
that all terms on the rhs of the reduced SDE system, (\ref{eq:new2.44vcdmred}), 
encapsulate
the radiative depolarization effect. 
Moreover the terms on the rhs of (\ref{eq:new2.44vcdmred})  correspond
in the full spin-orbit SDE system, (\ref{eq:new2.44vcdmfinal}),
to the terms on the rhs of the first equation of (\ref{eq:new2.44vcdmfinal}),
and to the second term on the rhs of the second equation of (\ref{eq:new2.44vcdmfinal})
(note that the reduced SDE system is the restriction of the full spin-orbit SDE system to the
reduced setup).
Thus these terms in (\ref{eq:new2.44vcdmfinal})
encapsulate the 
radiative depolarization effect.
Therefore and since (\ref{eq:new2.44vcdmfinal}) is an Ito SDE system associated
with the Fokker-Planck equation, (\ref{eq:new2.44vvanewfinal}),
the terms on the rhs of the first equation in (\ref{eq:new2.44vcdmfinal})
in combination with the second term on the rhs of the second equation in (\ref{eq:new2.44vcdmfinal})
correspond to the first, third and sixth terms on the rhs
of (\ref{eq:new2.44vvanewfinal}). Thus these terms in 
(\ref{eq:new2.44vvanewfinal}) encapsulate the 
radiative depolarization effect.
This completes the account of how the full spin-orbit SDE system
and the full spin-orbit Fokker-Planck equation encapsulate the 
radiative depolarization effect.

We next recall from Section 15.1 that the second and fourth terms on the
rhs of the BKS equation, 
(\ref{eq:new2.44vcdrnewfinal}), encapsulate the Sokolov-Ternov effect.
Clearly these two terms correspond to
the first and third terms on the rhs of the second equation of (\ref{eq:new2.44vcdmfinal}).
Thus these two terms in (\ref{eq:new2.44vcdmfinal})
encapsulate the Sokolov-Ternov effect.
Also since (\ref{eq:new2.44vcdmfinal}) is an Ito SDE system associated
with the Fokker-Planck equation, (\ref{eq:new2.44vvanewfinal}),
the first and third terms on the rhs of the second equation of (\ref{eq:new2.44vcdmfinal})
correspond to the second and fourth terms on the rhs of
(\ref{eq:new2.44vvanewfinal}).
Thus these two terms in (\ref{eq:new2.44vvanewfinal})
encapsulate the Sokolov-Ternov effect. 
This completes the account of how the full spin-orbit SDE system
and the full spin-orbit Fokker-Planck equation encapsulate the Sokolov-Ternov effect.

We now recall from Section 15.1 that the third term on the
rhs of the BKS equation, (\ref{eq:new2.44vcdrnewfinal}), encapsulates the 
Baier-Katkov correction to the Sokolov-Ternov effect.
Clearly this term corresponds to
the fourth term on the rhs of the second equation of (\ref{eq:new2.44vcdmfinal}).
Thus this term in (\ref{eq:new2.44vcdmfinal})
encapsulates the Baier-Katkov correction to the Sokolov-Ternov effect.
Also since (\ref{eq:new2.44vcdmfinal}) is an Ito SDE system associated
with the Fokker-Planck equation, (\ref{eq:new2.44vvanewfinal}),
the fourth term on the rhs of the second equation of (\ref{eq:new2.44vcdmfinal})
corresponds to the fifth term on the rhs of
(\ref{eq:new2.44vvanewfinal}).
Thus this term in (\ref{eq:new2.44vvanewfinal})
encapsulates the Baier-Katkov correction to the Sokolov-Ternov effect. 
This completes the account of how the full spin-orbit SDE system
and the full spin-orbit Fokker-Planck equation encapsulate the Baier-Katkov correction to the
Sokolov-Ternov effect.

We now address the kinetic polarization effect. 
To begin with, we note from \cite{DK75} that the sixth term on the rhs of the full Bloch equation,
(\ref{eq:new2.44vcdc0newnew}),  
is that term which is responsible for
the kinetic polarization part in the Derbenev-Kondratenko formula
for the equilibrium polarization.
Moreover we know from our derivation, in Section 9,
of the full spin-orbit Fokker-Planck equation 
that the sixth term on the rhs of the full Bloch equation,
(\ref{eq:new2.44vcdc0newnew}),  
corresponds to the seventh term
on the rhs of the full spin-orbit Fokker-Planck
equation, (\ref{eq:new2.44vvanewfinal})).
Furthermore since (\ref{eq:new2.44vcdmfinal}) is an Ito SDE system associated
with the Fokker-Planck equation, (\ref{eq:new2.44vvanewfinal}),                                                                        
the seventh term on the rhs of (\ref{eq:new2.44vvanewfinal}) 
corresponds to the white-noise terms in  (\ref{eq:new2.44vcdmfinal}), i.e., corresponds to
the orbital white-noise term ${\cal B}_{orb}(t,Q)\nu(t)$ and to the spin white-noise term
$\vec{\cal B}_{spin}(t,Q)\nu(t)$.
Thus the two white-noise terms in  (\ref{eq:new2.44vcdmfinal}) encapsulate the kinetic polarization effect.
This completes the account of how the full spin-orbit SDE
system and the full spin-orbit Fokker-Planck equation
encapsulate the kinetic polarization effect.

\subsection{Interpreting the full Bloch equation}

Using the derivation of the full spin-orbit Fokker-Planck equation, in Section 9, 
we see by the interpretation of (\ref{eq:new2.44vvanewfinal}) in Section 16.1 
that the first and second terms on the rhs of (\ref{eq:new2.44vcdc0newnew})
encapsulate the 
radiative depolarization effect, that the third and fifth terms on the rhs  of 
(\ref{eq:new2.44vcdc0newnew}) 
encapsulate the
Sokolov-Ternov effect and that the fourth term on the rhs  of (\ref{eq:new2.44vcdc0newnew}) 
encapsulates the Baier-Katkov correction whereas
the sixth term on the rhs  of (\ref{eq:new2.44vcdc0newnew}) 
encapsulates the kinetic polarization effect.

\subsection{Discussing the main physical effects. Special features of the kinetic polarization effect}

First of all, if one would neglect at least one of the two white-noise terms in 
(\ref{eq:new2.44vcdmfinal}) then, by Section 16.1, the seventh term
on the rhs of the full spin-orbit Fokker-Planck
equation, (\ref{eq:new2.44vvanewfinal}), 
would vanish and thus the kinetic polarization effect would disappear. 
Thus the cooperation between the two white-noise terms encapsulates the kinetic polarization effect.
The kinetic polarization effect disappears
for example in the reduced setup, considered in Section 11,
where $\vec{\cal B}_{spin}(t,Q)\nu(t)$ is absent. 
The kinetic polarization effect also disappears
in the BKS setup, considered in Section 15.1,
where ${\cal B}_{orb}(t,Q)\nu(t)$ is absent.

Secondly,  the aforementioned cooperation between ${\cal B}_{orb}(t,Q)\nu(t)$ and
$\vec{\cal B}_{spin}(t,Q)\nu(t)$, which encapsulates the kinetic polarization effect,
may be interpreted as an SDE version of
the interference effect which was credited in \cite{MSY05}
with being the QED mechanism underlying the kinetic polarization effect.
For the QED origin of the kinetic polarization effect, see also
\cite{Mon84}.

Thirdly, while the existence of the term $\vec{\cal B}_{spin}(t,Q)\nu(t)$ is necessary 
for the kinetic polarization effect it follows from Section 16.1 that it impacts neither the
radiative depolarization effect nor the
Sokolov-Ternov effect and its Baier-Katkov correction.
In other words, the only main physical effect which depends on the 
spin white-noise term in (\ref{eq:new2.44vcdmfinal})
is the kinetic polarization effect.

Fourthly, the orbital white-noise term, ${\cal B}_{orb}(t,Q)\nu(t)$, 
in (\ref{eq:new2.44vcdmfinal}) plays a double role: In fact it follows from Section 16.1 that 
it contributes to the kinetic polarization effect and to the radiative depolarization effect.

Fifthly, having just identified in Section 16.1
how the full spin-orbit SDE system
and the full spin-orbit Fokker-Planck equation encapsulate the various physical effects,
we can imagine new roads  of studying these effects, in particular the kinetic polarization effect.
We already mentioned in Section 1.3 one avenue namely numerically solving 
the full spin-orbit SDE system.
Here we mention an analytical avenue
 whereby one obtains, via an appropriate approximation of 
the full spin-orbit Fokker-Planck equation, an ODE system for the polarization vector
$<\vec{S}>_{W_f}(t)$ and for $<Q_j>_{W_f}(t),<Q_j Q_k>_{W_f}(t)$ where $j,k=1,...,6$.
For the definition of $<\cdots>_{W_f}$, see Section 12.
The dependence of this ODE system on the kinetic polarization term,
$\frac{\partial}{\partial s_k}\frac{\partial}{\partial q_l}({\cal B}_{orb,l} {\cal B}_{spin,k}f)$, in 
the full spin-orbit Fokker-Planck equation, (\ref{eq:new2.44vvanewfinal}), leads to
a dependence of its solutions on this term
which can be used to quantitatively studying the kinetic polarization effect.
Many other analytical approximations of (\ref{eq:new2.44vvanewfinal}) (including even
higher moments of $\vec{S}$ and $Q$) can be pursued as well.

Sixthly, not only the kinetic polarization effect results from a
cooperation between terms in (\ref{eq:new2.44vcdmfinal}).
In fact, by Section 16.1,
the Sokolov-Ternov effect is the result of the cooperation between 
the first and third terms on the rhs of the second equation of (\ref{eq:new2.44vcdmfinal}).
However, in contrast to the situation for the kinetic polarization effect, the neglect of one of these two terms
does not delete the Sokolov-Ternov effect but creates a new (and unphysical) effect.
In fact, if one neglects the first term on the rhs of the second equation of (\ref{eq:new2.44vcdmfinal}),
then the Sokolov-Ternov effect mutates to a kind of spin damping effect.
If one neglects the third term on the rhs of the second equation of (\ref{eq:new2.44vcdmfinal}),
then the Sokolov-Ternov effect mutates to a kind of driven undamped spin motion.
Similar remarks can be made on the radiative depolarization effect which is tied
to both terms on the rhs of the first equation of (\ref{eq:new2.44vcdmfinal})
and to the second term on the rhs of the second equation of (\ref{eq:new2.44vcdmfinal}), i.e.,
the cooperation of these three terms creates the 
radiative depolarization effect.  
If one neglects one of these three terms then one has
not deleted the 
radiative depolarization effect but one has created a new (and unphysical) effect (we
leave the details to the reader).
Recall also from Section 16.1 that the Baier-Katkov correction to the Sokolov-Ternov effect is encapsulated
by the fourth term on the rhs of the second equation of (\ref{eq:new2.44vcdmfinal}) and thus it
cooperates with the two aforementioned Sokolov-Ternov terms 
(we leave the details to the reader).

Seventhly, we recall from Sections 16.1 and 16.2
that the interpretation of the terms which
encapsulate the kinetic polarization effect has rested on the observation from
\cite{DK75} that the sixth term on the rhs of the full Bloch equation, (\ref{eq:new2.44vcdc0newnew}),  
is that term of (\ref{eq:new2.44vcdc0newnew}) which is responsible for
the kinetic polarization part in the Derbenev-Kondratenko formula
for the equilibrium polarization. 
Recalling that the Derbenev-Kondratenko formulas belong to the realm of the non-kinetic approach
we thus have translated in Sections 16.1 and 16.2
the kinetic polarization effect from the non-kinetic to
the kinetic realm. 
Thus one may view the kinetic polarization effect independently of the non-kinetic
approach in terms of the kinetic approach, namely via the 
sixth term on the rhs  of (\ref{eq:new2.44vcdc0newnew}) (we leave this as a suggestion for the reader!).
Analogously one may view the Sokolov-Ternov effect independently of the non-kinetic
approach namely in terms of the kinetic approach (via the third and fifth terms on the rhs  of 
(\ref{eq:new2.44vcdc0newnew})).
Moreover one may view the Baier-Katkov correction independently of the non-kinetic approach
namely in terms of the kinetic approach (via the
fourth term on the rhs  of (\ref{eq:new2.44vcdc0newnew})).
Furthermore one may view the radiative polarization effect independently of the non-kinetic approach
namely in terms of the kinetic approach (via the
first and second terms on the rhs of (\ref{eq:new2.44vcdc0newnew})).
In summary, one may view the main physical effects independently of the non-kinetic
approach (namely via the aforementioned terms
on the rhs of (\ref{eq:new2.44vcdc0newnew})).

Eighthly, we recall from Sections 16.1 and 16.2
that the terms on the rhs (\ref{eq:new2.44vcdc0newnew})
correspond to terms on the rhs of (\ref{eq:new2.44vvanewfinal})
and to terms on the rhs of (\ref{eq:new2.44vcdmfinal}).
In other words one may view the main physical effects independently of the non-kinetic
approach namely in terms of the kinetic approach (via terms
on the rhs of (\ref{eq:new2.44vvanewfinal}) or via terms on the rhs of (\ref{eq:new2.44vcdmfinal})). In summary, we got three equivalent
ways of viewing the main physical effects independently of the non-kinetic
approach (and thus independently of the notion of invariant spin field).

\section{Acknowledgements}
The authors would like to thank Georg Hoffstaetter de Torquat, David Sagan, and Matthew
Signorelli for fruitful discussions.
The work of J.A. and K.H. 
has been 
supported by U.S. Department of Energy, Office of Science,
under Award Numbers DE-SC0018008 and DE-SC0025476. 
The work of J.P.D. has been supported 
under the Tigner Traineeship by U.S. Department of Energy
under award No. DE-SC0024907.
The work of E.H. has been supported 
by U.S. Department of Energy, Office of Science,
under Award Number DE-SC0012704.
%

%


\end{document}